\documentclass[review]{elsarticle}
\usepackage{graphicx}
\usepackage[left=2cm,top=2cm, right=2cm, nohead,nofoot]{geometry}
\usepackage{multirow}
\usepackage{nameref}
\usepackage{amssymb, amsfonts, amsmath}
\usepackage{array}
\usepackage{stfloats}
\usepackage{booktabs, tabularx}
\usepackage{subfig}
\usepackage{alphalph}
\usepackage{ragged2e,placeins}
\graphicspath{{../}}
\usepackage{rotating}

\usepackage[table, dvipsnames]{xcolor}
\definecolor{light-gray}{HTML}{E5E4E2}

\usepackage{setspace}
\usepackage{nicefrac}
\usepackage{lineno,hyperref}
\hypersetup{
     colorlinks   = true,
     citecolor    = blue
}

\biboptions{authoryear}

\journal{Journal of Computers in Human behaviour}

\begin{document}
\begin{frontmatter}
\title{Predicting Demographics, Moral Foundations, and Human Values from Digital Behaviours}

\author[ISI]{Kyriaki Kalimeri\corref{mycorrespondingauthor}}
\author[UBA]{Mariano G. Beir\'{o}\tnoteref{myfootnote1}}
\author[ISI]{Matteo Delfino}
\author[Path]{Robert Raleigh}
\author[ISI]{Ciro Cattuto}

\address[ISI]{Data Science Laboratory,  ISI Foundation,  Turin,  Italy}
\address[UBA]{Universidad de Buenos Aires, Facultad de Ingenier\'{i}a, INTECIN (CONICET), CABA, Argentina}
\address[Path]{PathSight Predictive Science Inc.,  New York City,  US}

\tnotetext[myfootnote1]{Work done when in Data Science Laboratory,  ISI Foundation,  Turin,  Italy}
\cortext[mycorrespondingauthor]{Corresponding author. Email: kalimeri$@$ieee.org}

%
%

\begin{abstract}

Personal electronic devices including smartphones give access to behavioural signals that can be used to learn about the characteristics and preferences of individuals. In this study, we explore the connection between demographic and psychological attributes and the digital behavioural records, for a cohort of 7,633 people, closely representative of the US population with respect to gender, age, geographical distribution, education, and income.
Along with the demographic data, we collected self-reported assessments on validated psychometric questionnaires for moral traits and basic human values, and combined this information with passively collected multi-modal digital data from web browsing behaviour and smartphone usage.
A machine learning framework was then designed to infer both the demographic and psychological attributes from the behavioural data. In a cross-validated setting, our models predicted demographic attributes with good accuracy as measured by the weighted AUROC score (Area Under the Receiver Operating Characteristic), but were less performant for the moral traits and human values. These results call for further investigation, since they are still far from unveiling individuals' psychological fabric. This connection, along with the most predictive features that we provide for each attribute, might prove useful for designing personalised services, communication strategies, and interventions, and can be used to sketch a portrait of people with similar worldview.


\end{abstract}


\end{frontmatter}



\section{Introduction}

Strategic demographic attributes, namely gender and age, have been traditionally employed to gain insights into the population; today scientists, policymakers, and practitioners from diverse disciplines are paying an ever-growing attention to digital data as they offer an alternative, complementary view of the society.  
Computational social science and digital humanities fields emerge exactly from this interdisciplinary interest
to address societal questions timely and in greater scale.
The research questions asked in these fields are complex;  cultural, psychological, and other human factors are entangled within the social phenomena under investigation.

Seeing society through the lens of the digital data provides a unique opportunity to capture accurately the complexity of the human psychometric attributes, shedding light to deeper, socio-culturally relevant descriptors, which yield a more useful understanding of the human behaviour. 
Digital data act efficiently as a proxy to demographic attributes with important applications
ranging from the gender gap tracking (\citealp{Fatehkia2018}) to monitoring (\citealp{Zagheni2017}) and assimilation of migrants (\citealp{Dubois2018}), to unemployment (\citealp{Burke2013,Liorente2015,Bonanomi2017}),
and health monitoring (\citealp{Ginsberg2009}). 
The need for understanding human factors via digital data sparkled an ever-growing interest in automatic recognition of personality traits due to its association with important life aspects, including ``happiness, physical and psychological health, occupational choice, satisfaction, and performance,  community involvement, criminal activity, and political ideology''  (\citealp{Ozer2006}). 
Over the last years, researchers proposed a series of computational personality recognition methods based on a varied set of digital data (see \citealp{Vinciarelli2014,Finnerty2016,Farnadi2016} for a review) demonstrating the feasibility of the approach.

In this study, we aim to advance an integrative view of the person, attempting to infer the moral traits as described by the Moral Foundations Theory (MFT) (\citealp{Haidt2003}) and personal values by the Schwartz theory of human values (SHV) (\citealp{Schwartz1992}).
Such psychological constructs are associated with attitudes and behaviours in complex situations like politics (\citealp{Miles2015}), charitable donations (\citealp{Winterich2012,Hoover2018}), climate change (\citealp{Wolsko2016}), poverty (\citealp{Low2015}), vaccination (\citealp{Amin2017}), or even violent protests  (\citealp{Mooijman2018}); where personality dispositions alone do not suffice to explain our judgements.
But how much do shallow digital records reveal us? Which data sources are the most informative? What ethical implications might emerge?

Attempting to tackle these questions,  we present a large-scale study consisting of spontaneous, ``in-the-wild'', observations of digital behaviours.
A cohort of 7,633 participants, nearly representative to the US census with respect to several demographic variables, was recruited. Our participants allowed access to their digital data  - either desktop web browsing or smartphone data -  for one month. 
Often in computational social science, large-scale studies are based only on observational data; here, we obtained from all participants self-assessments on a series of validated psychometric questionnaires regarding their moral views,  human values, and demographic information. 
We employed the individual digital traces to train classification models for predicting moral traits, human values and demographic attributes. The reported results are cross-validated on the information provided directly by the participants.

Such experimental design allows us to provide a clear view of the attributes, both demographic and psychological, that are easier to predict from proxy digital data while shedding light to the ones that instead remain still challenging.
At the same time, we present a comparative study on the predictive power of each digital data source and modality (web browsing, mobile browsing, application usage) as well as their combination.
The quantitative comparison of performance among different modalities demonstrates the informative power of the mobile web browsing, whose role is essential in bridging online and offline worlds (\citealp{Macy2014}). 

Additionally to the quantitative performance of the predictive models for demographic and psychometric attributes, we take a deeper look into the models, and in particular at the top predictors per attribute, providing a qualitative analysis of the behavioural proxies associated with each attribute. 
We provide interesting insights on the individuals, sketching cyber-cultural profiles based on their interests, cultural elements, and habitual actions, essential for an in-depth understanding of the perception and attitudes of people towards important social issues. 

We tackle discrimination and privacy-preserving issues, pointing out the exact digital proxies employed to differentiate between two individuals both for the demographic and psychological attributes. 
Even though further investigation is required in this domain, we highlight how well attributes can be inferred and by which digital proxies, contributing to the transparency of the predictions and raising awareness around ethical implications (\citealp{Chen2017}).
 
Undoubtedly, disclosing the connection between digital human behaviour and psychometrics gives not only the possibility of endowing machines with a notion of social intelligence but also paves the way to a fairer personalised user experience.
With the important penetration of smartphones in society, the effects of automatically predicting demographics and moral values at scale may help to better and timely understand attitudes and judgements towards current or upcoming issues, preventing phenomena like gender or ethnicity discrimination, while improving the efficiency of communication campaigns in the digital humanities field.

\section{Literature Review on Psychometric and Demographic Attribute Inference}

Researchers across various disciplines including sociology, demography and public health have always been keen on examining how society functions observing populations at scale. Here, we present the most influential studies that tackled the issue of demographic and psychometric attribute prediction employing data sources similar to ours.

\subsection{Inferring Demographics}

There is a huge body of literature regarding demographic attribute prediction from digital data. 
Early studies performed by means of traditional interviews revealed differences in the way men and women use the Internet, showing for example that women were more prone to use email while males preferred web browsing (\citealp{Jackson2001}). 
Attributes like gender and age of Internet users can be predicted from their Web browsing behaviours. Among the most influential studies on the topic, \cite{Hu2007} treated web browsing data as a hidden variable propagating demographic information between different users, \cite{Weber2010} employed a Yahoo! query database to infer gender, income, and ethnicity using web browsing query data from a closely representative sample of the US population, while \cite{Weber2011} analysed the queries submitted from different ZIP codes enriched by US census data and exploited them to highlight differences in user behaviour and search patterns among several demographic groups.

Gender and age can also be inferred using call detail records from smartphone devices over large populations (\citealp{Ying2012}, \citealp{Dong2014}, \citealp{Felbo2015}),  exploiting features as the number of unique contacts, the number of calls, text messages, and the total duration of calls for incoming and outgoing interactions. 
Moreover, several demographic attributes can be predicted from the installed applications on smartphones (\citealp{Malmi2016}).
\cite{Seneviratne2015} predicted the user's gender based only on the list of apps installed by 218 users. 
The automatic prediction of demographic characteristics allows for a deeper understanding of online communication and behavioural patterns, and has various applications: in particular, understanding gender differences when interacting with sponsored search results (\citealp{Jansen2010}) can help build better recommendation systems for e-commerce websites, improving user experience and increasing sales (\citealp{Zhao2014,Wang2016}).

Social media are increasingly used for demographic attribute prediction (see \citealp{Cesare2017} for a review).
Recently, \cite{Zhong2015} investigated the predictive power of location check-ins for inferring users' demographics on the Chinese social network Sina Weibo.
\cite{Mislove2011} assessed the demographics of Twitter users in terms of gender and ethnicity, while others predicted the demographics of users from their Facebook data (\citealp{Bi2013,Kosinski2013,Youyou2015}).

\subsection{Inferring Personality Traits}

Over the last years, researchers attempted to infer personality traits from diverse data sources including text, video, audio, mobile and wearable devices, social media, web queries and others.
The most dominant paradigm in the computational social science community is the Big-Five (BF) or Five-Factor Model (FFM) (\citealp{Costa1992}).
Based on this model,  many recent studies exploited the relationship between automatically extracted behavioural characteristics derived from digital data and the Big-Five personality traits (see \citealp{Vinciarelli2014,Finnerty2016} for literature reviews).
\cite{Butt2008,Chittaranjan2012,deMontjoye2013} assessed the predictability of personal traits via calls and SMS data from smartphone usage while \cite{Xu2016} was based on behavioural patterns of smartphone application adoption.
\cite{Lepri2012} and \cite{Kalimeri2013a} explored the interplay between behavioural traits and the influence of the social context while \cite{Kalimeri2013,Rauthmann2015} assessed a dynamic view of the personality traits, focusing on the ever-changing nature of actual behaviour/personality states and the elusive role of situational factors from mobile sensor data.

More recently, a plethora of studies employed social media data for the prediction of personality traits from the massive popularity of these sites. 
The Twitter platform has been one of the first platforms employed for this task (indicatively \citealp{Golbeck2011, Quercia2011}) and remained popular ever since. \cite{Kosinski2012} studied the role of personality in website preferences, combining personality profiles and website choices from more than 160,000 users and investigating whether different websites attract audiences with different personalities. Facebook Likes have also been employed to predict the personality traits of the users by means of regression models (\citealp{Kosinski2013, Youyou2015}). In both of these works, Openness and Extraversion turn out to be the most predictable traits.  
\cite{Farnadi2016} presented a comparative analysis of state-of-the-art computational personality recognition methods on a varied set of social media ground-truth data from Facebook, Twitter and YouTube, while fusion of different social media data sources is also proven to be informative (\citealp{Skowron2016,Gou2014}).

Personality trait prediction is out of the scope of this study, however, we briefly presented the most critical studies on this topic for a fruitful discussion of our findings later on, since the obtained results prove the feasibility of the approaches but are still low regardless of the digital data employed and the scientific maturity on the topic.

\subsection{Inferring Moral Traits}

Here, we place the focal point on the prediction of moral traits which is an open issue for the research community.
With respect to the dispositional traits of personality, moral traits are considered to be a higher level construct   (\citealp{McAdams1995,McAdams2006}) clarifying how and when dispositions and attitudes towards interpersonal and intergroup processes relate with persuasion and communication narratives.
Studies have pointed out the connection between decision making and moral judgements (\citealp{Weber2015}), essential to understand the perception process of media communication.

In this study, we employ the Moral Foundation Theory (MFT) (\citealp{Haidt2003}), which proposes a set of moral principles that could be considered at a higher level with respect to the dispositional traits of personality expressed in the Five-Factor model (\citealp{Costa1992}).
It  provides insights on the characteristic adaptations of the individuals (\citealp{Haidt2009}) as described by Dan McAdams's three-level account of personality
(dispositional traits,  characteristic adaptations,  and life stories) (\citealp{McAdams1995, McAdams2006}). Focusing on the psychological basis of morality,  it identifies the following five moral foundations (see \citealp{Haidt2007,  Haidt2004});
 \emph{care/harm}, 
 \emph{fairness/cheating}, 
 \emph{loyalty/betrayal}, 
 \emph{authority/subversion}, 
and \emph{purity/degradation}, 
These foundations collapse into two superior foundations,  namely the (1) \emph{individualizing} and (2) \emph{binding} foundations.

Even if in its infancy,  MFT has already impacted not only psychology but various scientific domains (see \citealp{Graham2012} for a review). 
In the computational social science field, Moral Foundation Theory (MFT) (\citealp{Haidt2003}) has been studied only recently. 
\cite{Kim2013} investigated whether digital traces gathered from game play data can reveal fundamental aspects
of a person's sacred values or moral identity pointing out several associations. 
\cite{Lin2017} assessed the predictability of moral traits from linguistic features using a set of manually annotated texts from Twitter. Their task did not consist on predicting a high or low value related to the orientation of each trait (e.g., care vs. harm), but only its presence in the tweet. They reported satisfying results for care, fairness, and authority (average F-score of 74\%), while loyalty and purity were less performant (average F-score of 44\%). By comparing the performance of their models against that of a human annotator during the testing phase, they demonstrate the difficulty of the task also for humans.

It is expected that moral traits are expressed more clearly in verbal rather than non-verbal manner, for instance, direct answers to blog discussions, or personal opinions on the Twitter platform (\citealp{Mooijman2018,Hoover2018}), rather than the fact that the individual visited a certain blog.
The value of our approach is that we try to set a baseline of prediction based on readily available data, which any ISP, mobile carrier, or even mobile application has access to, without being tied to a specific platform such as Facebook or Twitter, while addressing potential implications emerging when making inference on such data.
To the best of our knowledge, the present study is one of the first attempts to predict the moral traits from web and smartphone data.



%

\subsection{Inferring Human Values}
The moral domain is also influenced by virtue ethics,  education,  religion and the ethical codes of specific cultures (\citealp{Haidt2004}).  
The Schwartz Theory of Human Values (SHV) (\citealp{Schwartz1992}) provides us with a rigorous framework which identifies the following ten basic human values:
\emph{self-direction},  
\emph{stimulation},  
 \emph{hedonism},   
\emph{achievement},  
 \emph{power},  
\emph{security},   
 \emph{conformity},  
 \emph{tradition},  
 \emph{benevolence},   
 \emph{universalism}.  
The above ten values can be clustered into four higher order value types,  so-called quadrant values and into two dimensions:
\emph{Openness to change} (self-direction,  stimulation) vs. \emph{Conservation} (security,  conformity,  tradition) and \emph{Self-enhancement} (universalism,  benevolence)  vs. \emph{Self-transcendence} (power,  achievement).
Therefore,  the first dimension captures the conflict between values that emphasise the independence of thought,  action,  and feelings and readiness for change and the values that emphasise order,  self-restriction,  preservation of the past,  and resistance to change. The second dimension captures the conflict between values that emphasise concern for the welfare and interests of others and values that emphasise the pursuit of one's own interests and relative success and dominance over others. Hedonism shares elements of both openness to change and self-enhancement.
According to \cite{Schwartz1992}, values are ``desirable, trans-situational goals, varying in importance that serve as guiding principles in people's lives''.
These values are shown to have cross-cultural validity (\citealp{Schwartz2005}) and have been found to correlate to several behaviours (\citealp{Schwartz2003,Bardi2003}), including among others consumer decisions (\citealp{Grunert1995}), pro-environmental behaviour (\citealp{Soyez2012}) as well as judgements (\citealp{Torelli2009}) while there is a demonstrated association with moral traits of the MFT (\citealp{Parks2015,Feldman2018}). 
In the computational social science domain, there have been a few attempts to predict human values from digital data.
\cite{Chen2014} presented an analysis of associations between human values and word use in online social media.
\cite{Youyou2015} assessed --among other attributes-- the prediction of Schwartz human values from Facebook Likes and only reported these results in the Supplementary Information of their study. Both studies reported low prediction scores.

\section{Methods}

%
\subsection{Data Collection} $7,633$ subjects participated in this study after being informed about its content,  purpose, and privacy policies,  during a probability based recruitment campaign carried out in the United States of America, conducted by a subcontracted marketing company. Informed consent was obtained from all participants enabling the collection,  storage, and treatment of data.
Upon acceptance of the privacy policy\footnote{Undisclosed due to blind review.},
the participants were asked to fill in a series of demographic information and validated psychometric questionnaires.
Moreover, they were requested to allow access to either their basic mobile or desktop traffic data, for a period of one month.
5,008 people (2,823 women) consented the access to their desktop data while 2,625 people (1,544 women) consented access to their mobile data,  hereafter referred to as ``Desktop'' and ``Mobile'' datasets respectively.
All were incentivised to participate in the study.

\subsubsection{Demographic Data}
The intake survey covered basic demographic factors (age,  gender,  ethnicity\footnote{We use the term ``African American'' as a shorthand for
the term ``black or African American'' as used in the official US census. See http://factfinder.census.gov/help/en/}),  geographic factors (home location,  expressed at zip code level),  socioeconomic factors (educational level,  marital status,  parenthood,  wealth,  income),   health-related factors (exercise,  smoke, and weight issues) and political orientation. Table \ref{tab:Demog} presents the complete list of the demographic information gathered, along with the respective range of values for all the 7,633 participants.
Our predictive models were trained to predict each of the distinct values for each demographic attribute collected and reported in Table \ref{tab:Demog}; for instance, for the classification of Political Party our classifier had to choose between four labels, ``democrats", ``republicans", ``libertarians" or ``independent". 

\begin{table}

\small
\centering
\renewcommand{\arraystretch}{1.2}

 \begin{tabularx}{\textwidth}{p{0.15\linewidth}p{0.16\linewidth}p{0.12\linewidth}p{0.15\linewidth}p{0.155\linewidth}p{0.12\linewidth}}\noalign{\smallskip}\noalign{\smallskip}
\toprule
\rowcolor{light-gray}\textbf{Attribute} &{\textbf{Demographic Variables}}&  \textbf{Sample size}&\textbf{Attribute} &{\textbf{Demographic Variables}}&  \textbf{Sample size}\\
\rowcolor{light-gray}  & \textbf{Range} &{($N=7,633$)}& & \textbf{Range} &{($N=7,633$)}\\
\midrule
\cellcolor{light-gray} \textbf{Age} &18-24 & 262 (3.4\%) & \cellcolor{light-gray} \textbf{Political Party}&Democrat & 2,973 (38.9\%) \\
&25-34 & 1,308 (17.1\%) & & Republican & 2,556 (33.5\%) \\
&35-49 & 2,009 (26.3\%) & &Libertarian & 215 (2.8\%) \\
&50-54 & 879 (11.5\%) & &Independent & 1,889 (24.7\%) \\
&55-64 & 1,759 (23\%) &&&\\
& 65Plus & 1,416 (18.6\%) &&&\\
\\
\cellcolor{light-gray} \textbf{Education}&College Graduate & 2,624 (34.4\%) & \cellcolor{light-gray} \textbf{Wealth}&50KLess & 2,246 (29.4\%) \\
&Post Graduate & 2,211 (29\%) & & 50K-100K & 915 (12\%) \\
&Some College & 1,680 (22\%) & &100K-250K & 1,213 (15.9\%) \\
&High-school & 619 (8.1\%) & &250K-500K & 1,303 (17.1\%) \\
&Trade School & 479 (6.3\%) & &500K-1000K&  1,067 (14\%) \\
&&& & 1000KPlus & 889 (11.6\%) \\
\\
\cellcolor{light-gray} \textbf{Ethnicity} &Asian  & 338 (4.4\%) & \cellcolor{light-gray} \textbf{Weight Issues} & No & 4,239 (55.5\%) \\
& African American & 485 (6.3\%) & & Yes & 3,394 (44.5\%) \\
&White & 6,359 (83.3\%) \\
& Hispanic & 347 (4.5\%) \\
\\
\cellcolor{light-gray} \textbf{Exercise}  & No & 3,328 (43.6\%) & \cellcolor{light-gray} \textbf{Parent} & No & 2,533 (33.2\%) \\
& Yes & 4,305 (56.4\%) & &Yes & 5,100 (66.8\%) \\
\\
\cellcolor{light-gray} \textbf{Gender} & Female & 4,367 (57.2\%) & \cellcolor{light-gray} \textbf{Smoker}                   &        No &7,027 (92.1\%) \\
& Male & 3,266 (42.8\%) & & Yes & 606 (7.9\%) \\
\\
\cellcolor{light-gray} \textbf{Income}&20KLess & 342 (4.5\%) & \cellcolor{light-gray} \textbf{Marital Status} &Divorced & 733 (9.6\%) \\
&20K-30K & 492 (6.4\%) & & Single & 1,281 (16.8\%) \\
&30K-50K & 1,222 (16\%) & &Married & 4,690 (61.4\%) \\
& 50K-75K & 1,613 (21.1\%) & &Living Together & 617 (8.1\%) \\
&75K-100K & 1,520 (19.9\%) &&&\\
&100K-150K & 1,581 (20.7\%) &&&\\
&150K-200K & 515 (6.7\%) &&&\\
& 200KPlus & 348 (4.6\%) &&&\\
\bottomrule
\end{tabularx}
\caption{Complete list of the demographic attributes collected and their respective ranges for the entire sample (7,633 participants) that took part in our study. Each option corresponds to a distinct ``label'' for the classification algorithm. For instance, for the classification of Political Party our classifier had to choose between four labels, ``democrats", ``republicans", ``libertarians" or ``independents". 
}
\label{tab:Demog}
\end{table}

\subsubsection{Representativeness of the recruited sample}
\label{sec:Repr}

\begin{figure*}[h]
\centering
\subfloat[Participants Geography]{\includegraphics[width=0.9\textwidth]{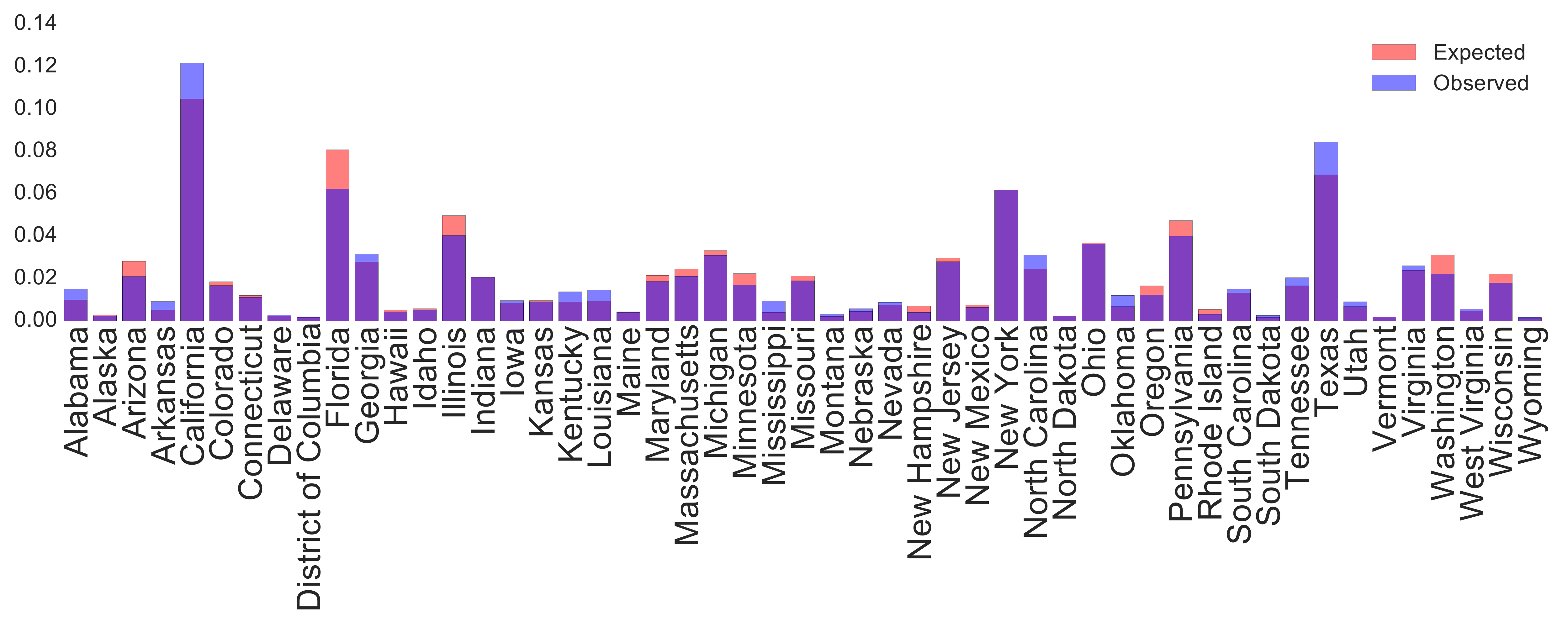}\label{fig:statesample}}\\
\vspace{-10pt}
\subfloat[Mobile Participants Geography]{\includegraphics[width=0.9\textwidth]{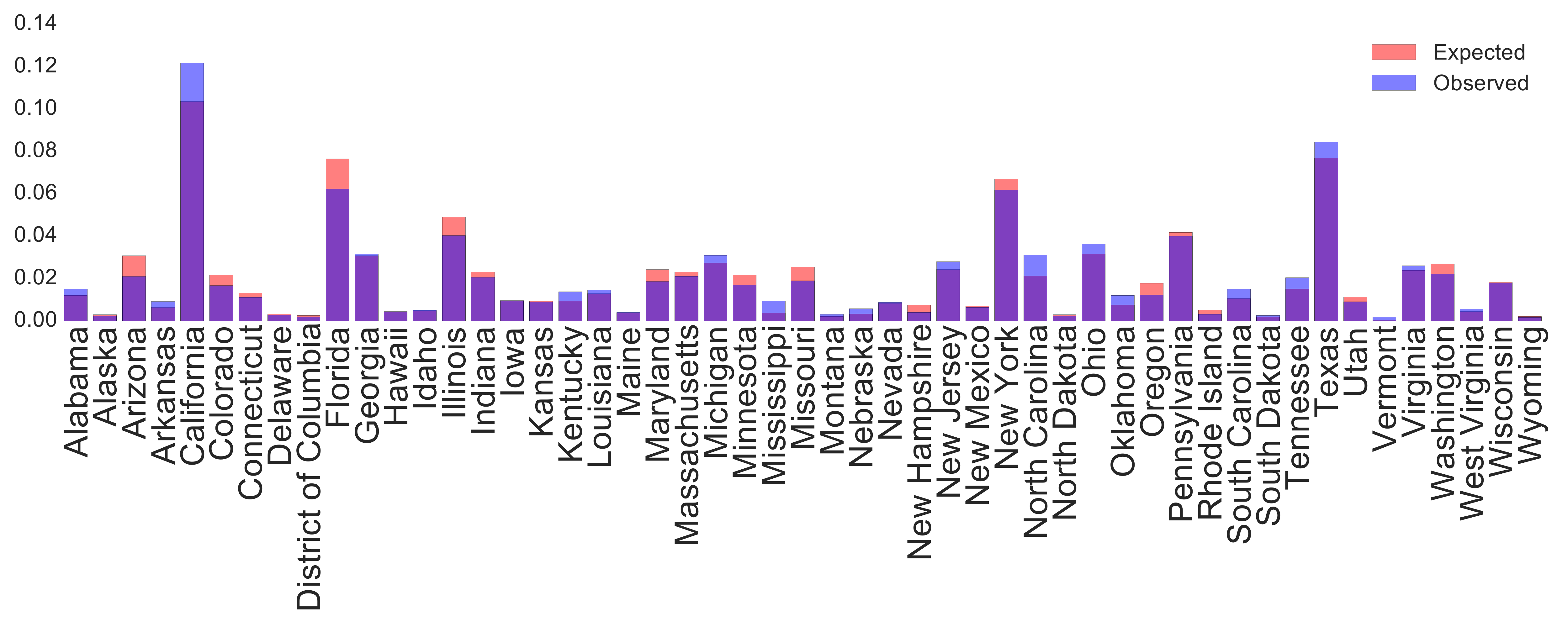}\label{fig:mstatesample}}\\
\vspace{-10pt}
\subfloat[Desktop Participants Geography]{\includegraphics[width=0.9\textwidth]{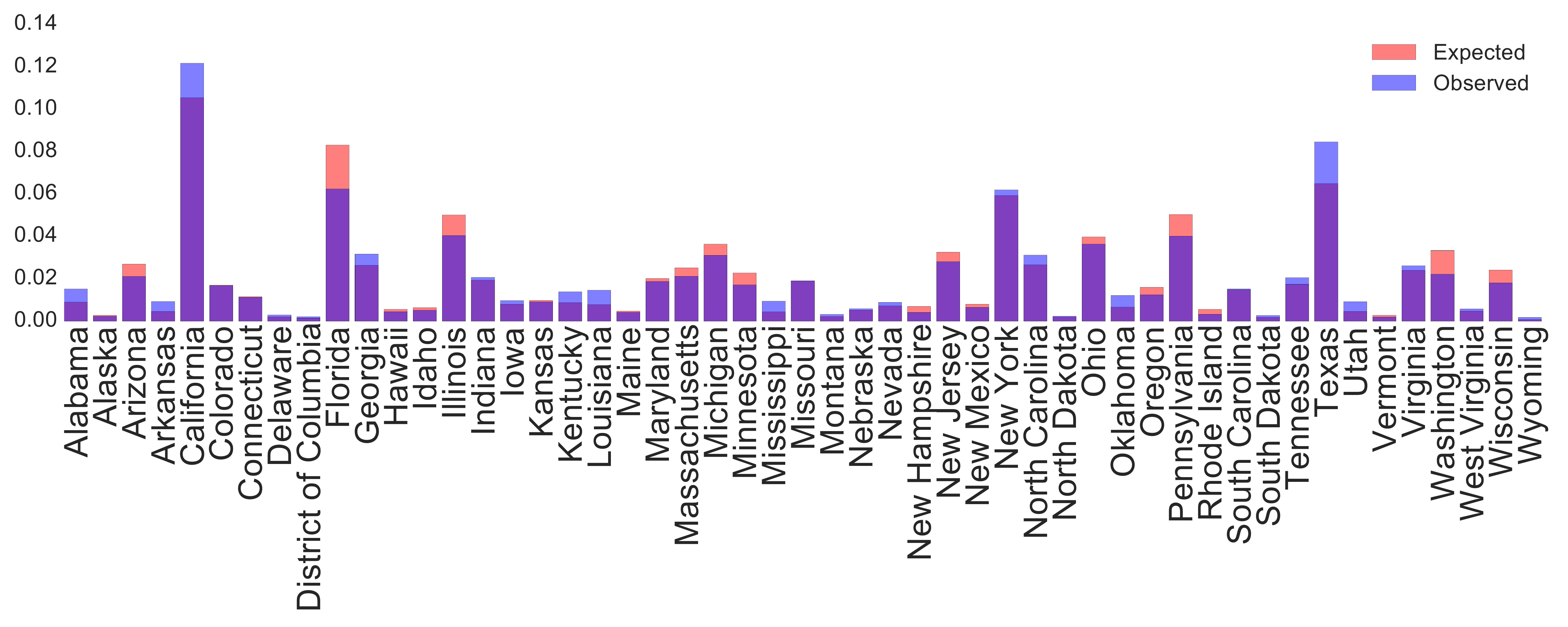}\label{fig:wstatesample}}
\caption{ Distribution of the geographical distribution of the recruited sample (observed values) in the entire dataset (expected values) as compared to the American census (Fig.~\ref{fig:statesample}). The geographical distribution of the mobile (Fig.~\ref{fig:mstatesample}) and the Desktop (Fig.~\ref{fig:wstatesample}) datasets follow closely the geographical distribution of participants.}
\vspace{-10pt}

\label{fig:exp3}
\end{figure*}

The corpus of this study closely follows the American census data not only in terms of gender distribution but also with respect to age groups, education, income and geographical distribution. 
We compared our sample distribution (7,633 participants) to the official American census data (\citealp{uscensus15}) provided by the US Census Bureau for the respective year of the study (2015) and containing information regarding age groups distribution by gender (Table~\ref{table:ageRepr}), education, ethnicity, and income (Table~\ref{tab:DemoCensus}), and geographical distribution (Figure~\ref{fig:exp3}).
We also report breakdown comparisons between the entire sample, the ``Mobile''  (2.625 subjects) and the ``Desktop" dataset (5,008 subjects):
Figure~\ref{fig:exp3}~\subref{fig:statesample} depicts the geographical distribution of the recruited sample (observed values) in the entire corpus as compared to the American Census (expected values). 
Figures~\ref{fig:exp3}~\subref{fig:mstatesample} and ~\ref{fig:exp3}~\subref{fig:wstatesample} present insights on the geographical distributions for the two subsets, ``Mobile''  and ``Desktop". Finally, Table~\ref{table:attribMobDesk} compares the age, gender, education, ethnicity and income distribution between both subsets.
Importantly, all the above attributes of the entire sample employed in this study, as well as the two subsets ``Mobile''  and ``Desktop", are approximating the expected distributions according to the US census.

\begin{table}
\centering
\small
\renewcommand{\arraystretch}{1.2}

\begin{tabular}{p{4.95cm}lcc}\noalign{\smallskip}\noalign{\smallskip}
\toprule
\rowcolor{light-gray} {\textbf{Age distribution by gender}}&  \textbf{US census}& \textbf{Recruited sample}\\
\rowcolor{light-gray}   & & \textbf{($N=7,633$)}\\
\midrule
\rowcolor{light-gray}  Entire sample &&\\
\midrule
	18 to 24				& 13.0\%		& 3.4\% \\
	25 to 34				& 17.6\%		& 17.1\% \\
	35 to 49				& 24.4\%		& 26.3\% \\
	50 to 54				& 9.2\%		& 11.6\% \\
	55 to 64				& 16.7\%		& 23.0\% \\
$\geq$65				& 19.2\%			& 18.6\% \\
\midrule
\rowcolor{light-gray}  Male subset &&\\
\midrule
	18 to 24				& 13.6\%		& 2.1\% \\
	25 to 34				& 18.2\%		& 12.0\% \\
	35 to 49				& 24.9\%		& 25.4\% \\
	50 to 54				& 9.2\%		& 11.2\% \\
	55 to 64				& 16.6\%		& 25.1\% \\
$\geq$65				& 17.5\%			& 24.1\% \\
\midrule
\rowcolor{light-gray}  Female subset &&\\
\midrule
	18 to 24				& 12.3\%		& 4.4\% \\
	25 to 34				& 17.0\%		& 21.0\% \\
	35 to 49				& 23.9\%		& 27.0\% \\
	50 to 54				& 9.1\%		& 11.7\% \\
	55 to 64				& 16.8\%		& 21.5\% \\
$\geq$65				& 20.8\%			& 14.4\% \\
\bottomrule
\end{tabular}
\caption{Age and Gender Representativeness. The distribution of the participants according to their age group and gender shows that the sample represents closely the demographics of the US. }
\label{table:ageRepr}
\end{table}

\begin{table}
\centering
\small
\renewcommand{\arraystretch}{1.2}

\begin{tabular}{p{4.95cm}lcc}\noalign{\smallskip}\noalign{\smallskip}
\toprule
\rowcolor{light-gray} {\textbf{Demographic Variables}}&  \textbf{US census ($\star$)}& \textbf{Recruited sample}\\
\rowcolor{light-gray}   & & \textbf{($N=7,633$)}\\
\midrule
\rowcolor{light-gray}  Education &&\\
\midrule
 High-school graduate or higher ($\star\star$)
  &  86.7\% & 93\%\\
 Bachelor's degree or higher ($\star\star$) &  29.8\% & 28\%\\
\midrule
\rowcolor{light-gray}  Ethnicity &&\\
\midrule
 Asian & 5.7\% & 4.4\%\\
 African American & 13.3\% & 6.3\%\\
 White &  76.9\% & 83.3\%\\
 Hispanic & 17.8\% & 4.5\%\\
 \midrule
\rowcolor{light-gray}  Income &&\\
\midrule
 Median  income ($\star\star\star$) & \$53K & \$50K - \$75K\\
\bottomrule
\end{tabular}
\caption{Comparative statistics for the demographic attributes,  education,  ethnicity, and income of the entire dataset employed for this study and the US census. US census data are provided by US Census Bureau (2011-2015 American Community Survey 5-Year Estimates ($\star$)). Education refers to the percent of persons age 25 years+
 2011-2015 ($\star\star$) and income is expressed in 2015 US dollars ($\star\star\star$).}
\label{tab:DemoCensus}
\end{table}

\begin{table}
\centering
\small
\renewcommand{\arraystretch}{1.2}

\begin{tabular}{p{4.95cm}lcc}\noalign{\smallskip}\noalign{\smallskip}
\toprule
\rowcolor{light-gray} {\textbf{Demographic Variables}}&  \textbf{Mobile subset}& \textbf{Desktop subset}\\
\midrule
\rowcolor{light-gray}  Age &&\\
\midrule
 	18 to 24		& 5.7\%				& 2.3\% \\
 	25 to 34		& 27.7\%				& 11.6\% \\
 	35 to 49		& 35.2\%				& 21.7\% \\
 	50 to 54		& 10.6\%				& 12.0\% \\
 	55 to 64		& 14.5\%				& 27.5\% \\
$\geq$65			& 6.4\%				& 24.9\% \\
\midrule
\rowcolor{light-gray}  Gender &&\\
\midrule
 	Female		& 58.8\%				& 56.4\% \\
 	Male			& 41.4\%				& 43.6\% \\
\midrule
\rowcolor{light-gray}  Education &&\\
\midrule
 	College graduate		& 36.2\%			& 33.4\% \\
 	Post graduate		& 26.8\%			& 30.1\% \\
 	Some college		& 23.4\%			& 21.3\% \\
 	High-school			& 6.7\%			& 8.8\% \\
 	Trade school		& 6.5\%			& 6.2\% \\
 	Less than high-school	& 0.3\%			& 0.2\% \\
\midrule
\rowcolor{light-gray}  Ethnicity &&\\
\midrule
 	Asian			& 4.6\%				& 4.4\% \\
 	African american	& 8.5\%				& 5.2\% \\
 	White			& 78.5\%				& 85.8\% \\
 	Hispanic		& 6.6\%				& 3.5\% \\
 Other			& 1.8\%				& 1.1\% \\
\midrule
\rowcolor{light-gray}  Income &&\\
\midrule
 	200K or more	& 3.9\%				& 4.9\% \\
 	150K to 200K	& 6.9\%				& 6.7\% \\
 	100K to 150K	& 20.8\%				& 20.6\% \\
 75K to 100K		& 18.7\%				& 20.6\% \\
 50K to 75K		& 21.5\%				& 20.9\% \\
 30K to 50K		& 16.5\%				& 15.7\% \\
 20K to 30K		& 7.0\%				& 6.1\% \\
 Less than 20K	& 4.6\%				& 4.4\% \\
\bottomrule
\end{tabular}
\caption{Comparison of age, gender, education, ethnicity and income distribution between the Mobile and Desktop subsets.}
\label{table:attribMobDesk}
\end{table}

\subsection{Psychometric Measures}


\subsubsection{Moral Foundations}

The five moral foundations from the Moral Foundation Theory (\citealp{Haidt2004, Haidt2007}) were assessed through the Moral Foundations Questionnaire (\citealp{Graham2011}),  a validated measure of the degree to which individuals endorse each of five dimensions:

\begin{itemize}
\item\emph{care/harm},  basic concerns for the suffering of others, including virtues of caring and compassion (Cronbach's alpha = 0.76);
\item\emph{fairness/cheating},  concerns about unfair treatment,  inequality,  and more abstract notions of justice (Cronbach's alpha = 0.74);
\item\emph{loyalty/betrayal},  concerns related to obligations of group membership,  such as loyalty,  self-sacrifice, and vigilance against betrayal (Cronbach's alpha = 0.73);
\item \emph{authority/subversion},  concerns related to social order and the obligations of hierarchical relationships
like obedience,  respect,  and proper role fulfilment (Cronbach's alpha = 0.73);
\item \emph{purity/degradation},  concerns about physical and spiritual contagion,  including virtues of chastity,  wholesomeness, and control of desires (Cronbach's alpha = 0.82).
\end{itemize}

The questionnaire is based on self-assessment evaluations and consists of 30 items, resulting in a unique numerical value from 0-30 per person. According to the MFQ, six items (on a 6-point Likert scale) per foundation were averaged to produce the individuals' scores on each of the five foundations.
All the obtained scores were quantised into two classes, ``Low'' and ``High'', with the threshold placed at the respective median value per foundation. We opted for a binary strategy for all the numeric variables, which is a common practice in the computational social science field to limit phenomena of extremely unbalanced classes.

Additionally, we assessed the two superior foundations into which those five collapse according to \cite{Haidt2007}: \emph{individualizing} (care and fairness) and \emph{binding} (loyalty, authority and purity).  The individualising foundation asserts that the basic constructs of society are the individuals and hence focuses on their protection and fair treatment,
defending their right to pursue their own goals.  In contrast, the binding foundation focuses on group-binding,  based on the respect of leadership and traditions, and the defence of the family as the nucleus of society (\citealp{Haidt2007}).

\subsubsection{Schwartz Basic Human Values}

\noindent The \textit{Schwartz human values}  were assessed by the Portrait Values Questionnaire (\citealp{Schwartz2012}), whose validity across cultures is validated in studies performed on 82 countries and samples belonging to highly diverse geographic,  cultural,  linguistic,  religious,  age,  gender,  and occupational groups. The questionnaire is based on self-assessments resulting in a numerical value per person for each of the ten basic values:

\begin{itemize}
\item \emph{self-direction},  independent thought,  action-choosing,  creating,  exploring (Cronbach's alpha = 0.67);
\item \emph{stimulation},  need for variety and stimulation to maintain an optimal level of activation (Cronbach's alpha = 0.85);
\item  \emph{hedonism},   related to organismic needs and the pleasure associated with satisfying them (Cronbach's alpha = 0.83);
\item \emph{achievement},  personal success through demonstrating competence according to social standards (Cronbach's alpha = 0.86);
\item  \emph{power},  the attainment or preservation of a dominant position within the more general social system (Cronbach's alpha = 0.62);
\item \emph{security},   safety,  harmony,  and stability of society,  of relationships,  and of self (Cronbach's alpha = 0.73);
\item  \emph{conformity},  restraint of actions,  inclinations,  and impulses likely to upset or harm others and violate social expectations or norms (Cronbach's alpha = 0.83);
\item  \emph{tradition},  symbols and practices or groups that represent their shared experience and fate (Cronbach's alpha = 0.60);
\item  \emph{benevolence},   concern for the welfare of close others in everyday interaction (Cronbach's alpha = 0.81);
\item  \emph{universalism},  this value type includes the former maturity value type;  including understanding,  appreciation,  tolerance,  and protection for the welfare of all people and for nature (Cronbach's alpha = 0.80).
\end{itemize}

The questionnaire is based on self-assessment evaluations on a 7-point Likert scale.
Following \cite{Schwartz2012}, we averaged the respective items per value and we accounted for individual differences.
The above ten values can be clustered into four higher order values,  so-called quadrant values and into two dimensions, as the sum of the individual items of which they consist:
\emph{Openness to change} (self-direction,  stimulation) vs. \emph{Conservation} (security,  conformity,  tradition) and \emph{Self-enhancement} (universalism,  benevolence)  vs. \emph{Self-transcendence} (power,  achievement).
Therefore,  the first dimension captures the conflict between values that emphasise the independence of thought,  action,  and feelings and readiness for change and the values that highlight order,  self-restriction,  preservation of the past,  and resistance to change. The second dimension captures the conflict between values that stress concern for the welfare and interests of others and values that emphasise the pursuit of one's own interests and relative success and dominance over others. Hedonism shares elements of both openness to change and self-enhancement.
The obtained scores were quantised into two classes, ``Low'' and ``High'', with the threshold placed at their respective median value as for the MFQ's values.

\subsection{Digital behaviours}

Additionally to the demographic and psychometric questionnaires, our participants were asked to provide us with their digital data for one month; either desktop or mobile and in both cases they were financially incentivised to do so.
As mentioned before, out of the 7,633 participants, 5,008 permitted us to assess their desktop web browsing data and 2,625 to their mobile data. In the following paragraphs, we describe in detail the information contained in these data.

\subsubsection{Desktop Data}

For the participants who permitted the logging of their desktops' web browsing data, 5,008 in total, we captured: (i) the domain names, and (ii) the average time spent online and (iii) the number of visits per day on each domain. All this information was aggregated by day, and only the domain names (and not the page or section of the websites) were stored, to ensure the privacy of the participants.
Having performed an initial exploratory analysis (see \textit{Effects of Quantity and Quality of Available Information} in Section~\ref{sec:effects}), we set the minimum of amount of web domain visits required per subject to $N = 30$; users with a lower number of visits than this, were discarded from the analysis leaving us with a total of 4,781 participants.

\subsubsection{Mobile Data}
\label{sec:mob_data}

Participants were asked to download an application which, upon agreement of the privacy policy, logged their web browsing activity and application usage.
\begin{itemize}
\item \emph{Applications' Data.} Application usage was captured whenever the application was running in the ``foreground''. Foreground usage means an application is open on someone's device,  regardless of whether the application is currently being engaged with or not.
Application usage data included for each participant,  records of the date and timestamp,  the local time zone, and time spent on the application (in seconds).
\item \emph{Mobile browsing Data.}  URL data was captured from the native browser on the subject's device (not any 3rd party browsers). URLs for both secure and non-secure traffic were captured,  though only the URLs'  domain was stored for privacy issues.  Similar to the desktop browsing data, users with a number of visits less than $N = 30$ unique domains were discarded from the analysis leaving us with a total of 2,406 participants. The storage of mobile web browsing data followed the same criterion as for desktop web browsing: only the domain names of the websites were stored to preserve the privacy of the subjects. 
\end{itemize}
 
Noteworthy is the fact that ``Mobile'' and ``Desktop'' browsing data provide the same information they only express different modes of web navigation i.e. mobile vs desktop. 

\subsection{Classification Method: Random Forest}
To automatically identify the morality and human values of the participants having access only to their digital data, we conceived this study as a supervised classification problem.
The digital behavioural traces obtained for each participant were used to predict their individual measures while the performance of the prediction was assessed based on the ground-truth provided by their responses in the questionnaires, and their demographic attributes. 
Ground-truth levels for the moral traits and human values were obtained by considering the self-assessed
questionnaires filled by the subjects without any additional labelling by external judges.
All psychometric measures were thresholded at the median point and treated as binary variables, namely \textit{Low, High}, in a single-class classification scenario.
Instead, for the demographic attributes, we followed a multi-class classification scheme with a one-against-all strategy (\citealp{bishop2006pattern}) whenever the range of possible values was more than two, for instance, the ``age'' attribute, and a single-class approach (similar to the psychometric one) when only two values were possible, for instance the ``gender'' attribute; Table~\ref{tab:Demog} reports all the possible classes per demographic attribute.

We designed a generic experimentation scheme,  aiming to assess and compare the predictive power, i.e. the ability to correctly anticipate unseen data, of (i)
the distinct behavioural sequences alone (desktop web browsing, mobile web browsing, mobile application usage) and (ii) the fusion of the separate mobile modalities (web browsing and apps for the ``Mobile'' dataset).
Our data fusion policy consisted of ``early'' fusion at a feature level;  for each mobile user we aggregated the two separate data vectors, the one expressing the web browsing activity and the one expressing the mobile application usage. 
We opted for a Random Forest (RF) classifier (\citealp{Breiman2001}) due to its ability to deal with sparse data (web browsing activity) and unbalanced labels (see Table~\ref{tab:Demog}).
The input to the RF classifier is a matrix whose rows (\textit{feature vectors}, hereafter) correspond to the activity of each participant,
 its columns to the visited domains or apps by all participants, while the values represent the frequency of visits to the specific domain, or app by the specific participant, if any. 
 Then, the classifier trained on the activity of a subset of subjects, predicts for the new unseen participants, based on their activity, the probability that they are closer to one or the other class label or else the target value (i.e., a specific attribute like ``gender'') as a continuous value between 0.0 (Low) and 1.0 (High).

In this study, we fitted a model for each target employing a five-fold cross-validation procedure in which 4-folds are used for model training, and the remaining one is used to evaluate the predictions, in a round-robin scheme. The aim of the cross-validation step is to use mutually exclusive subjects for the training (i.e., fitting) of the model and avoiding the common issue of overfitting during its consequent evaluation. The predictions were evaluated in terms of the weighted AUROC statistic (see Subsection~\ref{sec:auroc} for a definition of AUROC).
Inside each training stage, the hyper-parameter optimisation was performed using a grid-search procedure; the hyper-parameter combinations explored are reported in Table~\ref{tab:hyper}. 
For each hyper-parameter combination, a random forest model was trained by using an inner five-fold cross-validation inside the training set, and its performance was evaluated against the test set. 
Finally, we report the average AUROC performance over all folds. 
In this way, a data sample is used in the validation phase only once and was never seen during the training phase of the model.

As a final step, we estimated the most predictive behaviours (i.e., features) for each target variable by computing the relative rank (i.e., depth) of the features in the different trees that conform the random forest classifiers, according to the \textit{Gini impurity function} (\citealp{Breiman2001})\footnote{The implementation of the model was performed using code from the scikit-learn project~(\citealp{scikit-learn}).}. The higher the rank of a feature, the higher is its importance for the prediction of the specific target variable; hence the relation of the specific behaviour to the specific target variable.

\begin{table}[h]
\centering
\small
\renewcommand{\arraystretch}{1.2}
\begin{tabular}{p{4.8cm}p{3.8cm}lc}
\toprule
\rowcolor{light-gray} \small{Hyperparameter}&\small{Value Range}\\
\midrule

Number of trees &  150,  300,   600\\
Maximum number of features (N) & {\mbox \{$\frac{1}{2}$,  1,  2\} * $\sqrt{N}$}\\
Maximum tree depth &  5,  7,  15\\
Minimum number of samples in leaf & 5\\
Class weight function &  `entropy',  `gini'\\

\bottomrule
\end{tabular}
\caption{Random Forest hyper-parameter space and respective set of attributes exploited for each model in our experimental scenarios.}
\label{tab:hyper}
\end{table}

\subsection{Accuracy measurement: The weighted AUROC}
\label{sec:auroc}
Model performance was computed in terms of the {\em weighted Area Under the Receiver Operating Characteristic (AUROC)} statistic (\citealp{Hanley1982, fawcett2006introduction, Li2010}).
The AUROC is a performance measure for binary classifiers that uses a discrimination threshold to distinguish between a High and a Low class. Using a discrimination threshold implies that the classifier produces a continuous rank value for each sample, and the samples whose rank is above the threshold are classified as High, while the remaining ones are classified as Low. The AUROC was preferred over the commonly-used {\em accuracy} metric (i.e., the proportion of true positives and true negatives among the total number of samples) as it takes into account the effect of unbalanced labels,  which holds true for most of our demographic attributes (\citealp{Mason2002}). 
In the case of highly unbalanced classes, even if a classifier is unable to discriminate between High and Low, obtaining accuracies close to 1.0 is easier since they are easily achieved simply by predicting always the most common class label, which can lead to misleading results and interpretations.
Instead, in a similar scenario, the AUROC metric would give a result of 0.5 since taking into account the true positive (TPR) and  false positive (FPR) rates, each class would be inversely weighted with its relative frequency.

The AUROC resumes into a single quantity the trade-off between the false positive rate (FPR) or \textit{Type I} error and the true positive rate (TPR) or sensitivity, by measuring the variation of the TPR as a function of the FPR as the classifier's discrimination threshold moves from its minimum to its maximum. The curve followed by this function is called ROC curve, and the AUROC is the area under this curve, bounded between $0$ and $1$. The diagonal line between points $(0,0)$ and $(1,1)$ is called  \textit{no-discrimination line}, as a random classification algorithm (i.e., one that chooses between the High and Low-classes with equal probability) would move over it, and its expected AUROC value would be $0.5$\footnote{We conventionally refer to the AUROC values as percentages throughout this paper, using the notation 50\% instead of $0.5$}. The AUROC score for a perfect classifier is 1, and any AUROC value above $0.5$ is considered \textit{better than random}. 

We remark that the AUROC's definition is equivalent to the probability that a randomly-chosen sample among the High-class ones is given a higher rank than a randomly-chosen sample among the Low-class ones (\citealp{Hanley1982}).
For the multi-class classifications (for instance, income, education, wealth, etc.), we report the weighted AUROC statistic, which is the average of the AUROC's of single-class classifiers (one-against-all approach), weighted by the prevalence of each class (\citealp{fawcett2006introduction, Li2010}).

\section{Effects of Quantity and Quality of Available Information}
\label{sec:effects}
A common strategy followed in classification tasks with sparse data is to rank the feature vectors  according to their total frequency (i.e., the columns containing the most frequently used web domains and/or apps are placed first), improving the learning of the model and hence its prediction accuracy. In this way, the model learns a ``forced'' reality,  requiring at the same time access to large datasets of long-term observations. 
In cases where access to large datasets is not possible or when the training needs to be performed on a thin-slice of observations, the prediction of the model often drops.  
We present here an exploratory analysis,  focusing only on the ``gender'' prediction inferring from the desktop web browsing data.
In the following paragraphs, we aim to compare the predictive power of the Random Forest classification models evidencing the biases due to the quantity or the quality of information available each time.

\subsection{How much digital information is needed for a successful prediction?} \label{sec:thres}
\begin{figure}[h]
\centering
\subfloat[Distribution of unique domains seen by user (in log scale).]{\includegraphics[width=0.6\textwidth]{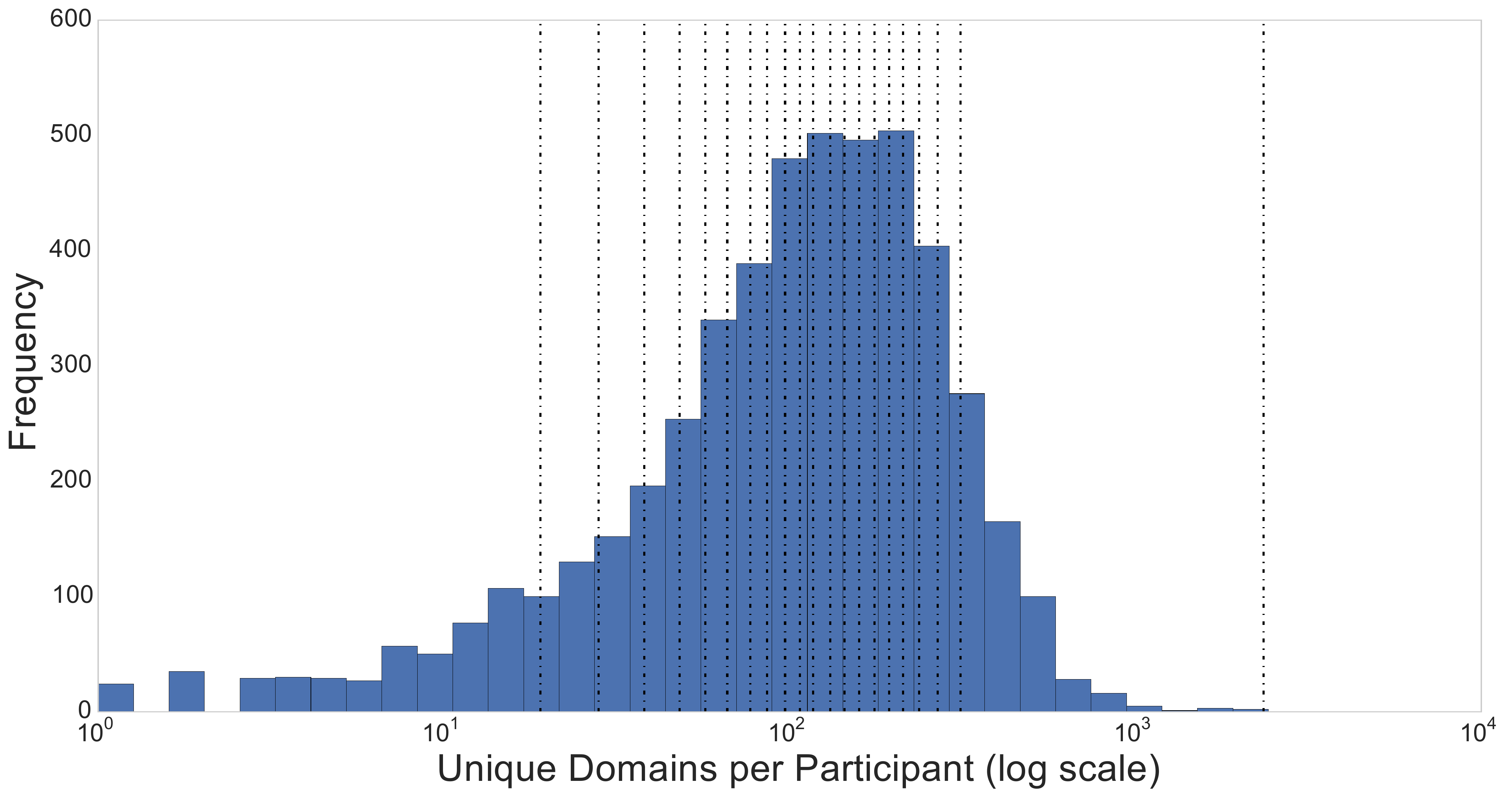}\label{fig:bins}}\\
\subfloat[AUROC values for increasing number of domains.]{\includegraphics[width=0.6\textwidth]{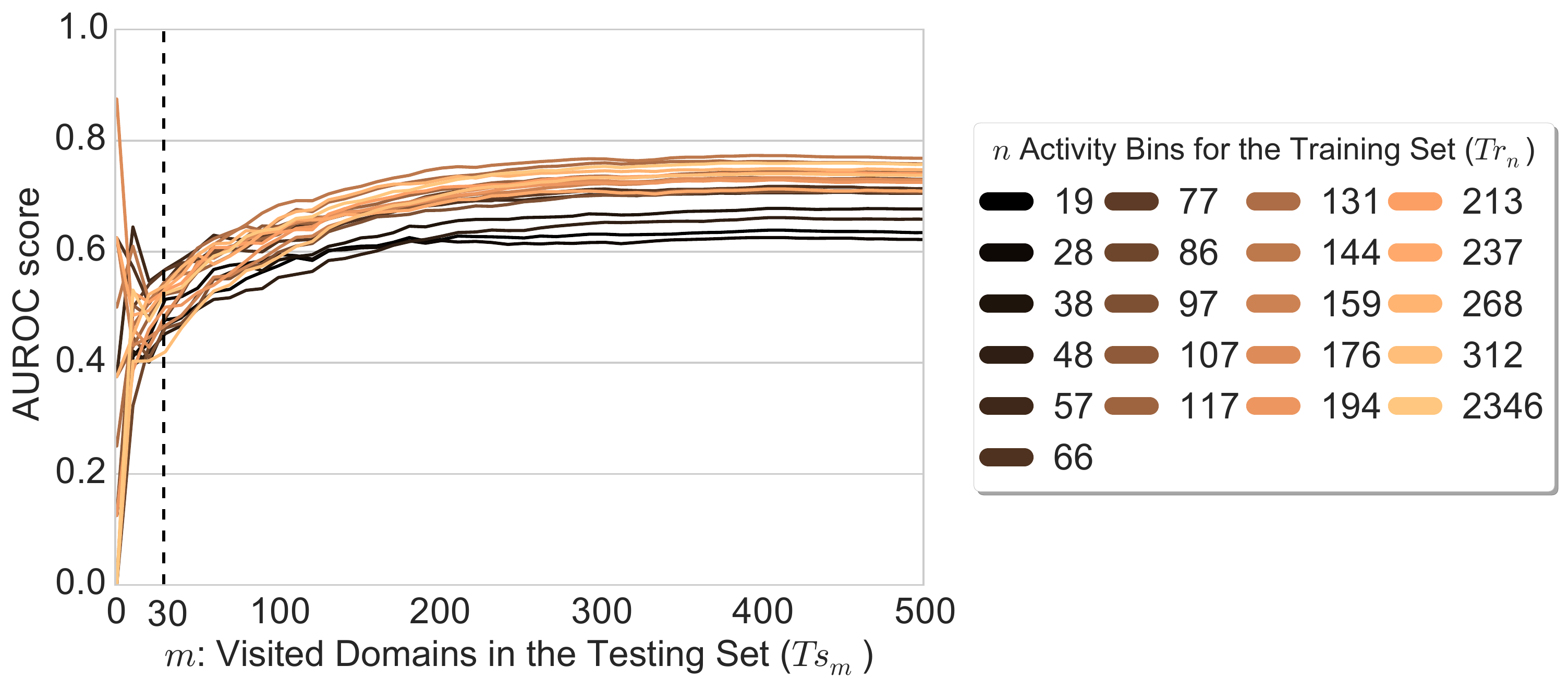}\label{fig:thres}}\\
\subfloat[Proportion of labels in the $Ts_m$.]{\includegraphics[width=0.4\textwidth]{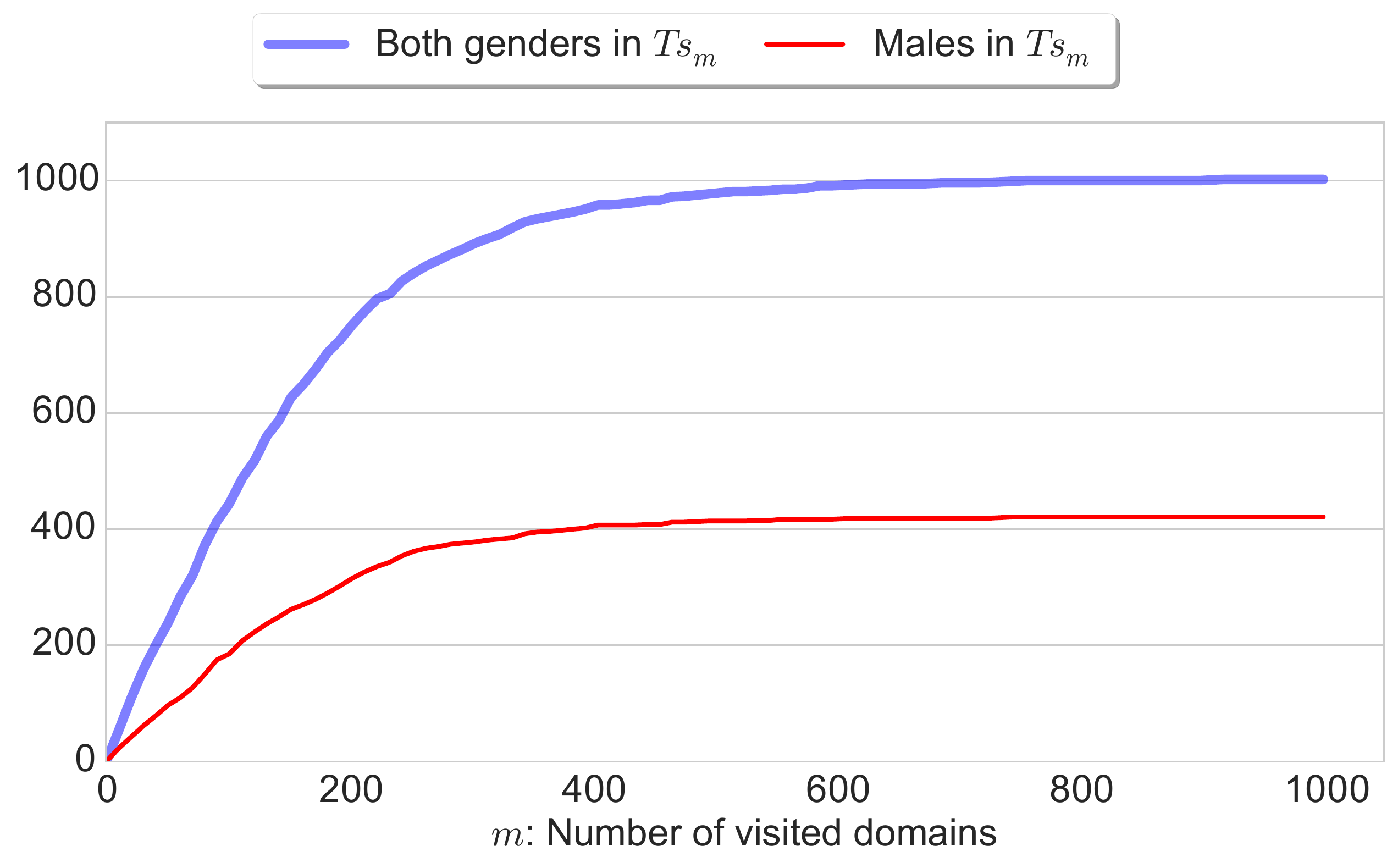}\label{fig:test}}
\subfloat[Proportion of labels in the $Tr_n$.]{\includegraphics[width=0.4\textwidth]{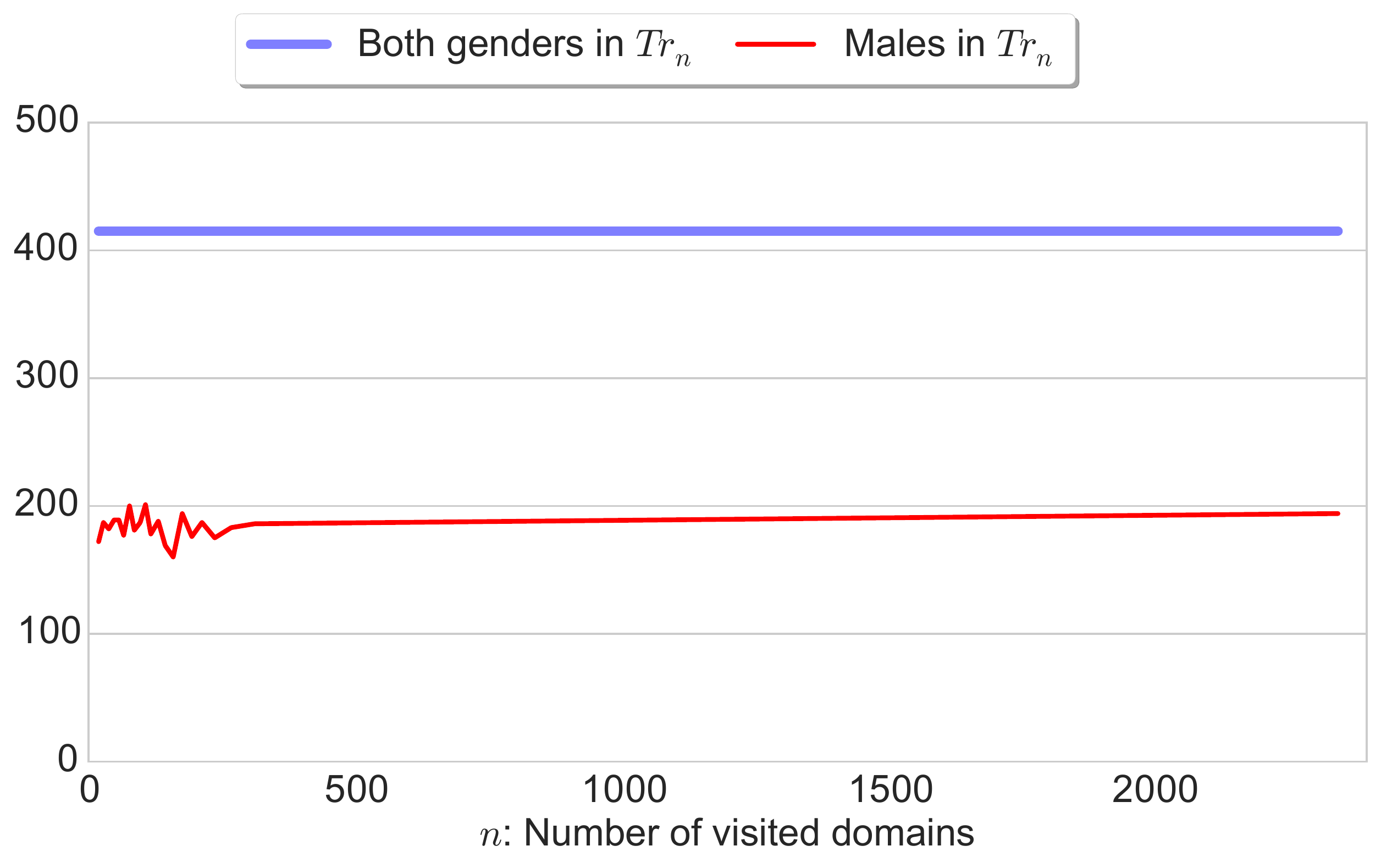}\label{fig:train}}
\caption{Figure~\ref{fig:bins} presents the activity bins in which we segmented the user activity in log scale. 
Figure~\ref{fig:thres} reports the AUROC performances for each trained model  $Tr_n$, with $n$ ranging from $20$ to $2346$, compared against all the testing sets $Ts_m$.  
Each model is represented by a coloured line ranging from darker to brighter colours for increasing values of $n$. 
Note in Figure~\ref{fig:thres}, for $Ts_m$, we plotted only the activity range between $0$ to $500$ to zoom in the initial interval of interest. For $n$ greater than $500$ the behaviour of the model remains invariant.
Figures~\ref{fig:test} and~\ref{fig:train} instead depict the size the testing and training sets respectively.
We note that the size of all training sets, $Tr_n$  is kept fixed for all models.
}
\label{fig:activity_curves}
\end{figure}

To address this question,  the focus of attention is placed on understanding the relationship between the prediction score and the amount of information required to achieve that score.
Our expectation is that the more information we have about an individual the better our models will perform until they reach a saturation level.
To exploit this trade-off we performed an extensive analysis for the prediction of the ``gender'' inferred from the desktop browsing activity, while increasing steadily the amount of information (number of domains) contained in the training set.
In the desktop dataset, there are users that have visited only one domain (referred to as minimum user activity) while others have visited 2,346 unique domains (referred to as maximum user activity).
We split the entire dataset into a training set $Tr$,  consisting of the 80\% of the samples and a testing, $Ts$,  containing the 20\% of the samples.
Leaving the testing set aside, we further split the training set creating 20 subsets one per activity bin $n$,  namely $Tr_n$.
We quantised the total activity of users considering  $n = 20$ activity bins that ranged from the minimum to the maximum user activity (with $n \in [1, 2346]$). 
Figure~\ref{fig:bins} shows in dotted lines the ranges of the activity bins. 
In this way, the first activity bin $n=1$ contains users that have visited up to 19 websites and so forth (see legend of Figure~\ref{fig:thres} for the exact number of maximum domains per activity bin $n$). 
This approach ensures that in each step there is a steady increase of the amount of information in terms of unique domains.
Therefore, in each subset $Tr_n$, are included only the users that have visited, in total, an amount of unique websites that are in the range defined by the activity bin $n$; for 
instance, $Tr_3$ includes all users that have visited less than 38 unique websites.
We trained one Random Forest (RF) model for each training set $Tr_n$,  employing a five-fold validation approach.

We validated each RF model against the testing set $Ts$ which was initially kept apart. 
The amount of information included in the testing set also influences the performance of a classifier, therefore, following a similar logic, we created a series of testing subsets, namely $Ts_m$.
In these subsets where only users that visited at most $m$ domains were included. Overall,  we considered 100 activity bins, $m$,  with $m \in [1, 1002]$ since the maximum user activity in the testing set $Ts$ was 1002 unique domains.

Figure~\ref{fig:thres} depicts the performance of each RF model, as a continuous line whose colour mapping ranges from darker to brighter tones according to the increasing activity bin $n$. Each RF model is trained on a specific $Tr_n$ and validated against all the testing subsets $Ts_m$. 
The more information is contained in the RF model in the training phase the brighter the colour mapping of its performance over the different testing sets.
From this comparison, we can set a minimum activity threshold of web browsing activity that the users should meet for the learning models to be robust.
This threshold may be set approximately at $m = 30$ domains, after which the fluctuations in AUROC metric scores seem to stabilise for all models (see dotted line in Figure~\ref{fig:thres}).
As expected,  the more information is present in the training set the higher the AUROC scores. 
The user activity improves the AUROC score from 70\% to 80\% (see Figure~\ref{fig:thres}); models trained on users that have only visited a few domains (darker colours) are systematically performing worse than the ones trained on users who have visited plenty of domains (models with brighter colours).
Models trained on less than 50 domains (darker lines) deviate significantly from the others, while they fluctuate more intensively when the testing set contains subjects with only few visited web browsing domains $m<30$. 
These observations indicate that a generic activity threshold set at 30 domains benefit the model both in the training and in the testing phase.

Figures~\ref{fig:test} and \ref{fig:train} depict the distribution of labels (males, females) for all the training and testing subsets. 
Worth-mentioning is the fact that the size of the training set is forced to be fixed for all the $Tr_n$ to avoid biases related to the size of the training/testing set instead of its informative content. 
We defined a maximum sample size for men and women, $C_{m}$ and $C_{w}$ respectively, to be included in the respective training sets. 
So for example if in $Ts_1$ we have approximately 200 women and 400 men, then for every $Ts_n$ we would randomly pick 200 women and 400 men. 
Figures~\ref{fig:train} and \ref{fig:test} depict the distribution of the labels for each training $Tr_n$ and testing $Ts_m$ subset.

\subsection{Sensitivity of the model to the quality of available information.}

This section presents an extensive analysis performed only on the most ``active'' users; as ``active'' we define a user that has visited at least $200$ unique domains,  given that the average number of unique domains visited by a user equals to $178$ in the entire desktop dataset. The scope of this analysis is to assess the sensitivity of the  models to the amount of informative content contained in the training data and its effect on the final prediction accuracy. 
We performed this exploration for all the demographic and psychometric attributes.

Initially,  we filtered our entire dataset with a minimum activity threshold set at $thresh = 200$ domains,  maintaining only the highly active users for the rest of our analysis (1,210 participants in total).
Sequentially, we split this into a training set, $Tr\;active$,  and a testing set, $Ts\;active$, consisting of the 80\% and 20\% of the samples, respectively. The $Ts\;active$ is only employed later in the validation phase. 
At this point we want to explore what is the effect of sample size in training in the following two cases; first (i) picking first the most active users (users with a plethora of visits to different websites) and second (ii) randomly picking users regardless of their activity, always from our $Tr\;active$ set.
For the first case, we ranked the users by their number of visits in unique domains, while for the second case the selection is random, hence, their probability of taking part in the training set is the same.

\emph{Case 1: Websites Ranked by Amount of Visits.} 
Having ranked our users according to their web browsing activity, we further split our training set $Tr\;active$ in 20 subsets, namely $Tr_{n}\;freq$, with $n$ uniformly distributed in the interval $[1, 200]$. I.e., for $Tr_{n}\;freq$, the feature vector for each user would contain his/her $n$ most visited websites.
For each attribute, we trained an RF model on each $Tr_{n}\;freq$ employing a five-fold validation approach. 
Successively,  we validated all $Tr_{n}\;freq$ models against the $Ts\;active$ set.

\emph{Case 2: Websites Selected Randomly.} 
Here, we split our initial training set $Tr\;active$ in 20 subsets, namely $Tr_{n}\;random$, with $n$ uniformly distributed in the interval $[1, 200]$. I.e., for $Tr_{n}\;active$, the feature vector for each user would consist on $n$ web domains randomly picked from his/her pool of domains.
For each attribute,  we trained an RF model on each $Tr_{n}\;random$ employing a five-fold validation approach. 
Successively,  we validated all $Tr_{n}\;random$ models against the $Ts\;active$.

\begin{figure*}[h]
\centering
\begin{minipage}{\linewidth}
\centering
\subfloat[Authority]{\includegraphics[scale=0.15]{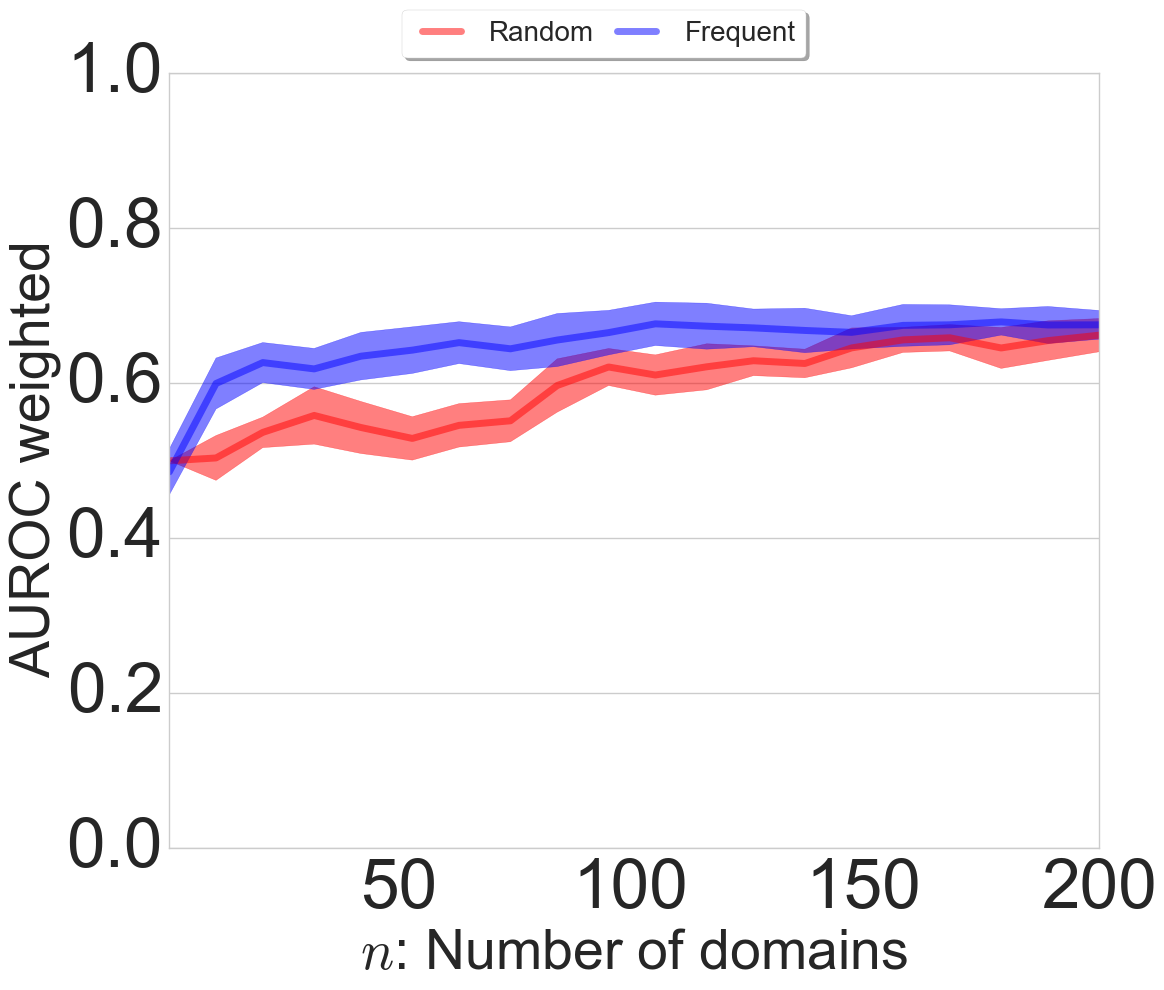}}
\subfloat[Care]{\includegraphics[scale=0.15]{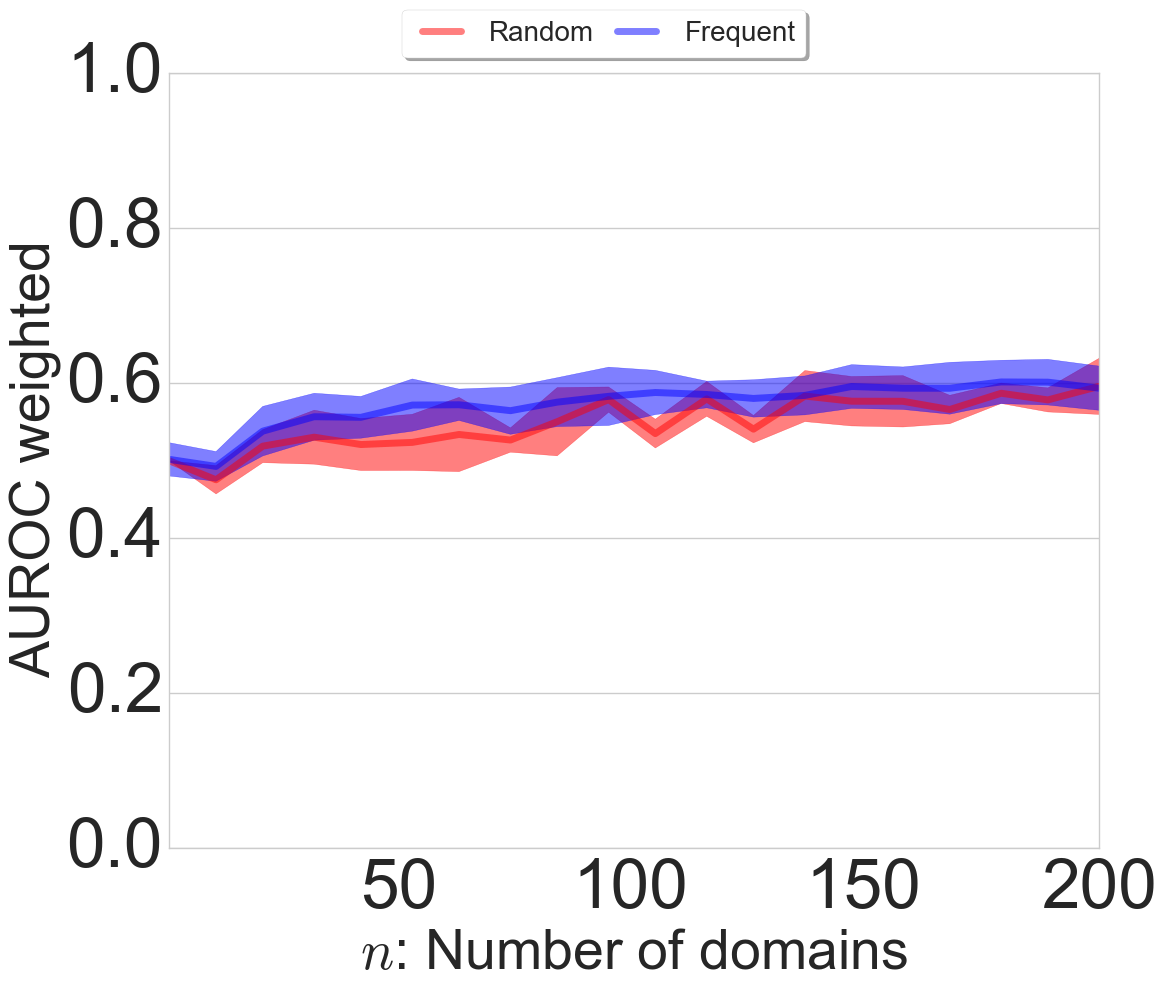}}
\subfloat[Fairness]{\includegraphics[scale=0.15]{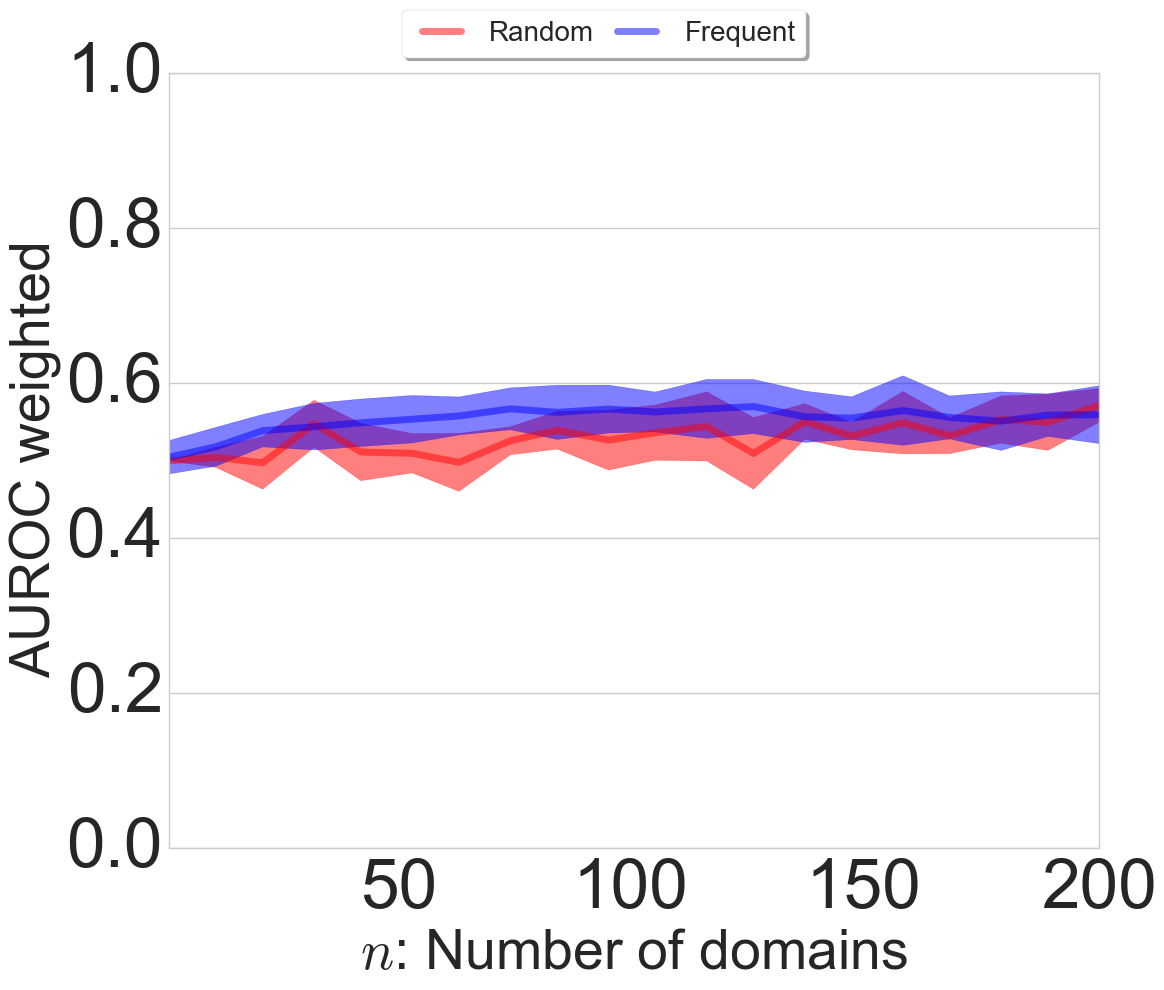}}
\subfloat[Loyalty]{\includegraphics[scale=0.15]{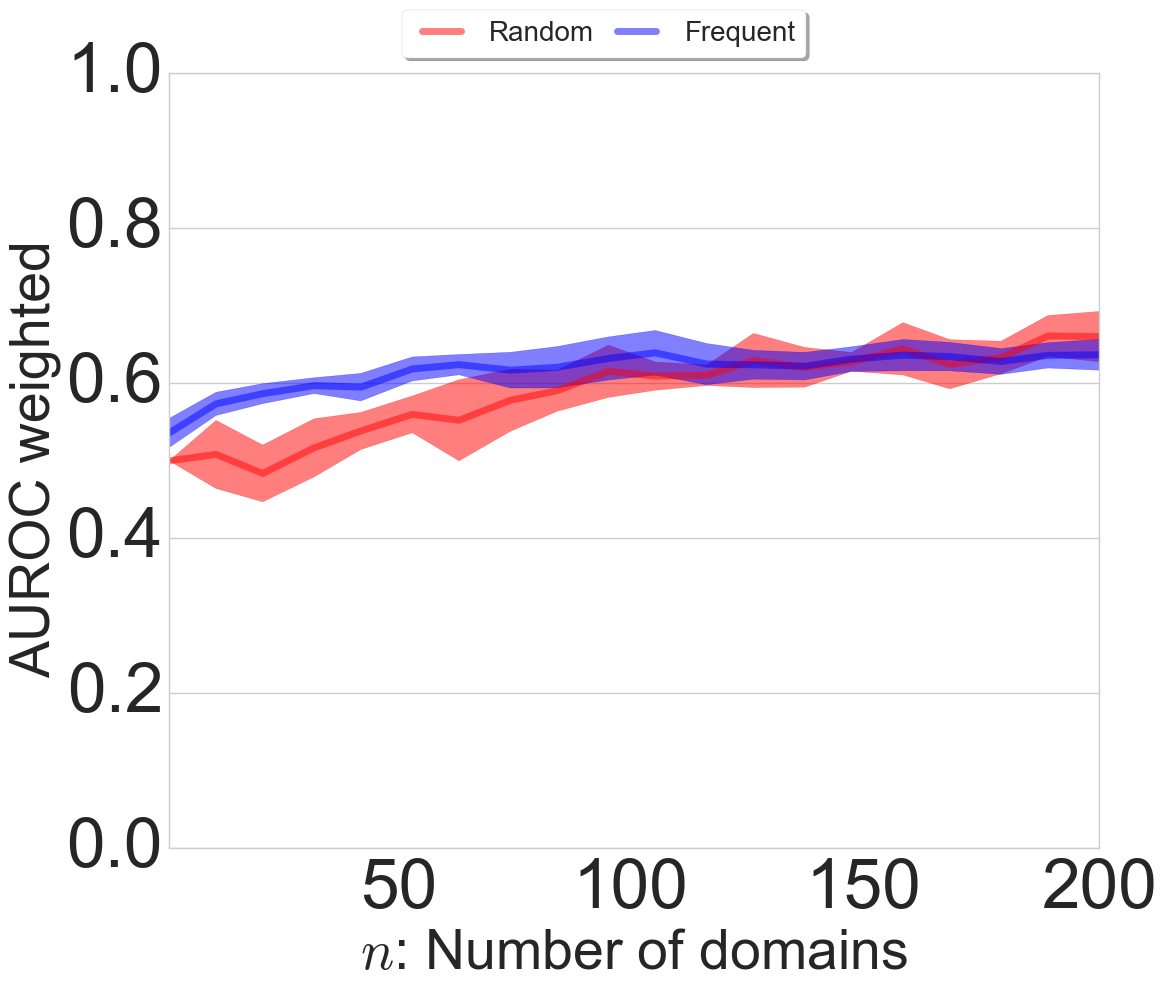}}\\
\subfloat[Purity]{\includegraphics[scale=0.15]{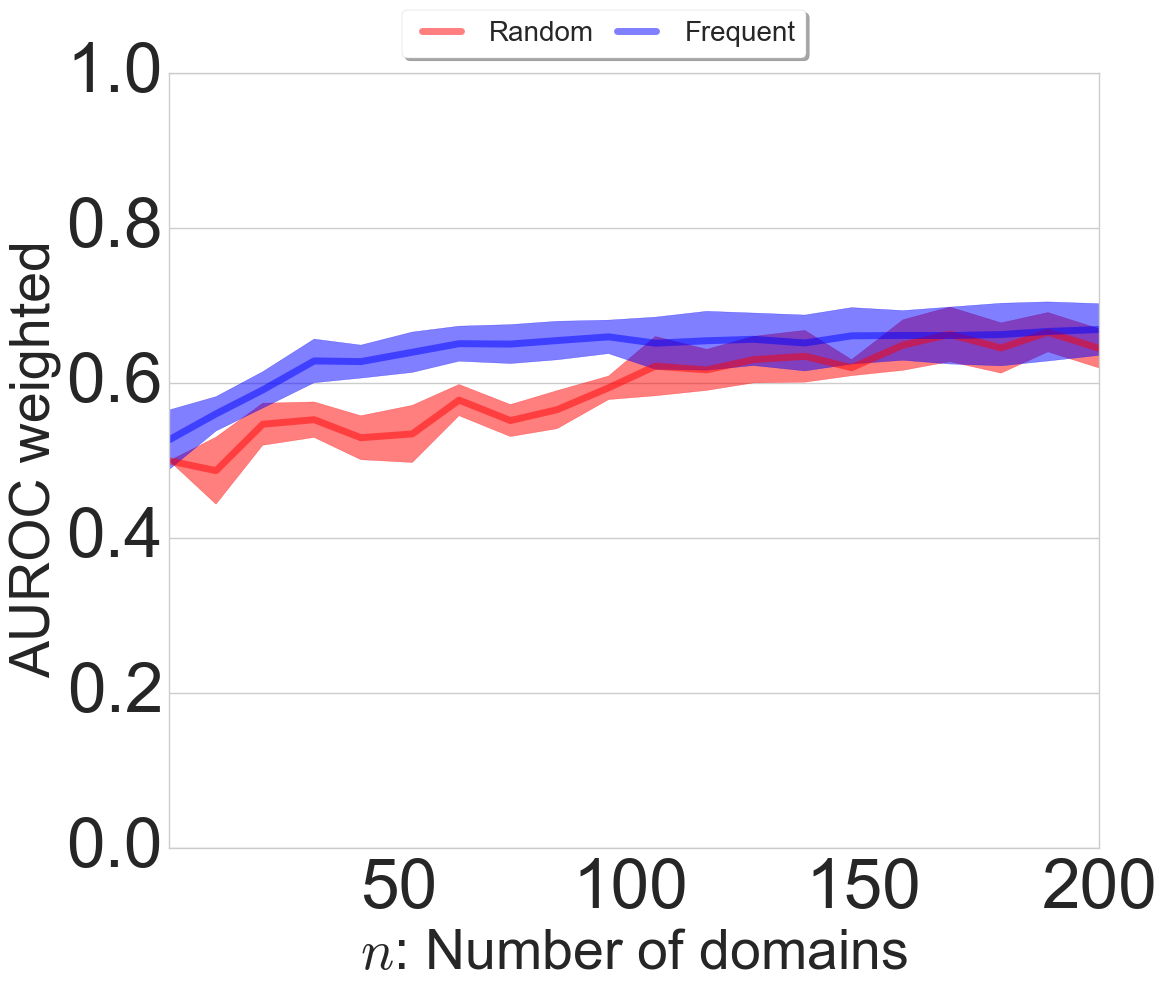}}

\caption{AUROC analysis for the MFT attributes. Increasing the number of domains visited according to their frequency of visit (blue line, \emph{Case 1}) and increasing the number of domains visited regardless of their frequency of appearance (red line, \emph{Case 2}), in the training sets, while always validated on the same testing set.}
\label{fig:RandomChoiceRoc0}
\end{minipage}
\end{figure*}

\begin{figure*}[h]
\begin{minipage}{\linewidth}
\centering
\subfloat[Conservation]{\includegraphics[scale=0.15]{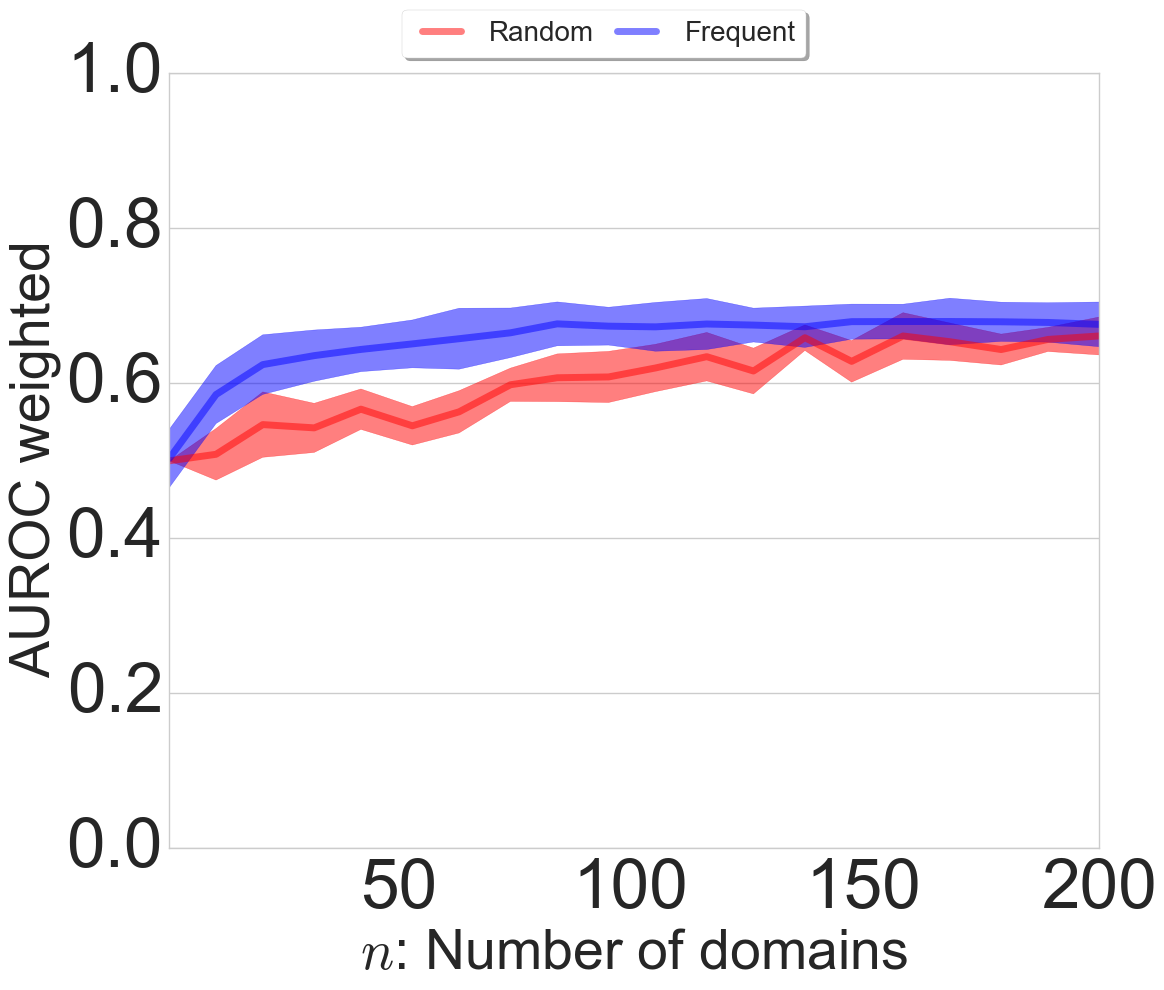}}
\subfloat[Openness]{\includegraphics[scale=0.15]{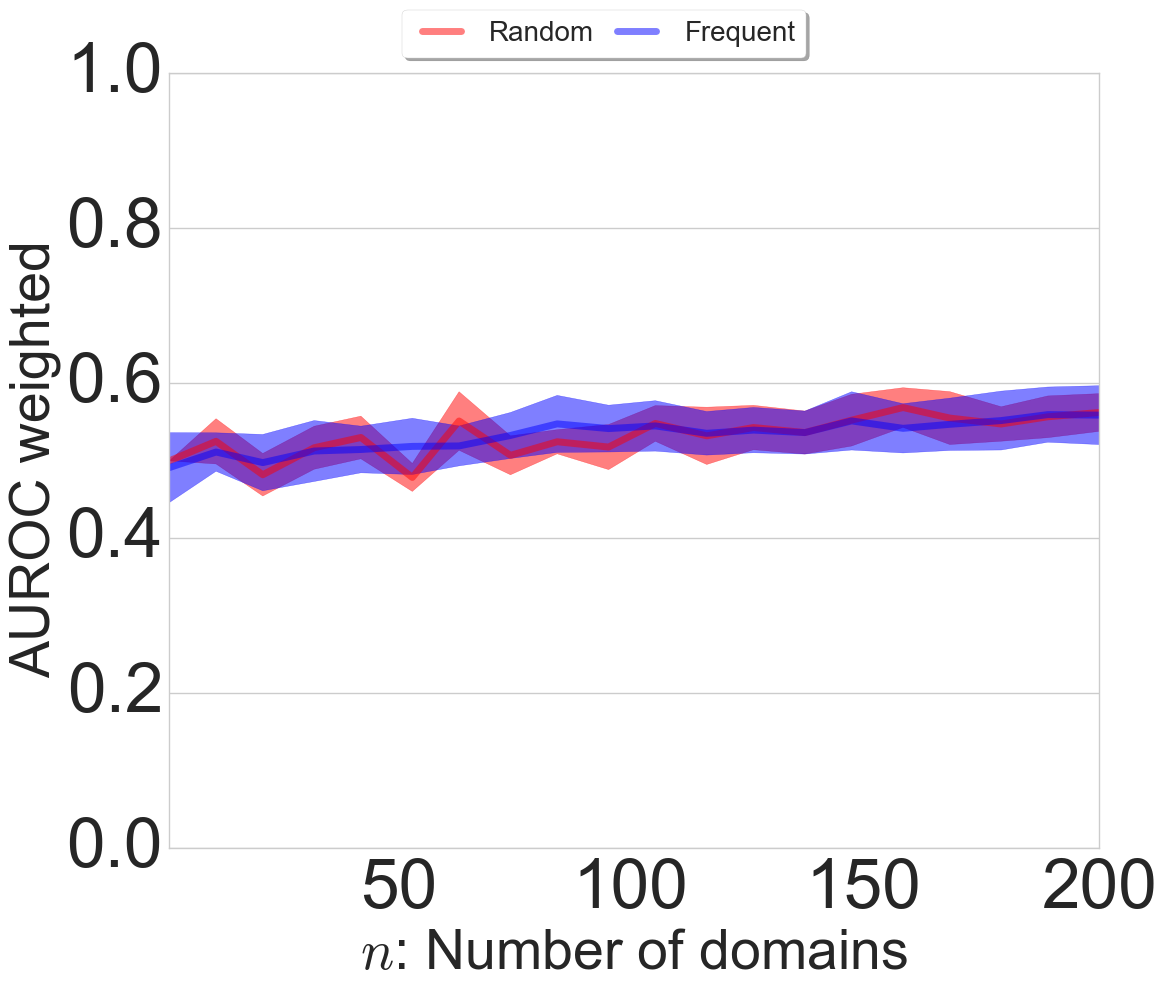}}
\subfloat[Self-enhancement]{\includegraphics[scale=0.15]{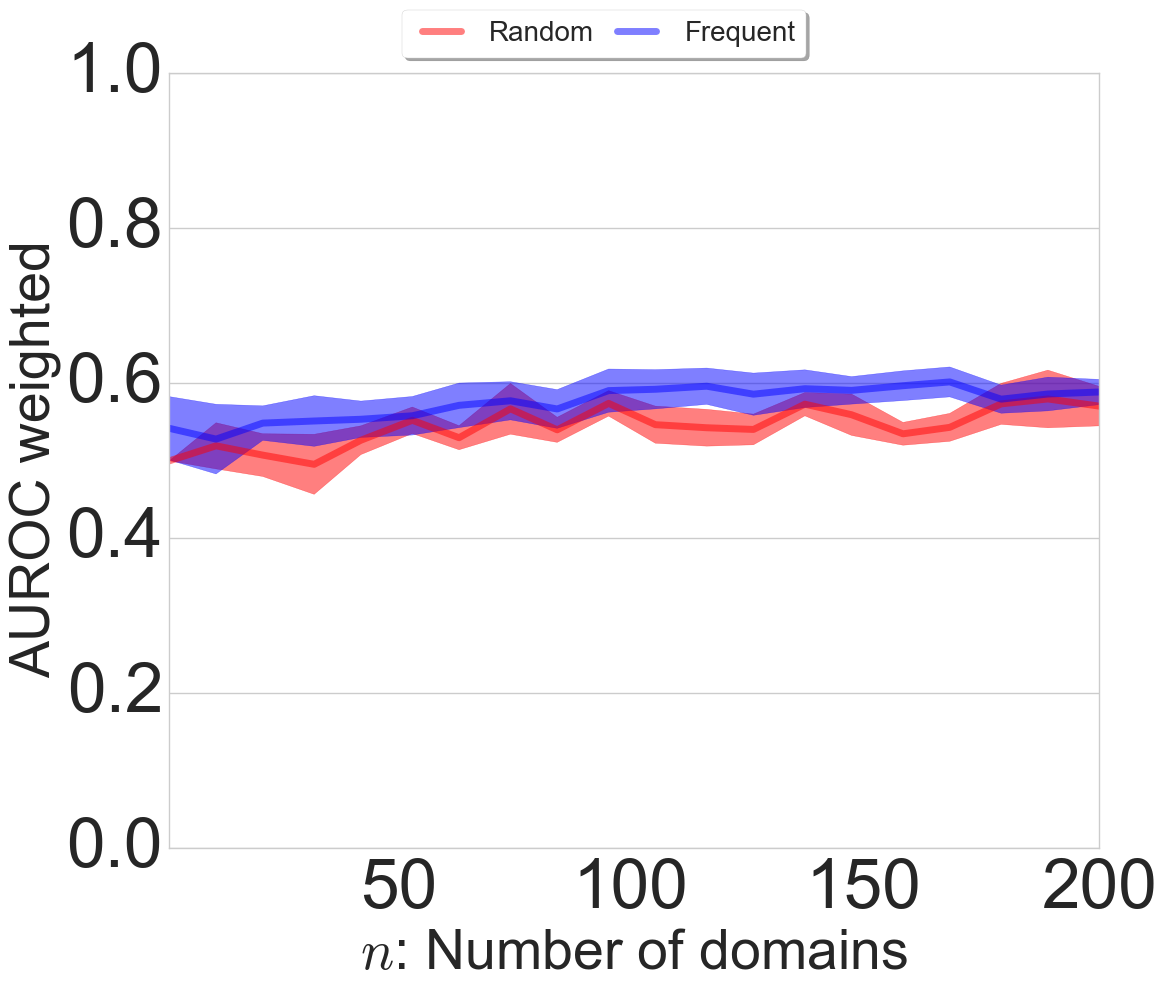}}
\subfloat[Self-transcendence]{\includegraphics[scale=0.15]{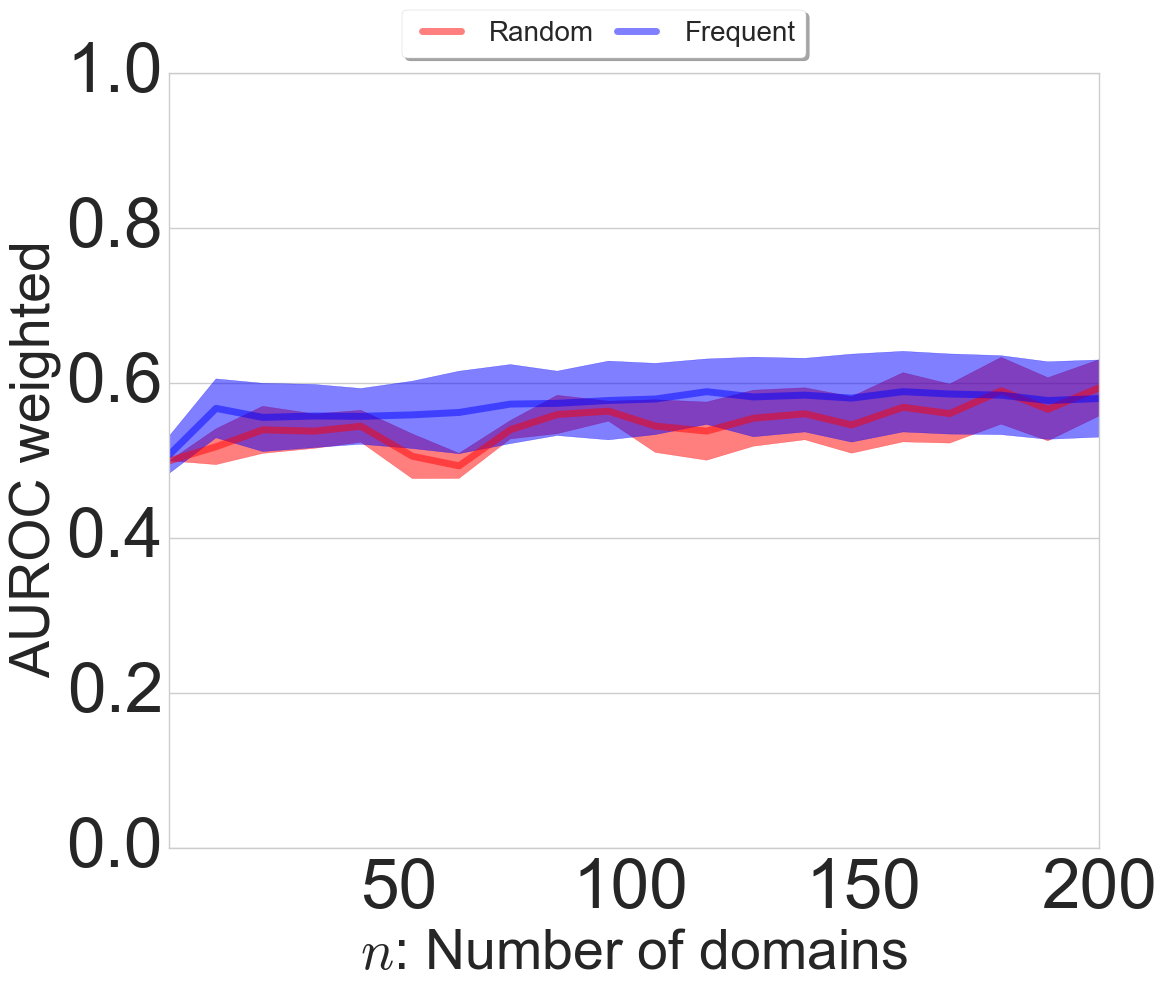}}\\
\caption{AUROC analysis for the Schwartz Quadrants. Increasing the number of domains visited according to their frequency of visit (blue line, \emph{Case 1}) and increasing the number of domains visited regardless of their frequency of appearance (red line, \emph{Case 2}), in the training sets, while always validated on the same testing set.}
\label{fig:RandomChoiceRoc1}
\end{minipage}

\end{figure*}

\begin{figure*}[h]
\centering
\subfloat[Achievement]{\includegraphics[scale=0.15]{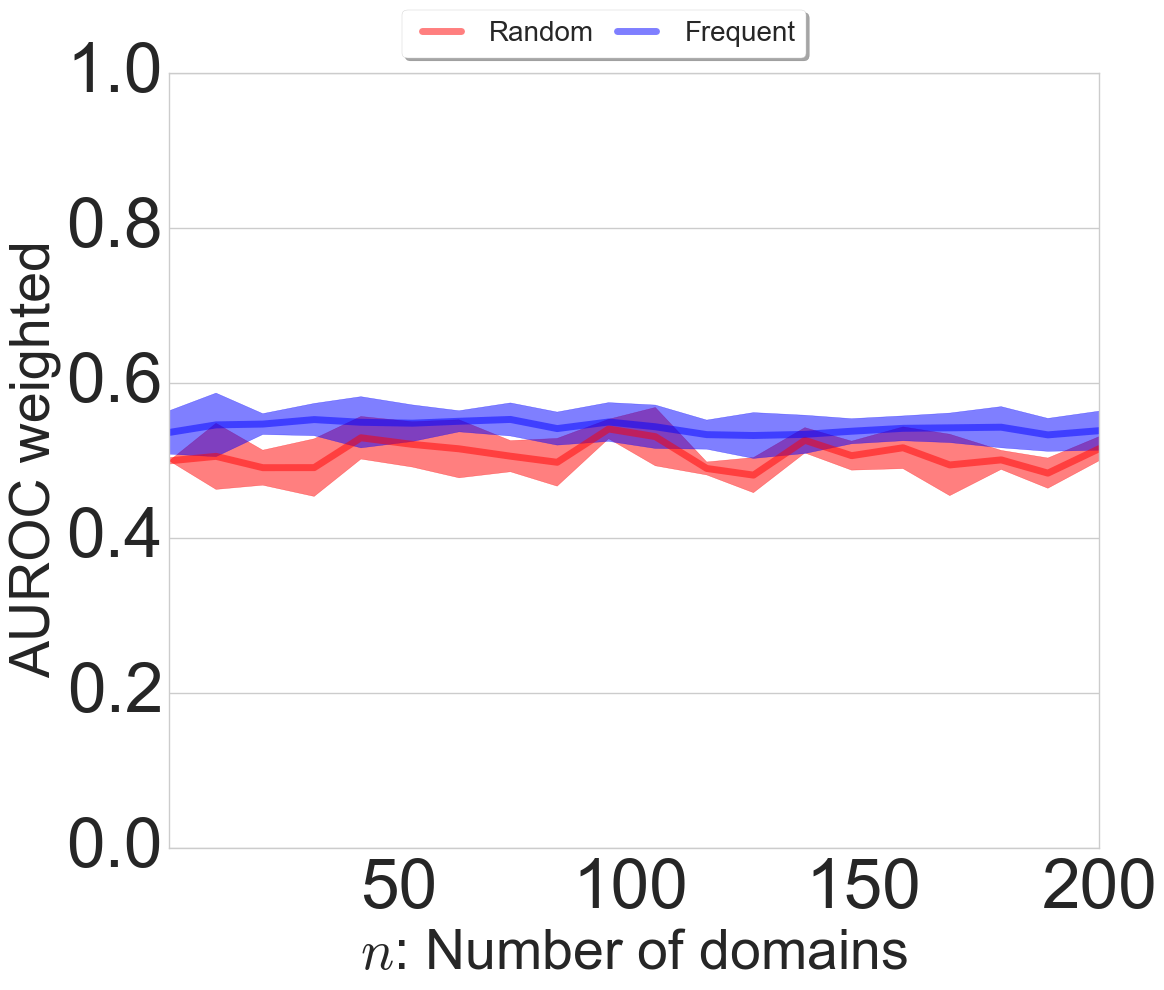}}
\subfloat[Benevolence]{\includegraphics[scale=0.15]{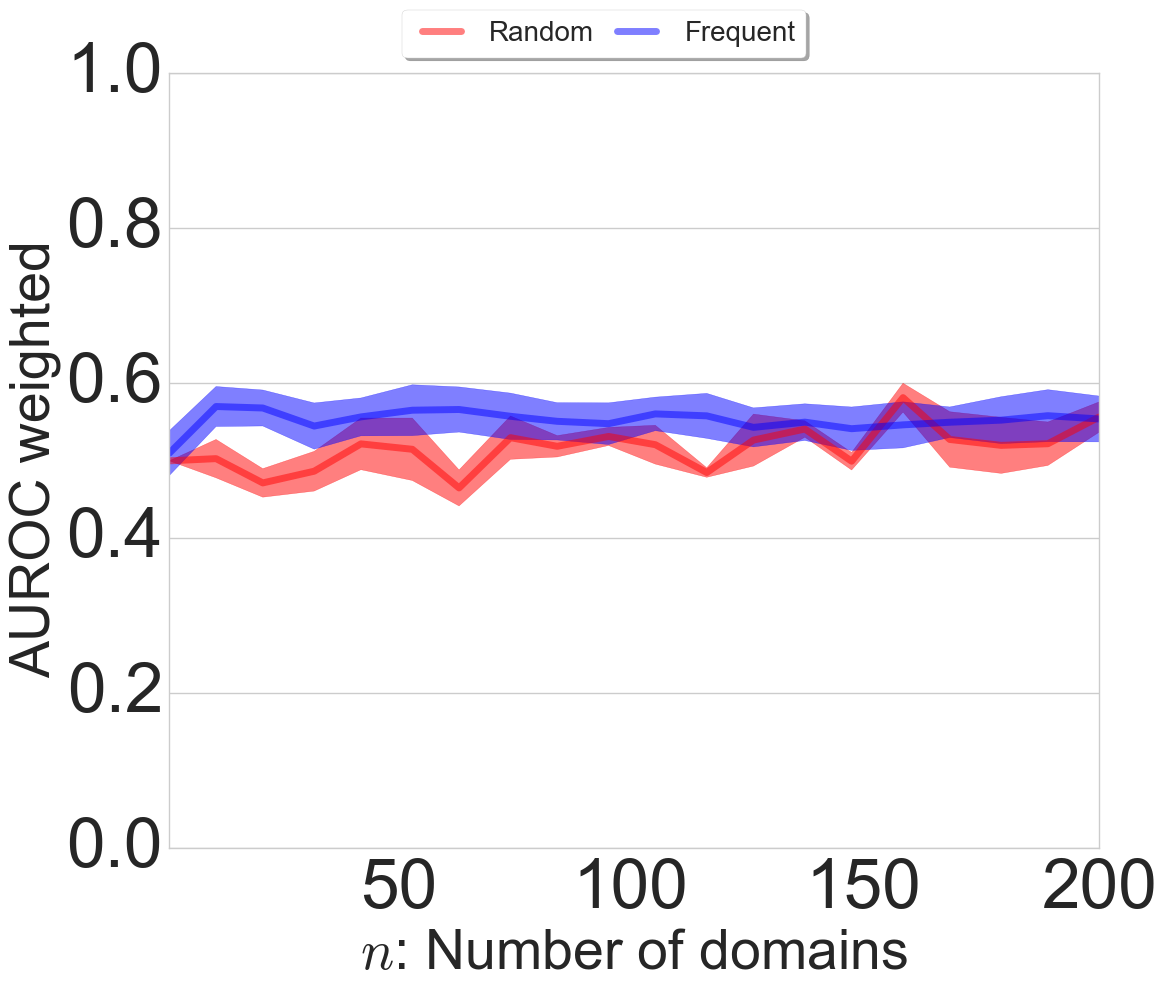}}
\subfloat[Conformity]{\includegraphics[scale=0.15]{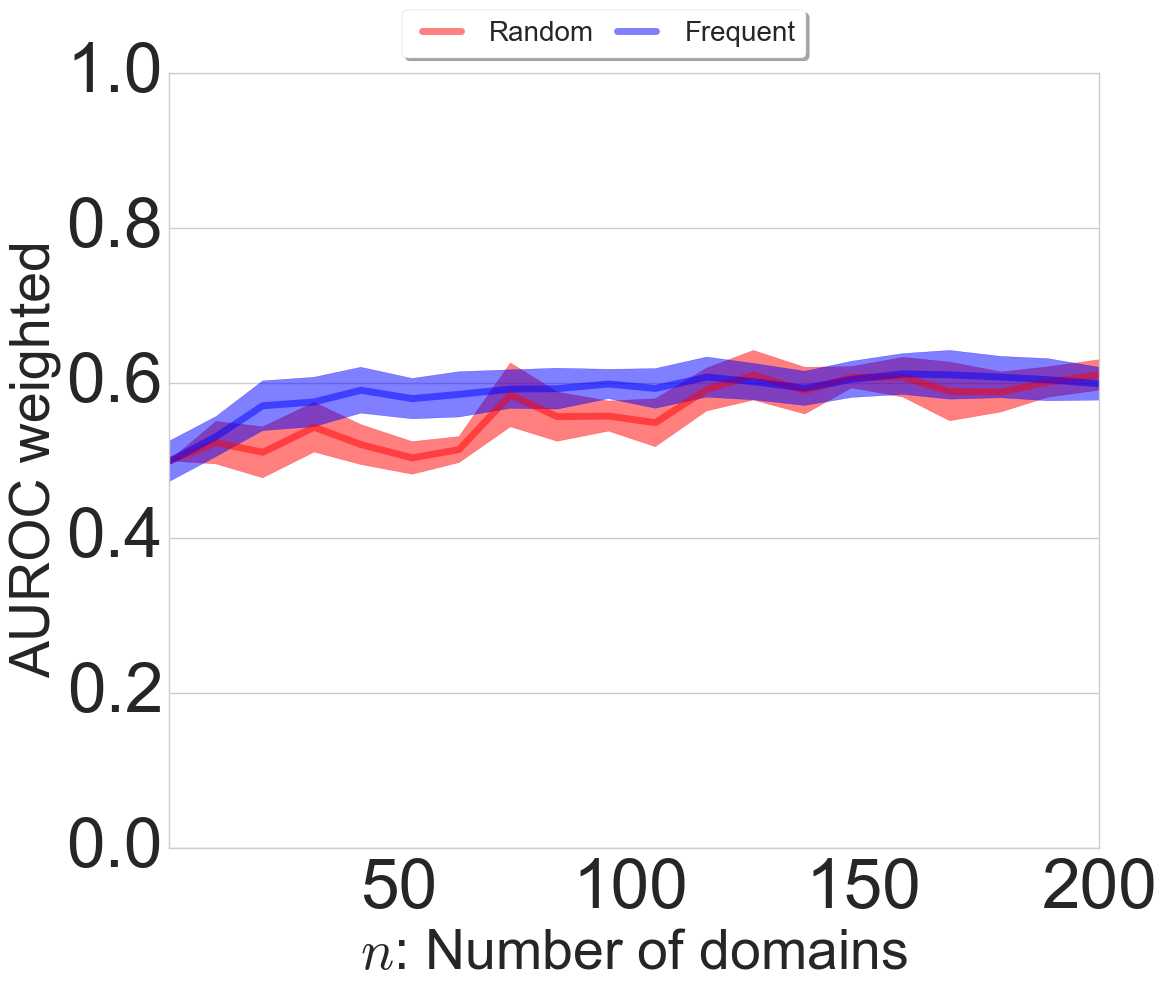}}
\subfloat[Hedonism]{\includegraphics[scale=0.15]{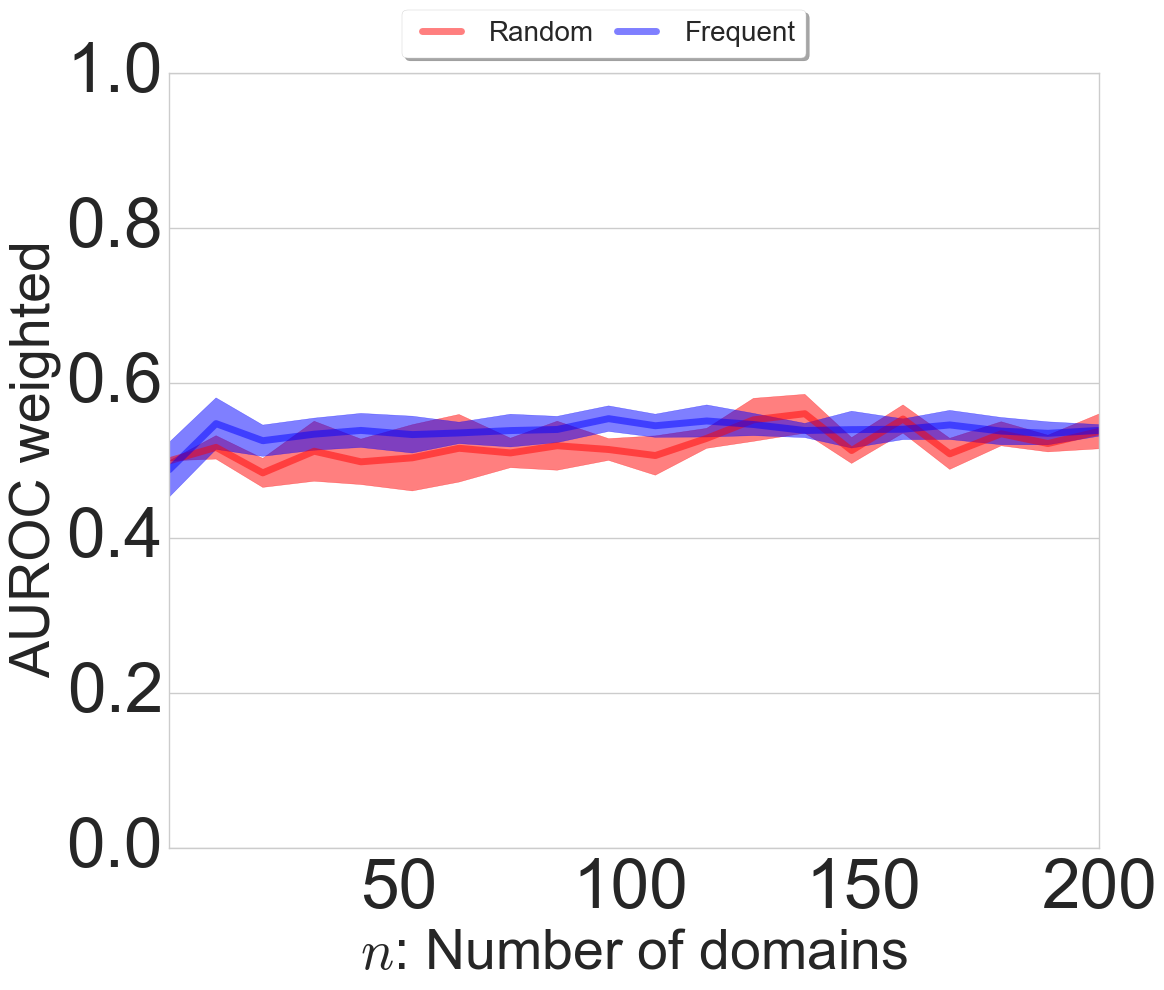}}\\
\subfloat[Power]{\includegraphics[scale=0.15]{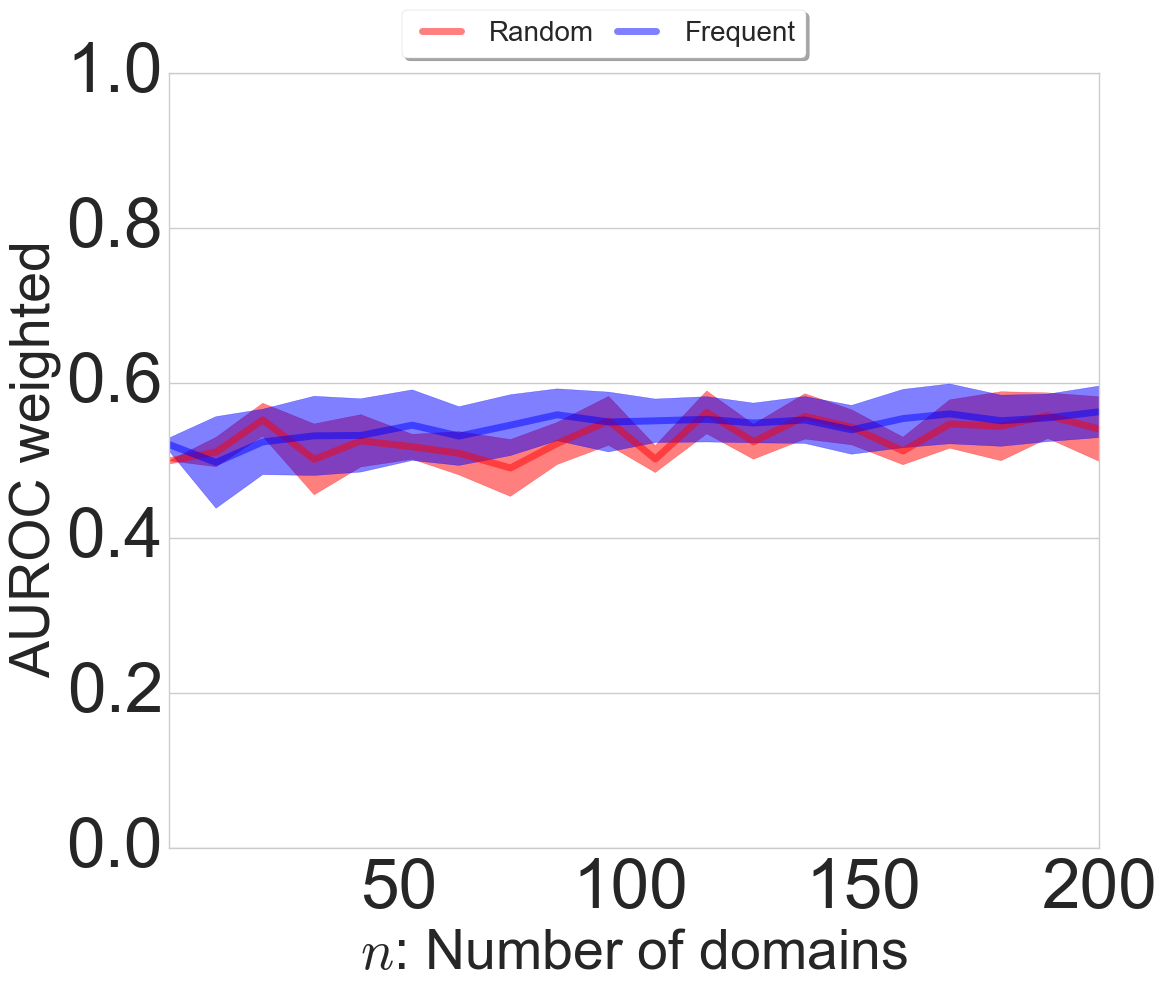}}
\subfloat[Security]{\includegraphics[scale=0.15]{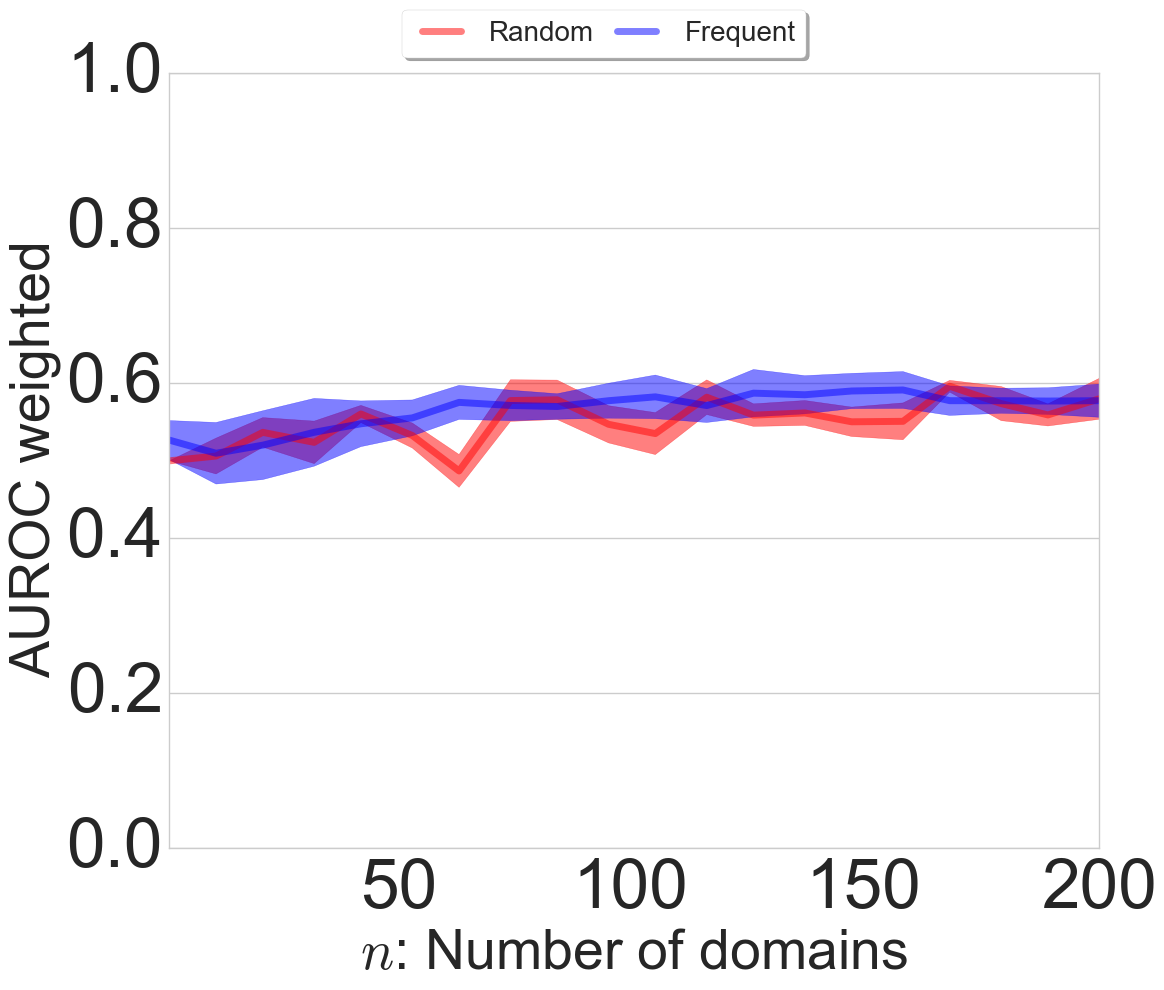}}
\subfloat[self-direction]{\includegraphics[scale=0.15]{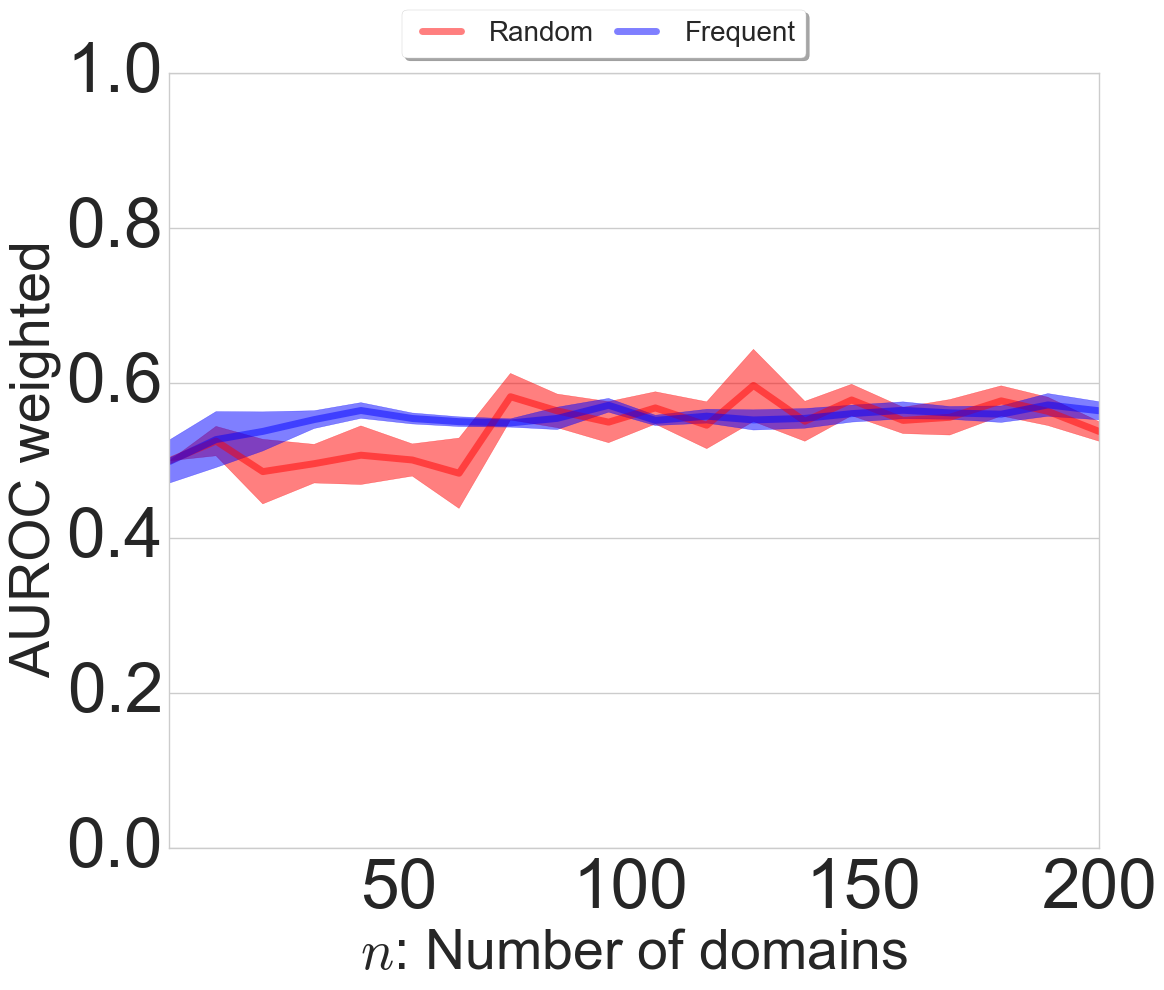}}
\subfloat[Stimulation]{\includegraphics[scale=0.15]{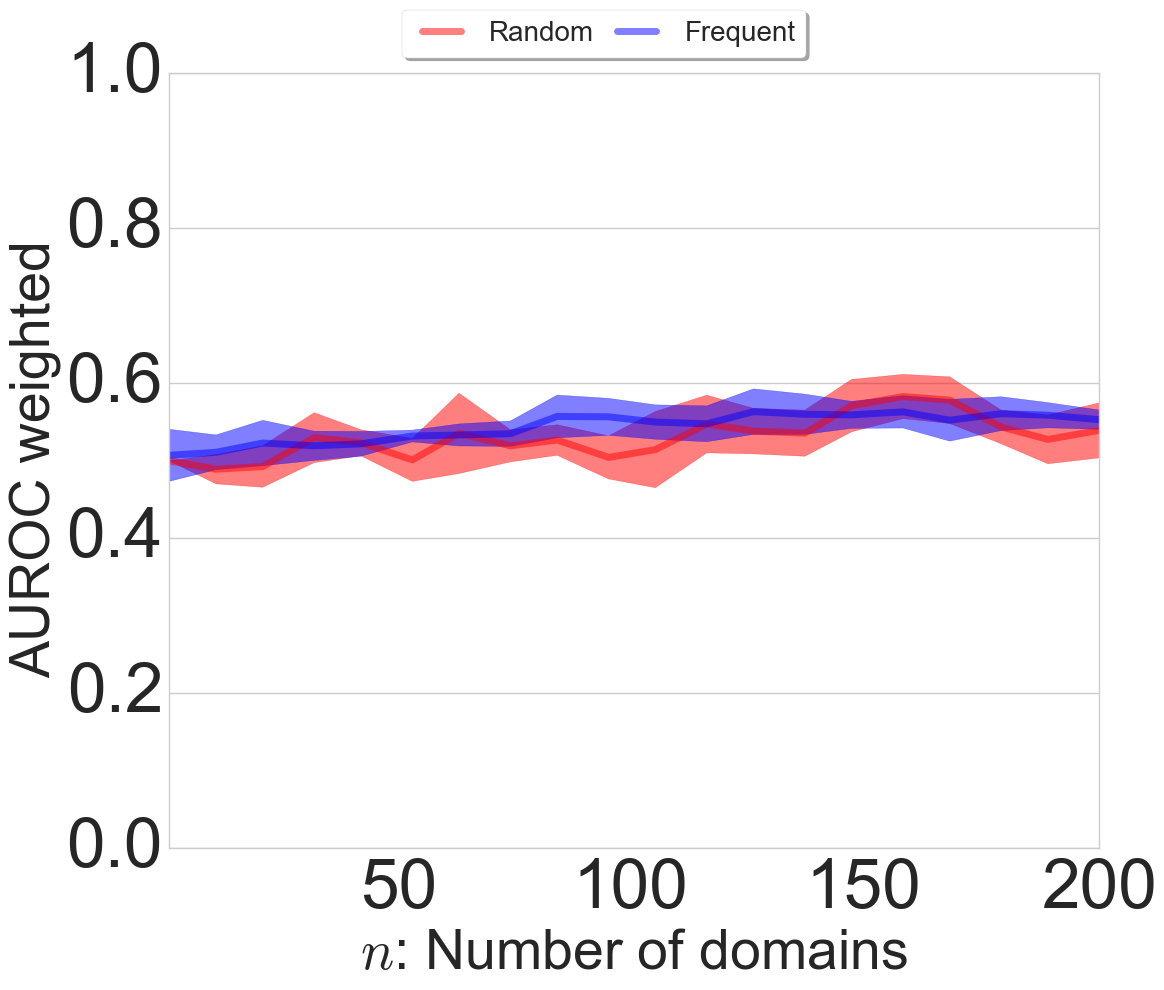}}\\
\subfloat[Tradition]{\includegraphics[scale=0.15]{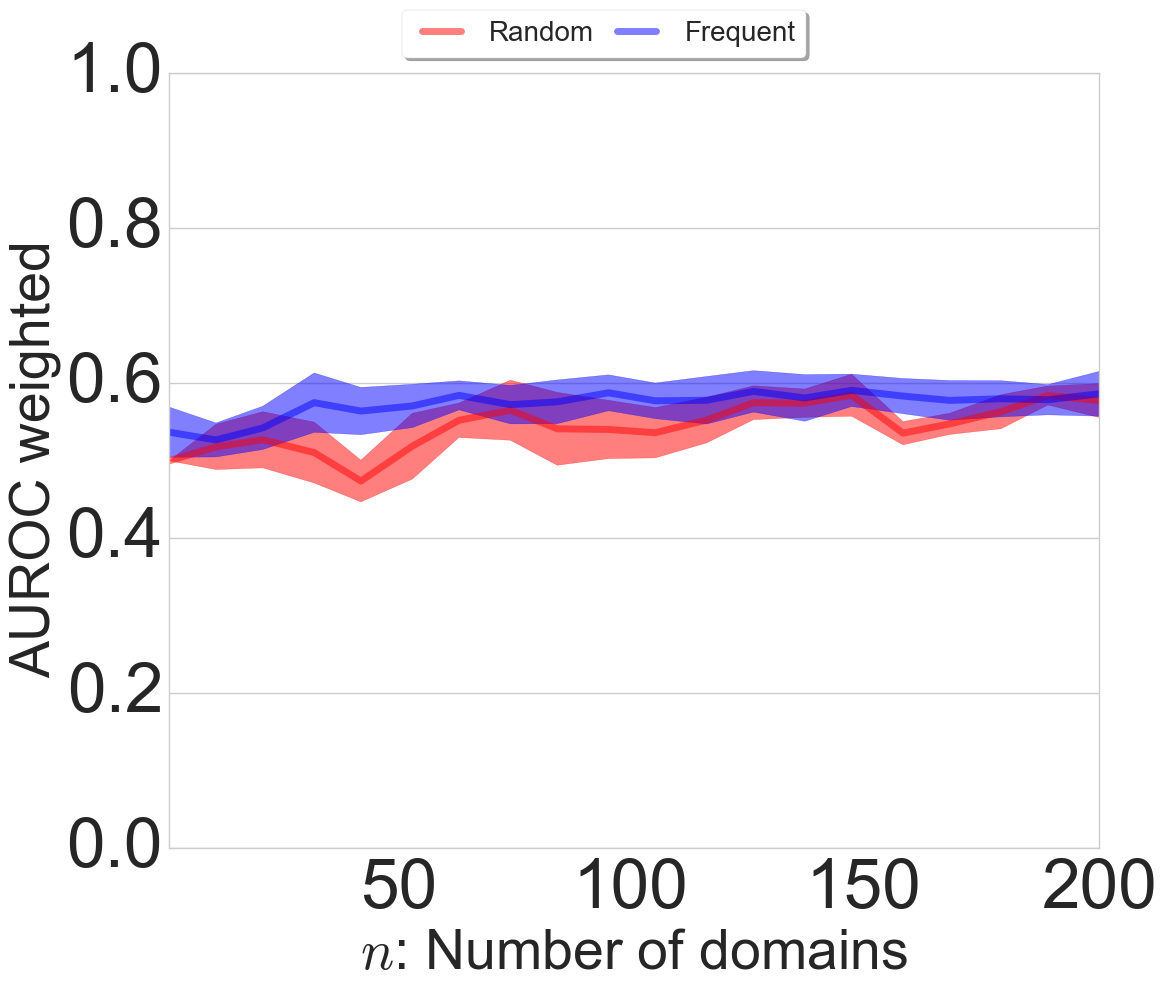}}
\subfloat[Universalism]{\includegraphics[scale=0.15]{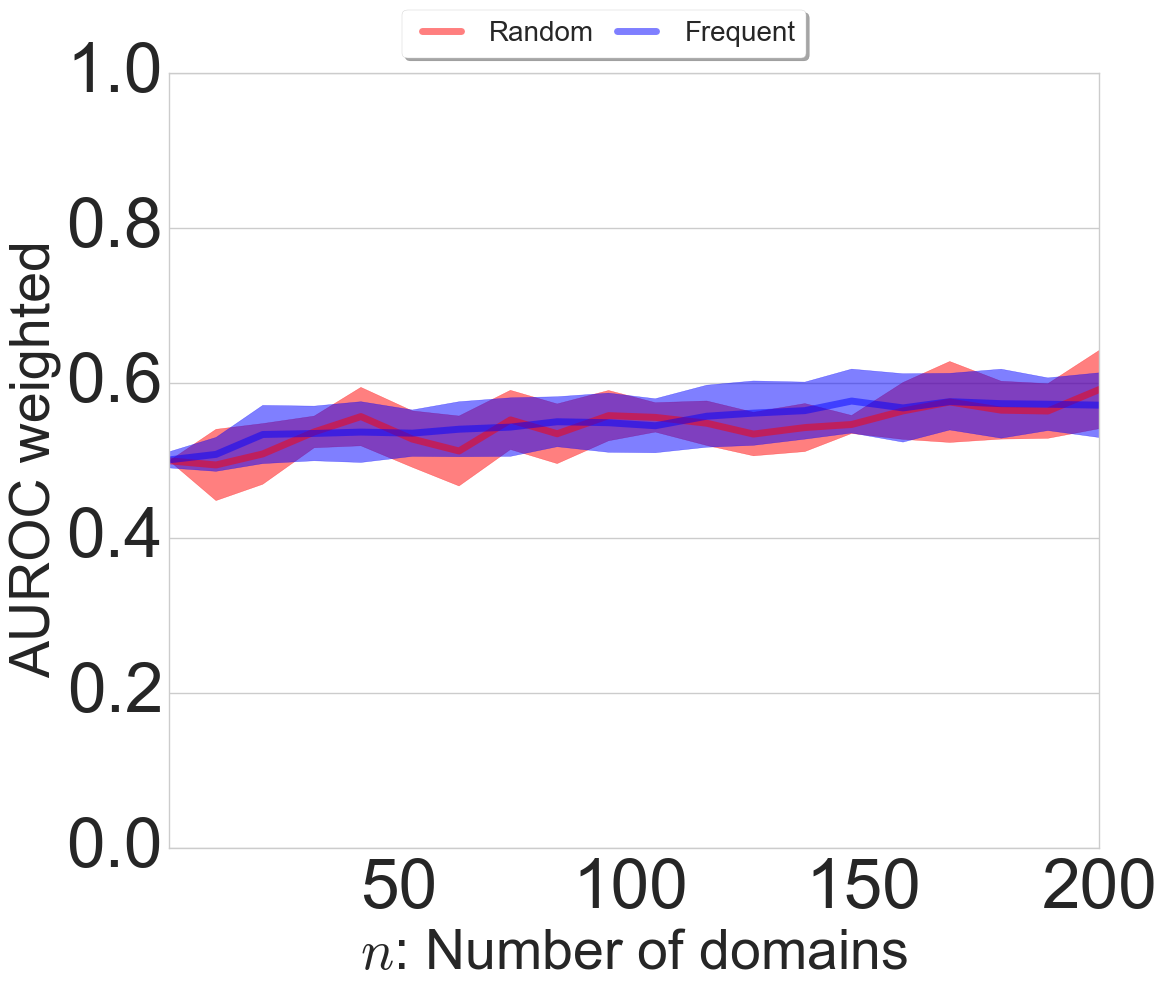}}
\caption{AUROC analysis for the Schwartz Values. Increasing the number of domains visited according to their frequency of visit (blue line, \emph{Case 1}) and increasing the number of domains visited regardless of their frequency of appearance (red line, \emph{Case 2}), in the training sets, while always validated on the same testing set.}
\label{fig:RandomChoiceRoc2}
\end{figure*}

\begin{figure*}[h]
\centering
\subfloat[Age]{\includegraphics[scale=0.15]{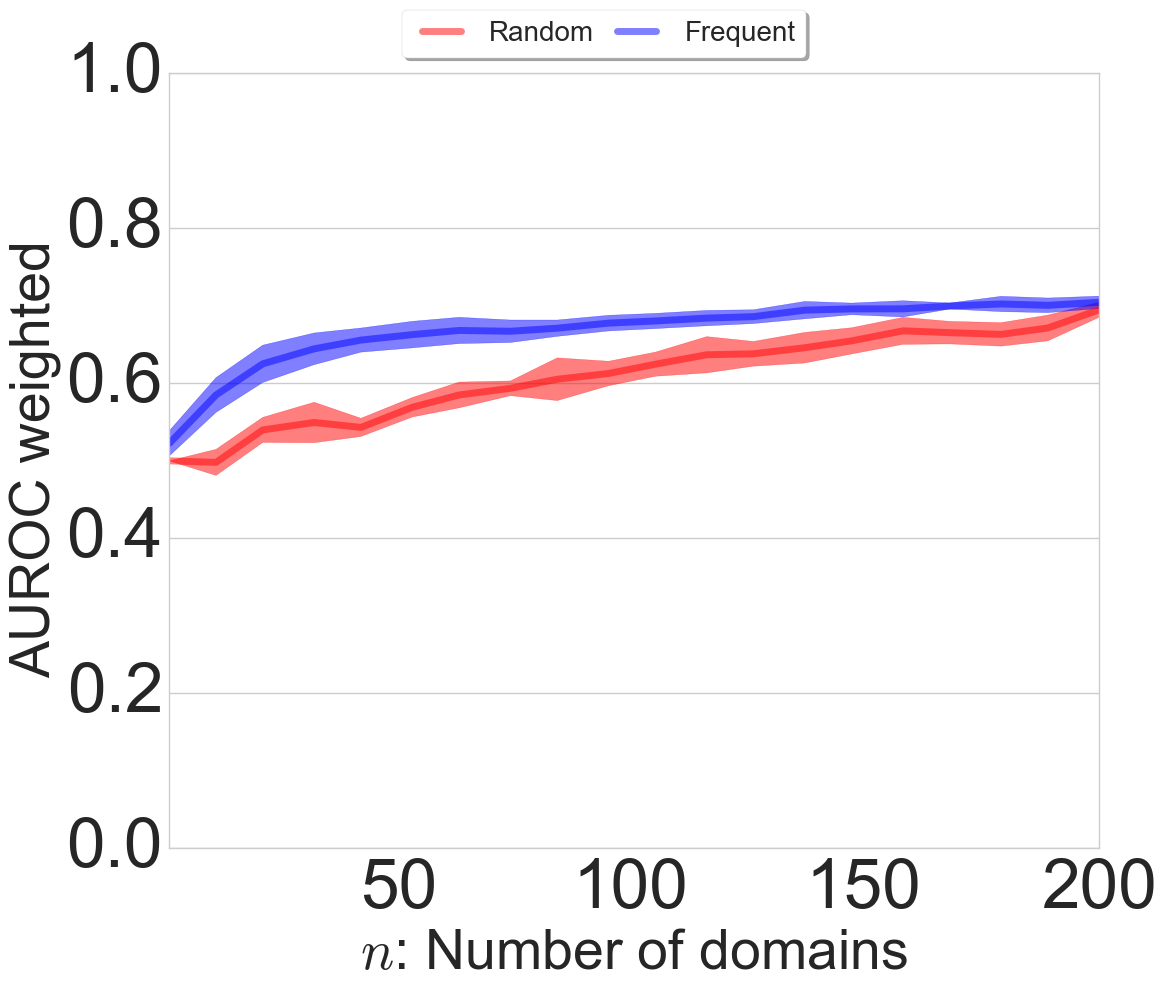}}
\subfloat[Education]{\includegraphics[scale=0.15]{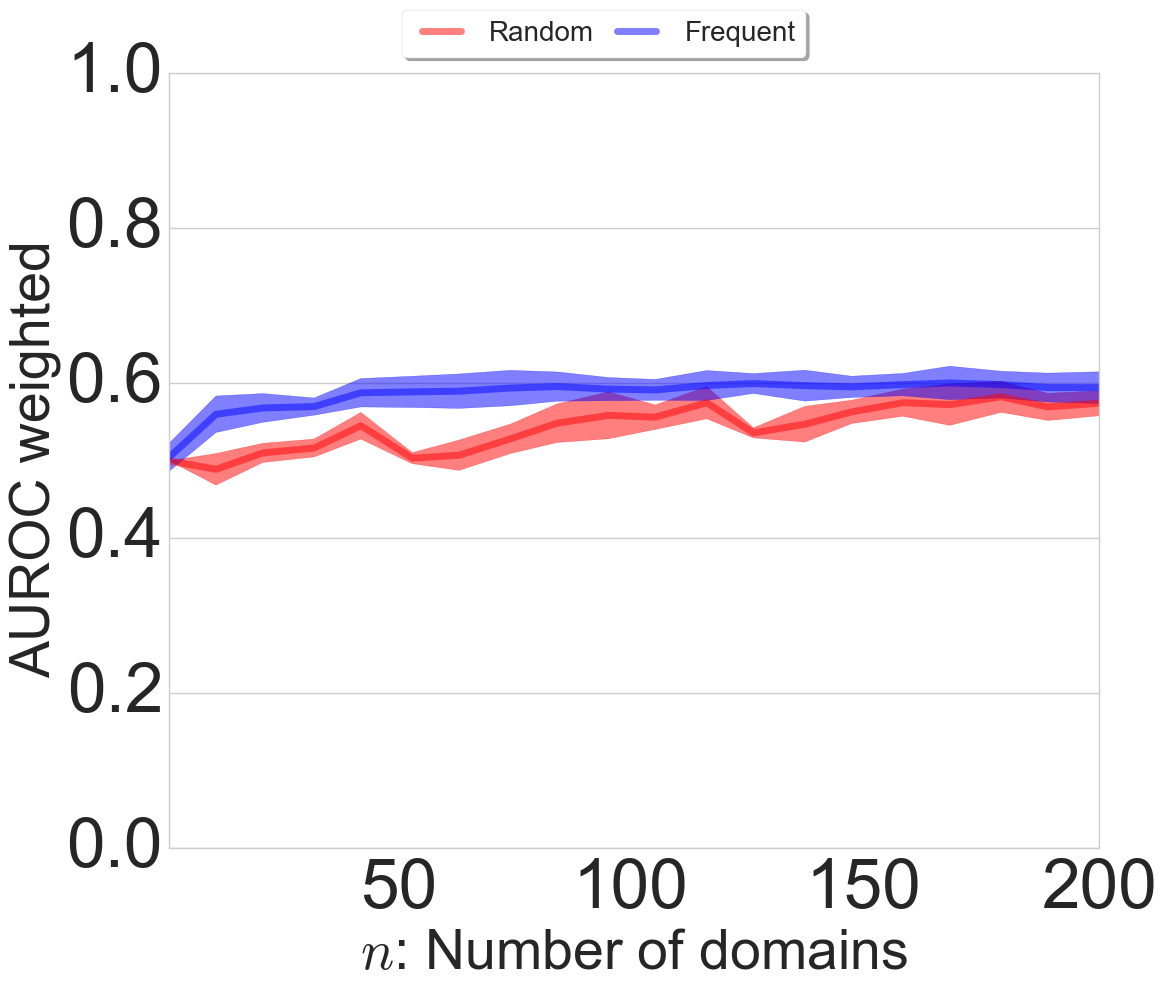}}
\subfloat[Ethnicity]{\includegraphics[scale=0.15]{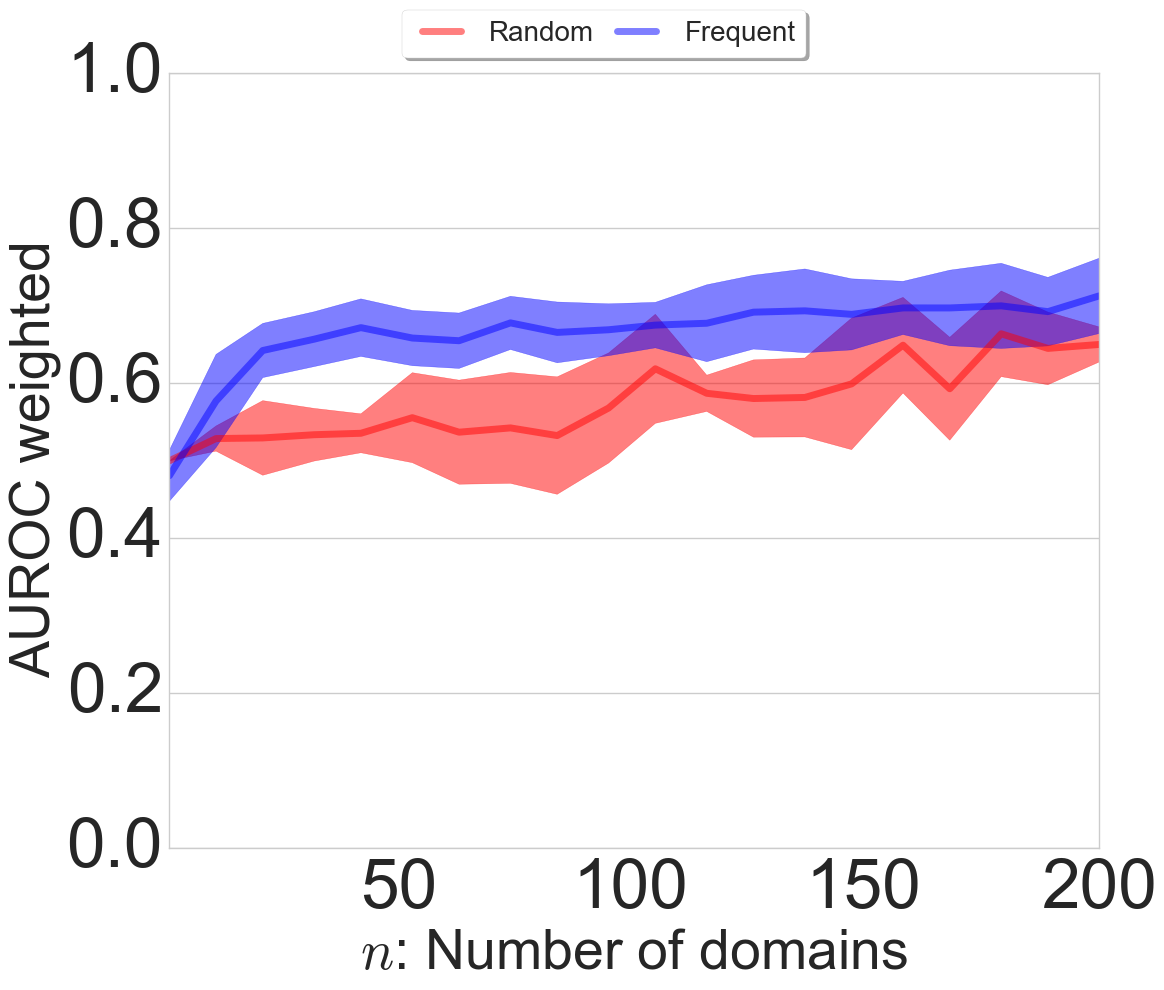}}
\subfloat[Exercise]{\includegraphics[scale=0.15]{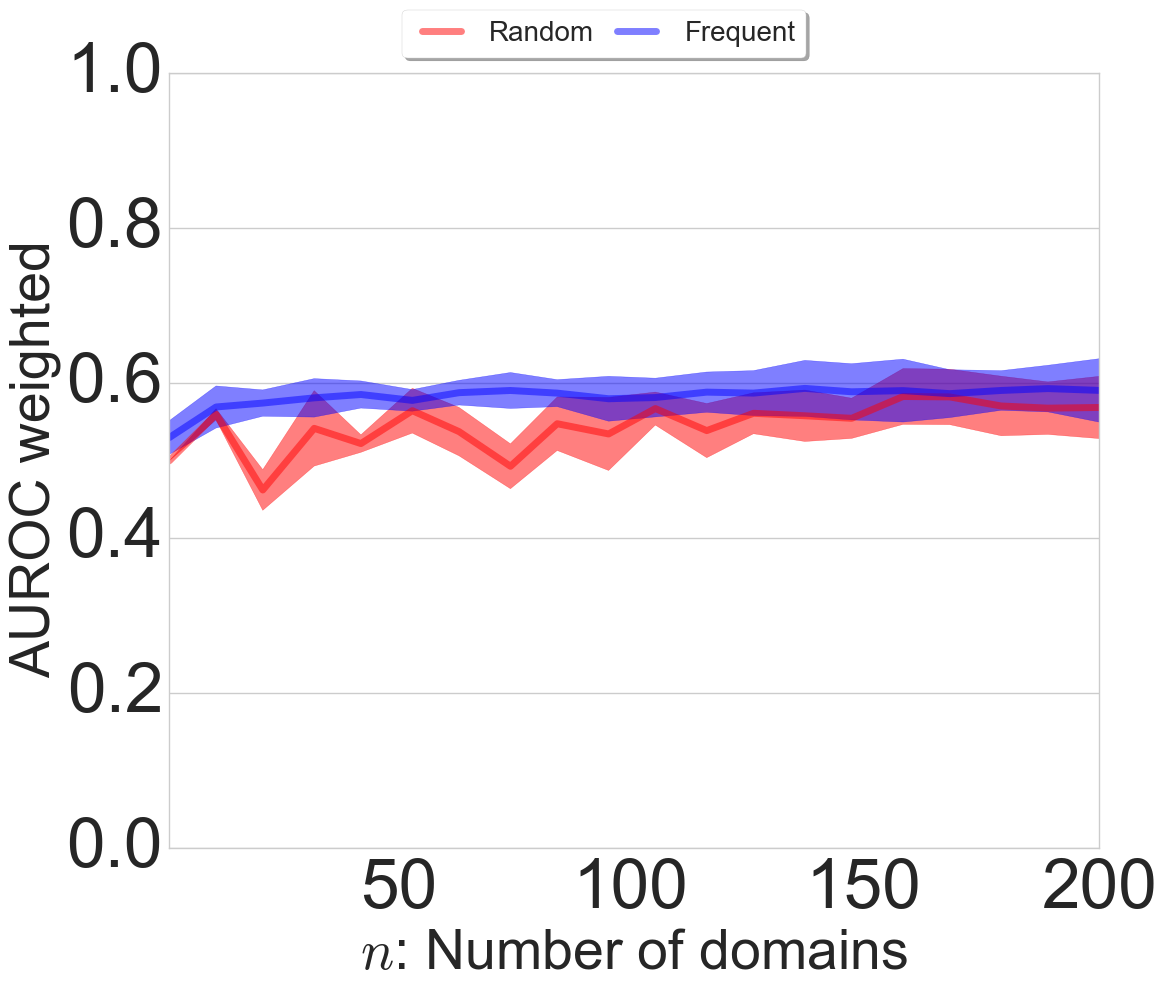}}\\
\subfloat[Gender]{\includegraphics[scale=0.15]{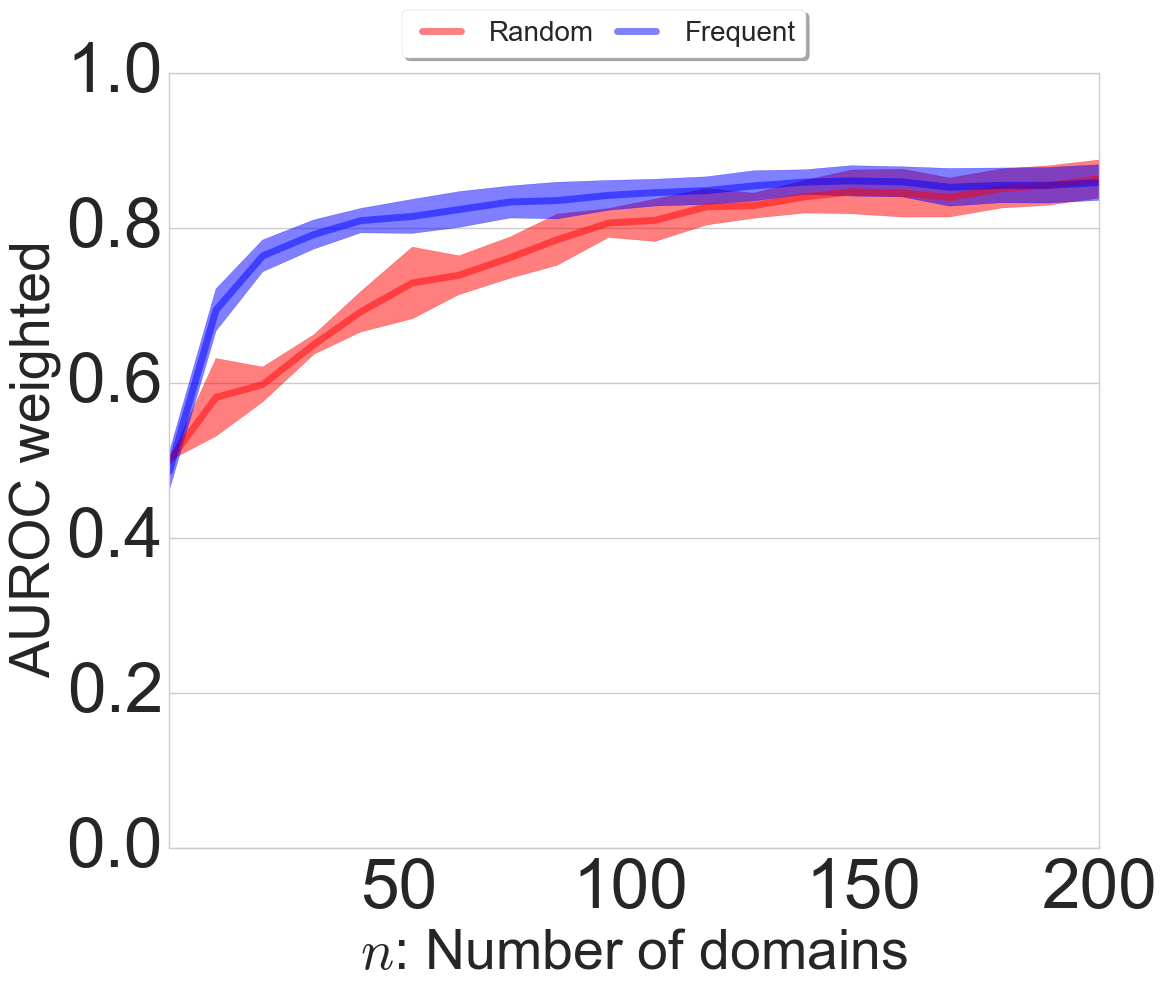}}
\subfloat[Income]{\includegraphics[scale=0.15]{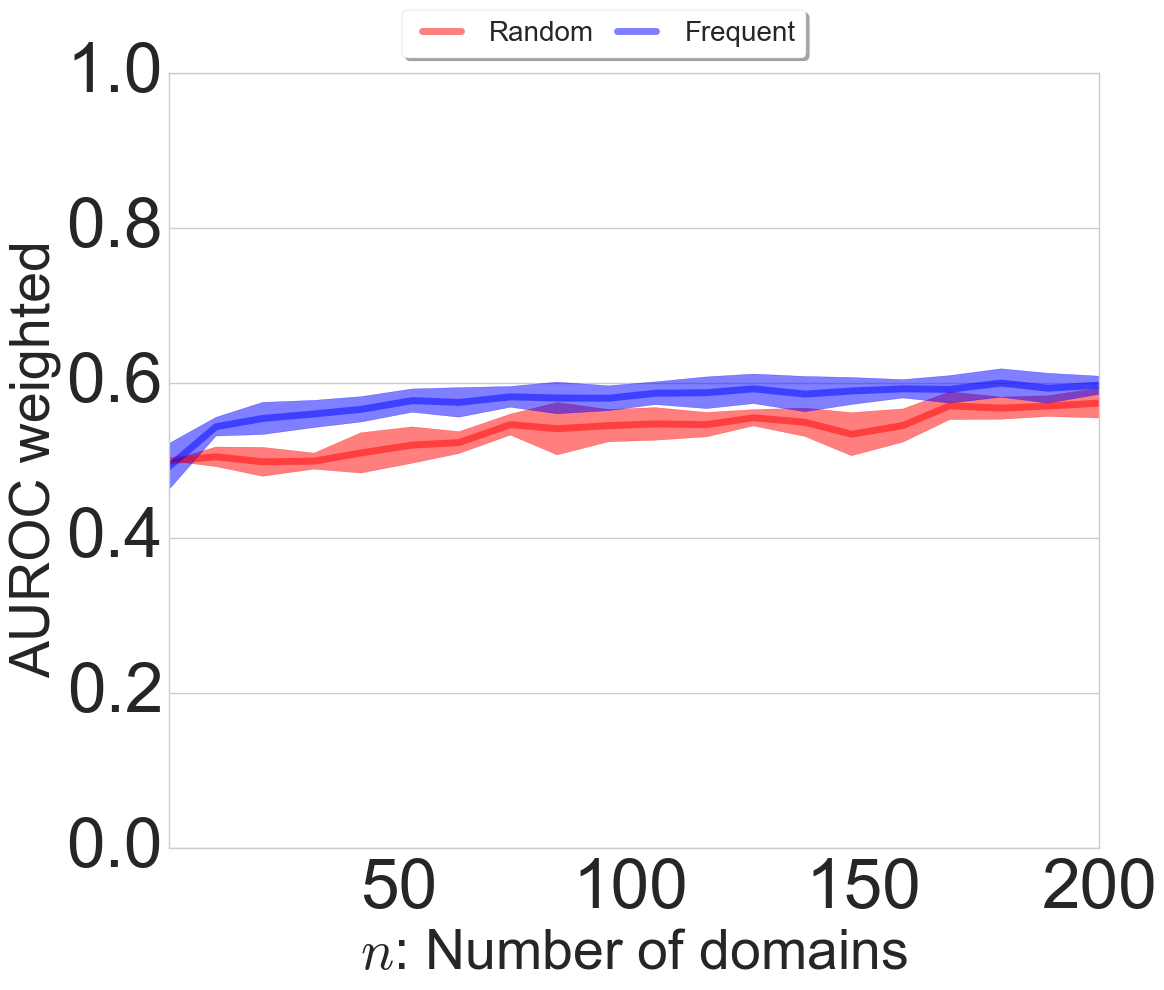}}
\subfloat[Marital Status]{\includegraphics[scale=0.15]{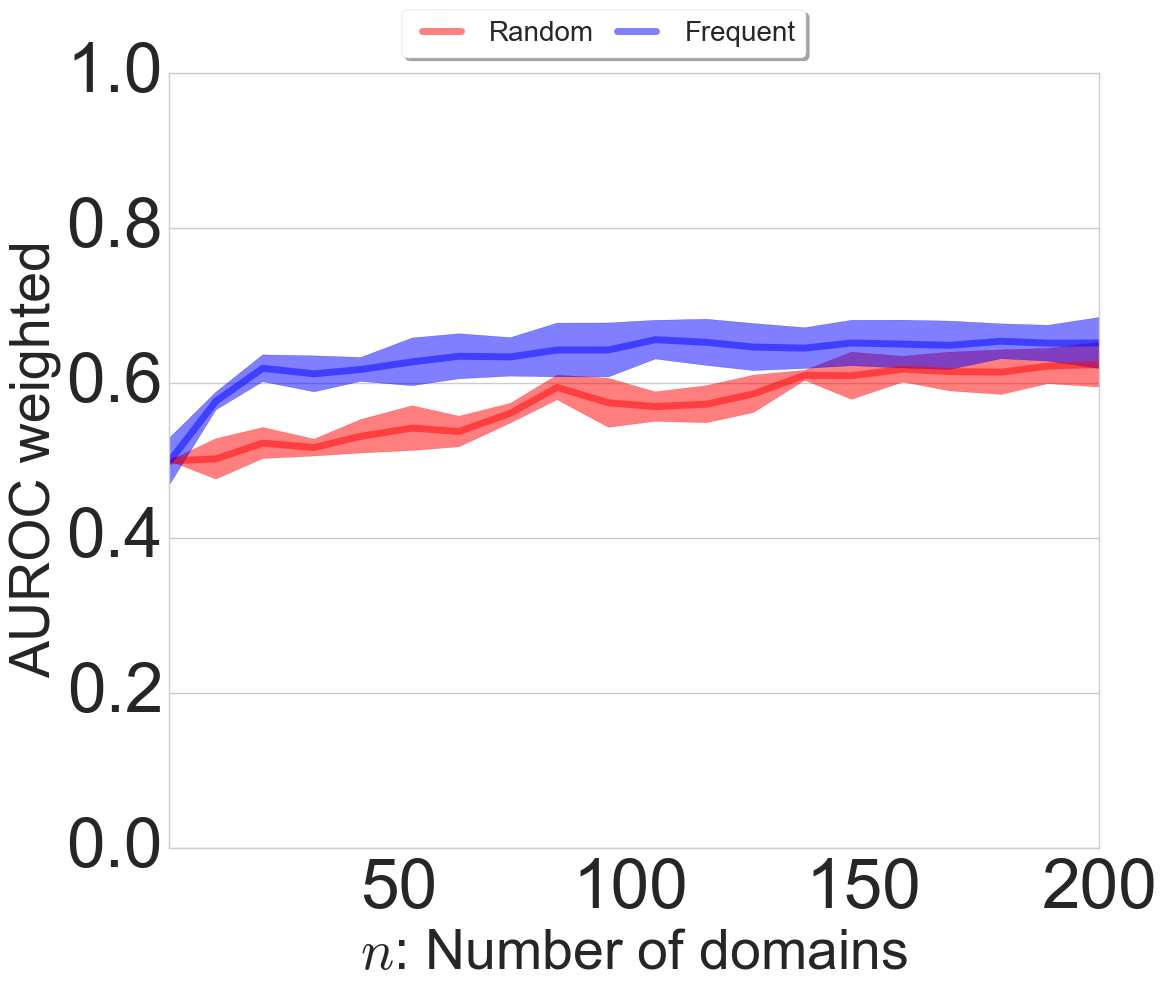}}
\subfloat[Parenthood]{\includegraphics[scale=0.15]{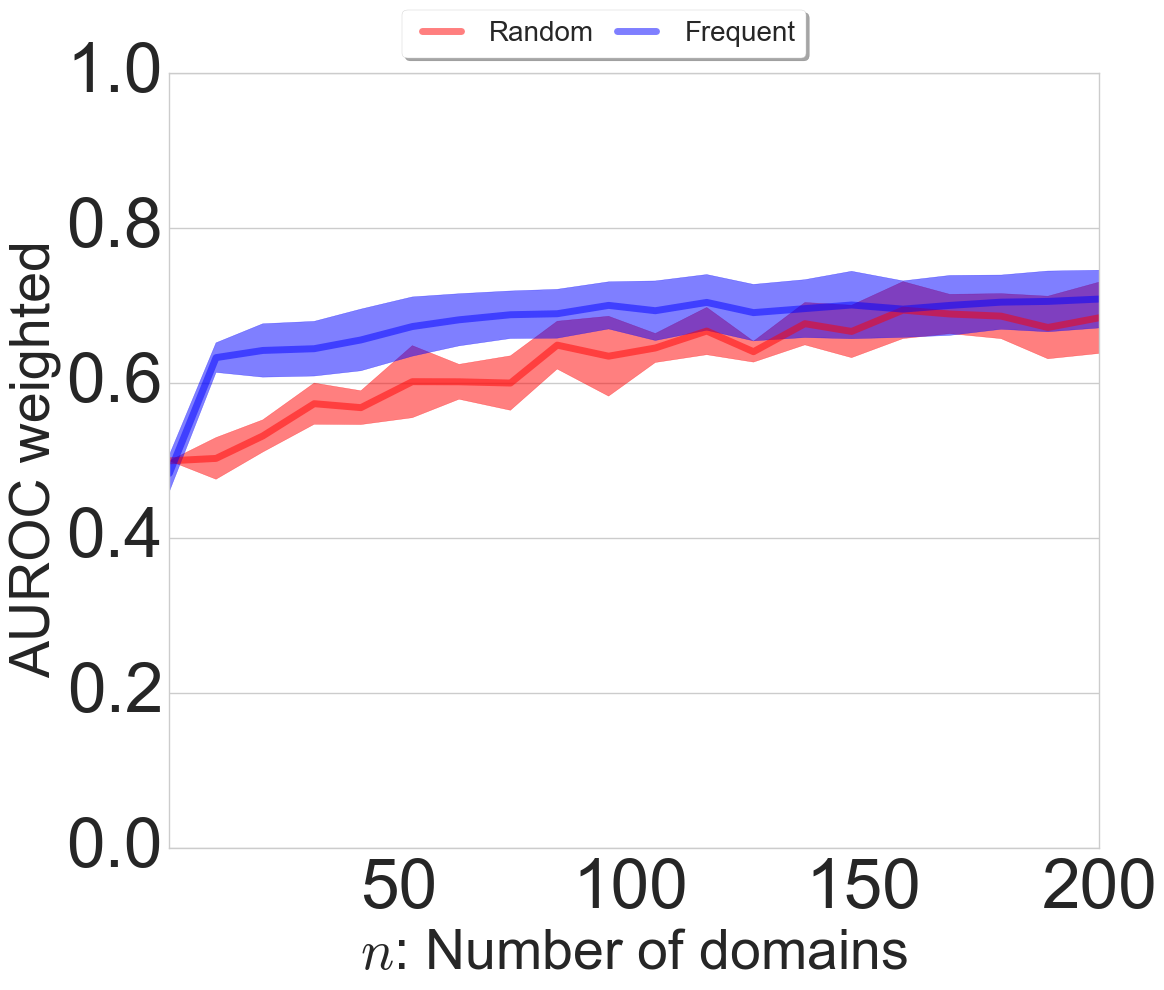}}\\
\subfloat[Political Orientation]{\includegraphics[scale=0.15]{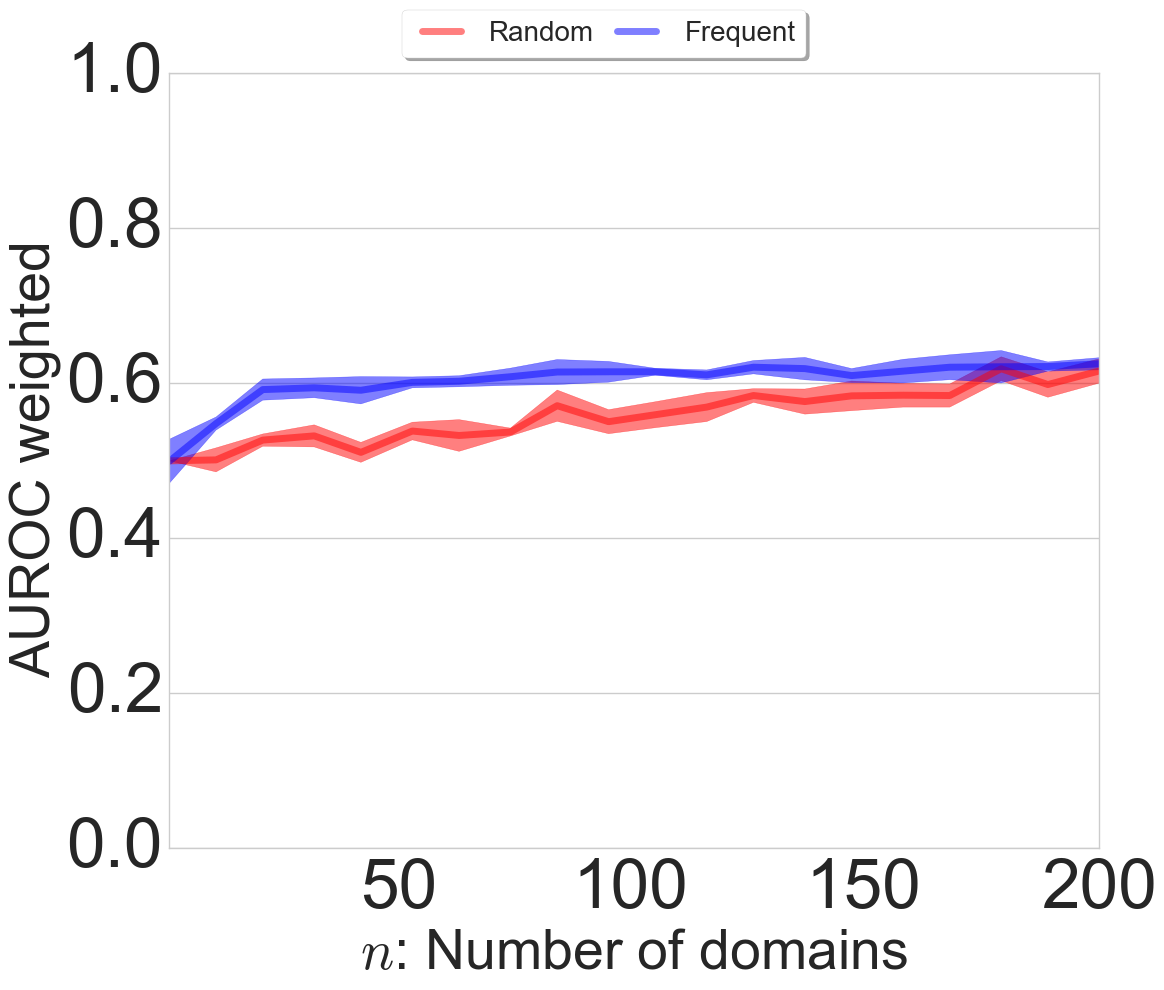}}
\subfloat[Smoking]{\includegraphics[scale=0.15]{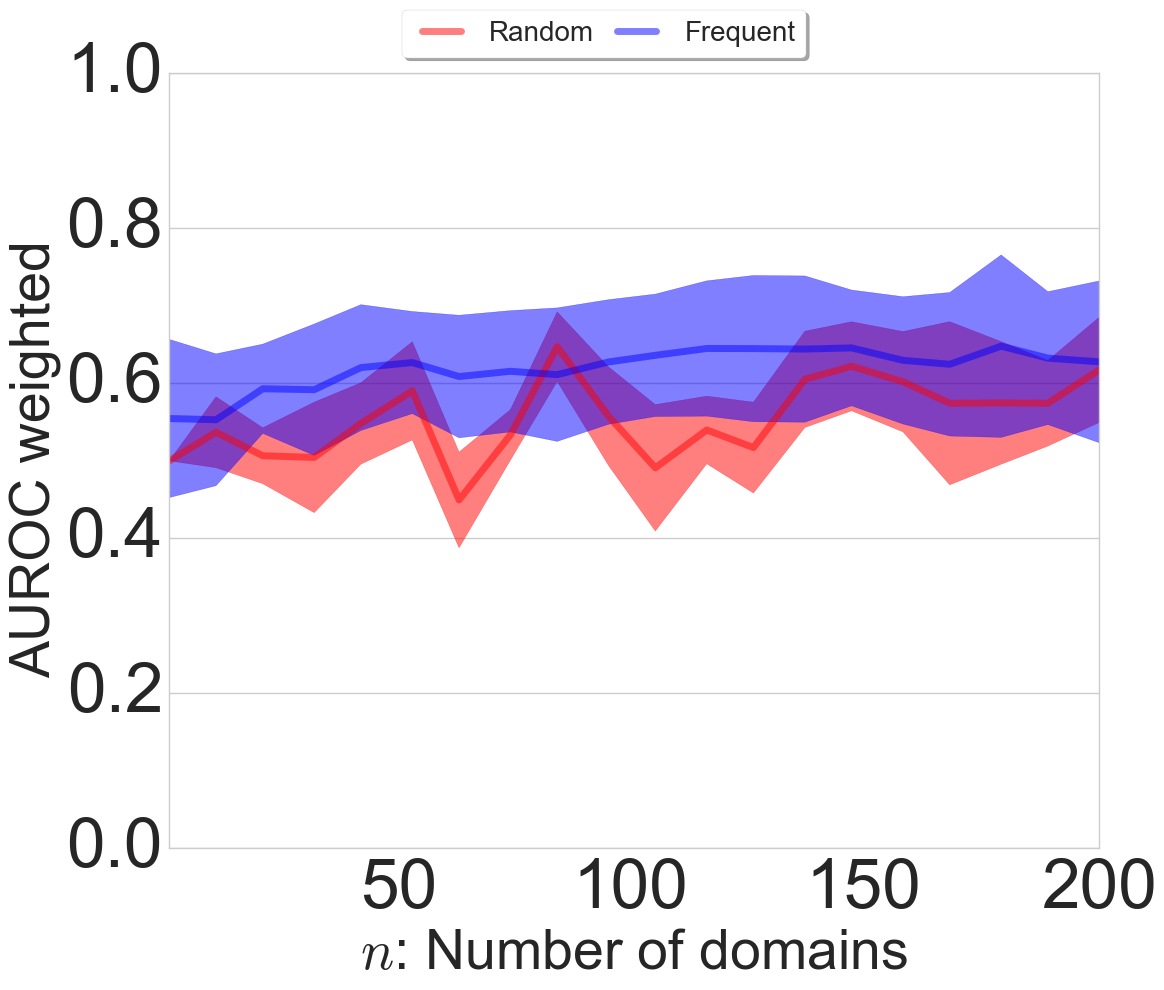}}
\subfloat[Wealth]{\includegraphics[scale=0.15]{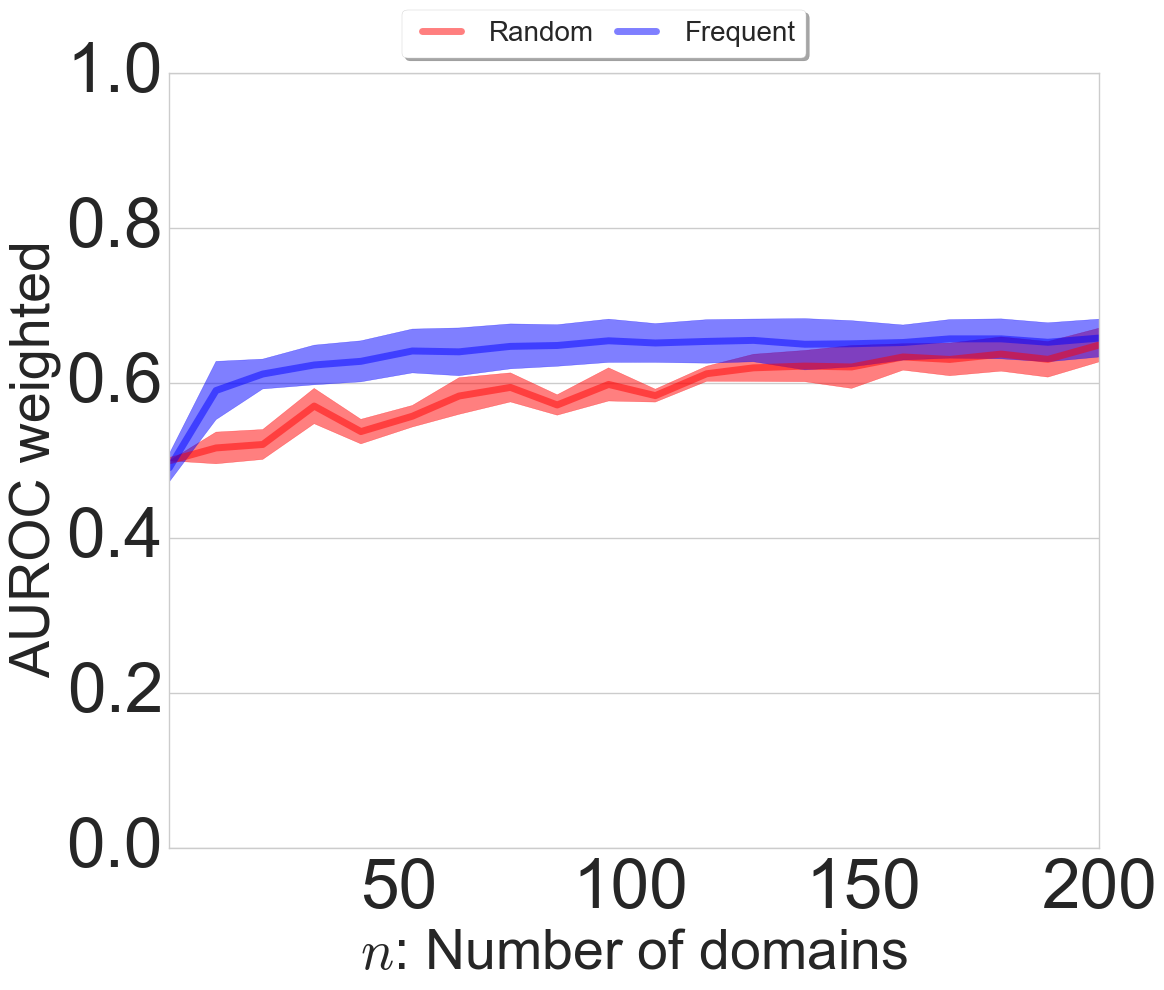}}
\subfloat[Weight Issues]{\includegraphics[scale=0.15]{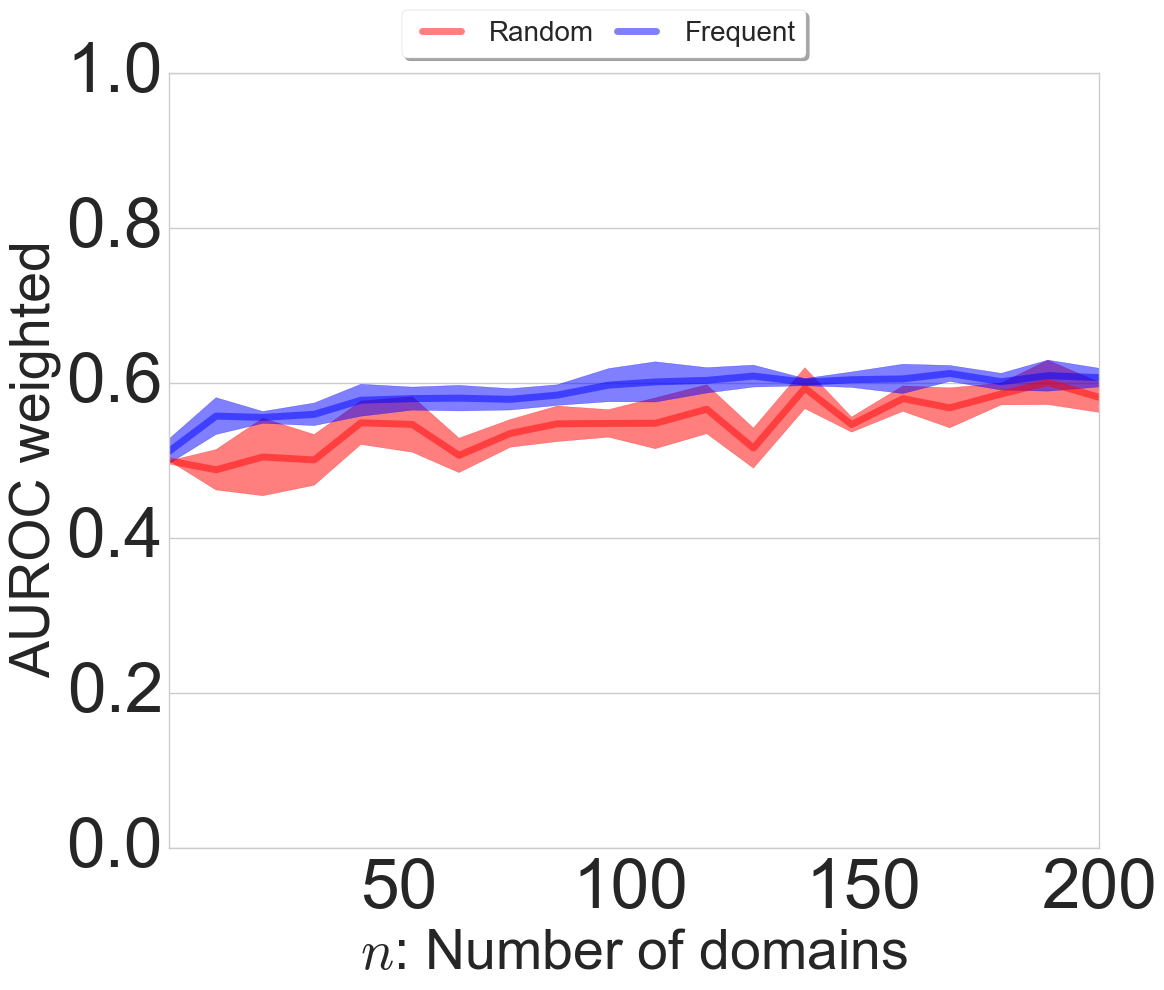}}

\caption{AUROC analysis for the demographic attributes. Increasing the number of domains visited according to their frequency of visit (blue line, \emph{Case 1}) and increasing the number of domains visited regardless of their frequency of appearance (red line, \emph{Case 2}), in the training sets, while always validated on the same testing set.}
\label{fig:RandomChoiceRoc3}
\end{figure*}

Blue curves in Figures~\ref{fig:RandomChoiceRoc0}, ~\ref{fig:RandomChoiceRoc1},  \ref{fig:RandomChoiceRoc2} and \ref{fig:RandomChoiceRoc3} depict the weighted AUROC metric for each considered attribute of the \emph{Case 1}, the most frequent domain selection.
Each point of the blue line indicates the average weighted AUROC metric, for the five-folds, for the model trained on the $Tr_{n}\;freq$ set and tested on the $Ts\;active$ set;  while the shaded interval indicates its standard deviation.
From this exploratory analysis,  we note that increasing the information contained in the training sets has a limited effect on the weighted AUROC metric after a brief stabilisation step noted approximately for $n<40$ domains.
Red curves in Figures~\ref{fig:RandomChoiceRoc0},  \ref{fig:RandomChoiceRoc1}, \ref{fig:RandomChoiceRoc2} and  \ref{fig:RandomChoiceRoc3} depict the weighted AUROC metric for each attribute of the \emph{Case 2}, the random domain selection.
Each point of the red line indicates the average weighted AUROC metric, for the five-folds, for the model trained on the $Tr_{n}\;random$  set and tested on the $Ts\;active$ set;  while the shaded interval indicates its standard deviation.

Interestingly, even with the experimentation with random selection of active users in the training set we noticed only a limited effect on the average weighted AUROC metric. There are, however, more fluctuations due exactly to the random amount of information included in each training set with respect to the first approach where the increase of the training set size, only improved the performances. 
This finding suggests, as expected, that models trained on the most frequent domains have higher prediction ability, even if trained on a small number of participants.

\section{Results}

\subsection{Moral Foundations}

For the moral foundations and the individualistic/binding hyper-cluster, we obtained the following prediction scores: 66\% for authority,  62\% for care,  58\% for fairness,  63\% for loyalty,  67\% for purity, and 66\% for individualists, respectively\footnote{All prediction scores refer to the weighted AUROC metric.}.
The web domains from Desktop browsing were more informative than mobile data for the prediction. 
The first part of Table~\ref{tab:demoprediction} reports the average weighted AUROC metric and the respective standard deviations over five-folds, for models trained on the web domains of the Desktop dataset (first column), on the web domains of the Mobile dataset (second column), on solely Application usage (third column), and finally on the fusion of web browsing and application usage of the Mobile dataset (fourth column).

In Table~\ref{tab:impo_MFT_MAPP} we indicatively report the ten most predictive domains (in typewriter style) and applications (in uppercase style) for each model. Finally,   Table \ref{tab:impo_MFT_IB} presents the most important predictors focusing on the individualistic versus binding foundations.

\subsection{Schwartz Human Values}

The prediction of Schwartz's basic human values and quadrants proved to be a more difficult classification problem with respect to the moral foundations one. The average weighted AUROC metric for all the attributes was around 60\%, with the most accurate predictions being the ones for the conservation 63\% and universalism 62\% values. As for the moral foundation prediction, we employed as predictors (i) the desktop browsing domains, (ii) the mobile browsing,  (iii) the application usage, and (iv) the fusion of mobile browsing and applications.
The second part of Table~\ref{tab:demoprediction} reports the weighted AUROC metric for each of the Schwartz's basic human values and quadrants, while Table~\ref{tab:impo_SWV_MAPP} reports the ten most predictive domains and applications for each model.

Comparing these findings with the state-of-the-art literature we see the performances obtained despite being low, are in line.
A study we can directly compare with, was performed by \citealp{Chen2014}, who employed supervised binary machine learning algorithms to classify human values in a binary scheme with above-median levels of each value dimension and reported their results using AUROC metric. From their best-performing classifiers, they obtained 60\% for Self-Transcendence, 56\% for Self- Enhancement, 59\% for Conservation, 61\% for Openness to change, and 61\% for Hedonism.
Except from the fact that \citealp{Chen2014} inferred on linguistic features from social media data, the performances they obtained are close to ours demonstrating the difficulty of predicting these values.

\subsection{Demographic Data}

The third part of Table~\ref{tab:demoprediction} reports in detail the weighted AUROC metric for all the demographic attributes, while Table~\ref{tab:impo_DEMO_MAPP} indicatively reports the ten most predictive domains and applications for each model.
We note that the demographic data were more accurately predicted with the ``gender'' attribute outperforming all others with a 90\%  AUROC score. 
Moving to less commonly studied demographic attributes as age 71\%, marital status 67\%,  parenting 72\%, and ethnicity 74\% our models have a satisfactory performance. 
Income and wealth were inferred with 60\% and 66\%, respectively. Table~\ref{tab:demo_predict} reports the findings of the state-of-the-art studies with which we can directly compare our findings since they reported also the AUROC metric. For the ``age'' attribute the prediction scores vary according to the number of classes considered; in our study we considered 6 classes and achieved 71\% which is a good performance. 
Moreover, we addressed health-related attributes like weight issues and the practice of physical exercise, achieving 63\% and 64\% weighted AUROC scores respectively. We were also able to capture the smoking behaviour of individuals with an accuracy of 64\%. These AUROC scores show that digital footprints can contribute to the health communication campaigns targeting individuals in need.

For the prediction of demographic variables, we considered all the categories (labels) reported in Table~\ref{tab:Demog}, which of course hardens the classification tasks. Complex attributes like education level 59\% and political orientation 60\% were difficult to infer, but if we consider only the two major political parties, namely Democrats and Republicans, the weighted AUROC prediction score reaches 70\%. Table~\ref{tab:Demog} provides a complete list of all the categories considered for each predicted attribute.

\begin{table*}[h]
\small
\centering
\begin{tabularx}{\textwidth}{p{0.2\linewidth}p{0.17\linewidth}p{0.17\linewidth}p{0.17\linewidth}p{0.17\linewidth}}
\toprule

\rowcolor{light-gray} \small{Morals} &    \small{Desktop Web} &  \small{Mobile Web } & \small{Mobile Apps } & \small{Mobile Apps}\\
\rowcolor{light-gray} &    \small{ Domains (STD)} &  \small{Domains (STD)} & \small{(STD)} & \small{Domains (STD)}\\
\midrule
Authority                    &    \textbf{0.66 (0.02)}& 0.63 (0.03) &     0.60 (0.02) &    0.64 (0.05) \\
Care                         &    \textbf{0.62 (0.02)} & 0.55 (0.05) &    0.57 (0.04) &    0.55 (0.03)\\
Fairness                     &   \textbf{0.58 (0.01)} &    0.57 (0.04)&    0.55 (0.01) &    0.55 (0.02)\\
Loyalty                      &    \textbf{0.63 (0.02)} &    0.60 (0.03) &    0.55 (0.04)&    0.58 (0.05)\\
Purity                       &    0.64 (0.01)  &    0.62 (0.02) &    \textbf{0.67 (0.04)}&    0.66 (0.04)\\
Individualist/Binding &    \textbf{0.66 ($<$0.01)}  &    0.65 (0.02) &    0.64 (0.02)&    \textbf{0.66 (0.02)} \\
 \\
\midrule
\rowcolor{light-gray} \small{Human Values} &    \small{Desktop Web} &  \small{Mobile Web } & \small{Mobile Apps } & \small{Mobile Apps}\\
\rowcolor{light-gray} &    \small{Domains (STD)} &  \small{Domains (STD)} & \small{(STD)} & \small{Domains (STD)}\\\midrule
Conservation                 &    \textbf{0.63 (0.02)} &    0.61 (0.05) &    0.59 (0.06)&    0.62 (0.05) \\
Openness                      &    0.56 (0.02)  &    0.56 (0.03) &    0.57 (0.03)&    \textbf{0.58 (0.04)} \\
Self-enhancement              &    \textbf{0.59 (0.01)}  &    0.51 (0.01) &    0.57 (0.04)&    0.56 (0.04)\\
Self-transcendence            &    0.58 (0.01) &    0.57 (0.02) &    0.58 (0.05)&    \textbf{0.59 (0.02)} \\
Achievement                  &    0.56 (0.03)&    0.54 (0.03) &    0.56 (0.04) &    \textbf{0.59 (0.04)}\\
Benevolence                  &    0.54 (0.01)  &    0.53 (0.04) &    0.50 (0.02)&    \textbf{0.56 (0.01)} \\
Conformity                   &    \textbf{0.59 (0.01)} &    0.57 (0.04) &    0.58 (0.03) &    0.58 (0.04)\\
Hedonism                     &    0.56 (0.02) &    0.57 (0.03) &    0.59 (0.05)&    \textbf{0.60 (0.03)} \\
Power                        &    0.56 (0.02) &    0.55 (0.03) &    0.55 (0.01)&    \textbf{0.58 (0.05)}\\
Security                     &    \textbf{0.58 (0.02)} &    0.56 (0.03) &    0.53 (0.04)&    0.55 (0.03) \\
Self-direction                &    0.56 (0.02) &    0.53 (0.03) &    \textbf{0.57 (0.03)}&    0.53 (0.02)\\
Stimulation                  &    0.56 (0.01) &    0.55 (0.03) &    0.56 (0.03)&   \textbf{0.58 (0.02)}\\
Tradition                    &    0.58 (0.02) &    \textbf{0.60 (0.03)}&    0.58 (0.06)&    \textbf{0.60 (0.03)}  \\
Universalism                 &    0.59 (0.01) &    0.57 (0.02)  &    \textbf{0.62 (0.06)}&    0.61 (0.04)\\
\\
\midrule
\rowcolor{light-gray} \small{Demographics} &    \small{Desktop Web} &  \small{Mobile Web } & \small{Mobile Apps } & \small{Mobile Apps}\\
\rowcolor{light-gray} &    \small{Domains (STD)} &  \small{Domains (STD)} & \small{(STD)} & \small{Domains (STD)}\\\midrule
Age                              &   \textbf{0.71 (0.01)}&    0.68 (0.03) &    \textbf{0.71 (0.03)}&    \textbf{0.71 (0.02)}\\
Education                        &    \textbf{0.59 (0.01)}&    0.57 (0.01)&    \textbf{0.59 (0.02)} &    \textbf{0.59 (0.01)}\\
Ethnicity                        &    0.73 (0.02) &    0.69 (0.02)&    0.72 (0.05)&    \textbf{0.74 (0.02)}\\
Exercise                         &    0.61 (0.02)  &    0.59 (0.02)&    \textbf{0.63 (0.04)}&    0.60 (0.03)\\
Gender                           &    0.86 (0.01) &    0.88 (0.02)&    \textbf{0.90 (0.02)}&    0.89 (0.02)\\
Income                           &    \textbf{0.60 (0.01)} &    0.55 (0.02) &    0.60 (0.02)&    0.58 (0.01)\\
Marital Status              &    \textbf{0.67 (0.02)} &    0.61 (0.02)&    0.64 (0.01)&    0.63 (0.03)\\
Parent                          &    0.71 (0.01) &    0.66 (0.02) &    \textbf{0.72 (0.04)} &    0.69 (0.04)\\
Political Orientation     &    0.58 (0.01)&    0.58 (0.02)&    0.59 (0.03)&    \textbf{0.60 (0.02)}  \\
Smoker                          &    0.63 (0.06)&    0.59 (0.06) &    0.62 (0.03)&    \textbf{0.64 (0.04)}\\
Wealth                           &    \textbf{0.66 (0.01)} &    0.60 (0.02)&    0.61 (0.02)&    0.62 (0.02)\\
Weight Issues           &    \textbf{0.62 (0.02)}&    0.58 (0.02)&    0.58 (0.03)  &    0.60 (0.02)\\
\bottomrule
\end{tabularx}

\caption{The prediction scores of moral attributes, human values and demographics expressed in average weighted AUROC scores. We trained a model for each attribute and for each category of predictors (i) features of desktop web browsing,   (ii) mobile browsing,  (iii) mobile application usage and (iv) fusion of mobile browsing and application usage. In parenthesis, the weighted AUROC's standard deviation (STD) of the five-fold cross-validation process is reported. The best-performing features are indicated in $\bf{bold}$.}
\label{tab:demoprediction}
\end{table*}

\begin{table*}[h]
\footnotesize
\begin{tabularx}{\textwidth}{p{0.18\linewidth}p{0.19\linewidth}p{0.18\linewidth}p{0.17\linewidth}p{0.16\linewidth}}
\toprule
\rowcolor{light-gray} \small{Authority} &  \small{Care} & \small{Fairness}&\small{Loyalty}&\small{Purity}\\
\midrule
 {\tt iflscience.com} ($-$)     & \MakeUppercase{Mobile HotSpot} ($+$)        & \MakeUppercase{Shopkick} ($-$)     & {\tt americanexpress.com} ($+$)   & \MakeUppercase{Bible} ($+$)
 \\
{\tt npr.org} ($-$)         & \MakeUppercase{Snapchat} ($-$)         & {\tt ijreview.com} ($-$)     & \MakeUppercase{Gmail} ($-$)            & \MakeUppercase{Yelp} ($-$)
 \\
\MakeUppercase{Hangouts} ($-$)        & {\tt theanimalrescuesite.com} ($+$)   & {\tt mookie1.com} ($-$)  & \MakeUppercase{Instagram} ($-$)        & {\tt google.com} ($-$)
\\
{\tt thinkprogress.org} ($-$)   & \MakeUppercase{Twitter} ($-$)          & \MakeUppercase{Mobile HotSpot} ($+$)    & \MakeUppercase{Pandora} ($-$)          & \MakeUppercase{Maps} ($-$)
\\
{\tt huffingtonpost.com} ($-$)  & \MakeUppercase{Papa John's} ($-$)      & {\tt accuweather.com} ($-$)  & {\tt wikipedia.org} ($-$)        & \MakeUppercase{LinkedIn} ($-$) \\
{\tt thoughtcatalog.com} ($-$)  & \MakeUppercase{Amazon Kindle} ($-$)        & {\tt facebook.com} ($+$)  & \MakeUppercase{Cloud} ($+$)             & {\tt iflscience.com} ($-$)
 \\
\MakeUppercase{Rewards} ($-$)         & \MakeUppercase{Bing} ($-$)             & {\tt marriott.com} ($+$)  & \MakeUppercase{YouTube} ($+$)           & \MakeUppercase{Pandora} ($-$)
\\
{\tt petco.com} ($-$)       & {\tt change.org} ($+$)        & {\tt change.org} ($+$)    & {\tt craigslist.org} ($-$)       & {\tt facebook.com} ($-$)
 \\
\MakeUppercase{Solitaire} ($+$)        & {\tt about.com} ($+$)         & {\tt imgur.com} ($-$)    & \MakeUppercase{Maps} ($-$)             & {\tt yp.com} ($+$)
\\
{\tt huffpost.com} ($-$)       & {\tt ihg.com} ($+$)           & {\tt care2.com} ($+$)     & {\tt yahoo.com} ($-$)        & {\tt wikipedia.org} ($-$)
 \\
\bottomrule
\end{tabularx}
\caption{Top ten websites and applications emerging as the best predictors for each moral foundation from mobile records (fusion of Web browsing and Application usage). Uppercase style indicates an \MakeUppercase{Application} while typewriter style indicates a {\tt{Web Domain}}. ($+$) and ($-$) signs refer to whether the specific behaviour was employed by the model to classify an individual as ``higher" or ``lower" in the specific moral.}
\label{tab:impo_MFT_MAPP}
\end{table*}

\begin{table}[!h]
\centering
\begin{tabular}{ll}
\toprule
\rowcolor{light-gray} \small{Individualist (Desktop)} &\small{Individualist (Mob and Apps)}\\
\midrule
 {\tt google.com} (I) &{\tt huffingtonpost.com} (I) \\
 {\tt foxnews.com} (B)&\MakeUppercase{Fox News} (B) \\
 {\tt dailykos.com} (I) & \MakeUppercase{Bible} (B)\\
 {\tt yelp.com} (I)& {\tt google.com} (I) \\
  {\tt imdb.com} (I) &{\tt accuweather.com} (B)\\
 {\tt cnsnews.com} (B)& \MakeUppercase{Hangouts} (I) \\
  {\tt wikipedia.org} (I) & \MakeUppercase{Emergency Alerts} (B)\\
 {\tt mrctv.org} (B)& {\tt facebook.com} (I) \\
 {\tt theblaze.com} (B)& \MakeUppercase{Gospel Library} (B) \\
  {\tt thepetitionsite.com} (I) &{\tt wikipedia.org} (I)\\
\bottomrule
\end{tabular}
\caption{Top ten websites and applications emerging as the best predictors of individualistic and binding foundations for two distinct models; the first one trained on Desktop browsing data and the second one on the fusion of mobile browsing and applications usage. The weighted AUROC metric is 66\% for both cases. We note that among the predictors we find many news websites which are in line with the expected profiles of the two foundations. Moreover, there is consistency in the correlation of certain predictors (e.g. {\tt wikipedia.org}) with a specific foundation indicating the stability of the approach.  ($B$) and ($I$) signs refer to whether the specific behaviour was employed by the model to classify an individual as ``Binders" or ``Individualists".}
\label{tab:impo_MFT_IB}
\end{table}

\begin{sidewaystable*}[h]
\footnotesize

\centering
\begin{tabular}{p{0.14\linewidth}p{0.14\linewidth}p{0.14\linewidth}p{0.14\linewidth}p{0.14\linewidth}p{0.14\linewidth}p{0.14\linewidth}}\noalign{\smallskip}\noalign{\smallskip}
\toprule
\rowcolor{light-gray} \small{Conservation} &\small{Openness} & \small{Self-enhancement} & \small{Self-transcendence}&\small{Achievement} &  \small{Benevolence} \\
\midrule
\MakeUppercase{Snapchat} ($-$)     & \MakeUppercase{Snapchat} ($+$) &  \MakeUppercase{WhatsApp} ($+$) &  \MakeUppercase{eBay} ($-$)      &  \MakeUppercase{Gmail} ($+$) &  \MakeUppercase{eBay} ($-$)\\
{\tt iflscience.com}  ($-$)             &\MakeUppercase{Instagram} ($+$) & \MakeUppercase{Southwest} ($+$) &  \MakeUppercase{GameStop} ($-$) &  \MakeUppercase{Instagram} ($+$) &  \MakeUppercase{Weather} ($+$)\\
\MakeUppercase{Maps}  ($-$)        &{\tt facebook.com}  ($+$)            & {\tt linkedin.com} ($+$)            &  {\tt bitecharge.com} ($+$)         &  \MakeUppercase{Facebook} ($+$) & \tt{mpstat.us} ($+$)\\
{\tt indiegogo.com}  ($-$)            & \MakeUppercase{Fox News}  ($-$)& \MakeUppercase{Snapchat} ($+$) &  \MakeUppercase{Mint} ($-$)        & {\tt kinja.com} ($+$)  & \tt{imdb.com} ($-$)\\
\MakeUppercase{Uber}  ($-$)        & \MakeUppercase{POLARIS Office}  ($+$)& \MakeUppercase{AccuWeather} ($-$) &  {\tt mlnwap.com} ($-$)     &  \MakeUppercase{Netflix} ($-$) & \tt{battle.net} ($-$)\\
{\tt lgbtqnation.com}  ($-$)        &  \MakeUppercase{Maps}  ($+$)    & \MakeUppercase{Scout} ($-$)     & {\tt kmart.com} ($-$)             &  \MakeUppercase{Bible} ($-$)& \tt{emgn.com} ($+$)\\
{\tt npr.org} ($-$)                & \MakeUppercase{Emergency Alerts}  ($-$) & {\tt xojane.com} ($+$)    & {\tt nytimes.com} ($+$)             &  {\tt target.com} ($+$) &  \MakeUppercase{Line} ($-$)\\
{\tt amazon.com} ($-$)            & \MakeUppercase{Sprint hotspot}  ($+$)& \MakeUppercase{Gmail} ($+$) &  \MakeUppercase{Fantasy Sports} ($-$) &  \MakeUppercase{Google Mobile} ($+$) &\tt{facebook.com} ($+$)\\
{\tt huffpost.com} ($-$)            &\MakeUppercase{Capital One}  ($+$)& \MakeUppercase{Twitter} ($+$) &  {\tt huffpost.com} ($+$)             & {\tt lds.org} ($-$) & \tt{sparkpeople.com} ($-$)\\
\MakeUppercase{Bible}  ($+$)    &\MakeUppercase{Pinterest}  ($-$)& \MakeUppercase{Mint} ($+$) &  \MakeUppercase{Facebook} ($-$)         &  \MakeUppercase{Twitter} ($+$) &  \MakeUppercase{Firefox} ($-$)\\
\\
\midrule
\rowcolor{light-gray} \small{Conformity} & \small{Hedonism} & \small{Power} & \small{Security}&\small{Self-direction} & \small{Stimulation} \\
\midrule
\tt{zimbio.com} ($-$) & \MakeUppercase{Instagram} ($+$) & \tt{aarp.org} ($-$) &  \MakeUppercase{Maps} ($-$) & \MakeUppercase{Facebook} ($-$) & \MakeUppercase{Maps} ($+$)\\
\tt{iflscience.com} ($-$) & \MakeUppercase{Yelp} ($+$) & \MakeUppercase{Tapped Out} ($+$) & \MakeUppercase{Regal} ($+$) & \MakeUppercase{YouTube} ($-$) & \tt{google.com} ($+$)\\
\MakeUppercase{Fox News} ($+$) & \MakeUppercase{YouTube} ($+$) &  \tt{simplyhired.com} ($-$) &\tt{papajohns.com} ($-$) & \MakeUppercase{Photo Grid} ($-$) & \tt{facebook.com} ($+$)\\
\MakeUppercase{Bible} ($+$) & \MakeUppercase{Pandora} ($+$) & \tt{att.com} ($-$) & \MakeUppercase{Yahoo! Mail} ($+$) & \MakeUppercase{Pinterest} ($-$) & \tt{uproxx.com} ($+$)\\
\MakeUppercase{YouTube} ($+$) & \MakeUppercase{Foursquare} ($+$) & \tt{nerdist.com} ($+$) & \tt{myfitnesspal.com} ($-$) & \MakeUppercase{Play Store} ($-$) & \MakeUppercase{YouTube} ($+$)\\
\MakeUppercase{Kroger} ($+$) & \MakeUppercase{Facebook} ($+$) & \MakeUppercase{Play Store} ($+$) & \MakeUppercase{Sprint hotspot} ($-$) & \tt{bitecharge.com} ($+$) & \MakeUppercase{Facebook} ($+$)\\
\tt{wikipedia.org} ($-$) & \MakeUppercase{ibotta} ($-$) & \MakeUppercase{Inkpad} ($-$) & \tt{instagram.com} ($-$) & \tt{amazon.com} ($+$) & \MakeUppercase{Shazam} ($+$)\\
\tt{yelp.com} ($-$) & \MakeUppercase{WhatsApp} ($+$) & \MakeUppercase{Spotify} ($+$) & \MakeUppercase{Spotify} ($-$) & tt{aa.com} ($+$) & \tt{ebay.com} ($+$)\\
\MakeUppercase{GALAXY Apps} ($+$) & \MakeUppercase{Maps} ($+$) & \tt{imdb.com} ($+$) & \MakeUppercase{Walmart} ($+$) &  \MakeUppercase{Candy Crush Saga} ($-$) &  \tt{buzzfeed.com} ($+$)\\
\tt{amazon.com} ($-$) & \MakeUppercase{Shazam} ($+$) & \tt{southwest.com} ($-$) & \tt{aarp.org} ($+$) & \tt{groupon.com} ($-$) &\MakeUppercase{Yelp} ($+$)\\
\\
\midrule
\rowcolor{light-gray}  \small{Tradition} &\small{Universalism}&&&&\\
\midrule
\tt{amazon.com} ($-$) & \MakeUppercase{Fox News} ($-$)&&&&\\
\MakeUppercase{Yelp} ($-$) &\MakeUppercase{Twitter} ($-$)&&&&\\
\MakeUppercase{YouTube} ($-$) &\MakeUppercase{Facebook} ($-$)&&&&\\
\tt{guff.com} ($-$) &\MakeUppercase{People} ($-$)&&&&\\
\MakeUppercase{Snapchat} ($-$) &\MakeUppercase{Instagram} ($-$)&&&&\\
\tt{iflscience.com} ($-$) &\MakeUppercase{eBay} ($-$)&&&&\\
\tt{facebook.com} ($-$) &\tt{twitter.com} ($+$)&&&&\\
\MakeUppercase{Facebook} ($-$) &\MakeUppercase{Evernote} ($+$)&&&&\\
\tt{wikipedia.org} ($-$) &\MakeUppercase{Groupon} ($-$)&&&&\\
\tt{google.com} ($-$) &\tt{couponsherpa.com} ($-$)&&&&\\
\bottomrule
\end{tabular}
\caption{Top ten websites and applications emerging as the best predictors for each of the Schwartz values and quadrants from mobile records (fusion of Web browsing and Application usage). Uppercase style indicates an \MakeUppercase{Application} while typewriter style indicates a {\tt{Web Domain}}. ($+$) and ($-$) signs refer to whether the specific behaviour was employed by the model to classify an individual as ``higher" or ``lower" in the specific value.}
\label{tab:impo_SWV_MAPP}
\end{sidewaystable*}

\begin{sidewaystable*}[h]
\footnotesize

\begin{tabularx}{\textwidth}{p{0.21\linewidth}p{0.18\linewidth}p{0.1765\linewidth}p{0.17\linewidth}p{0.17\linewidth}p{0.16\linewidth}}\noalign{\smallskip}\noalign{\smallskip}
\toprule
\rowcolor{light-gray}  \small{Education} & \small{Income}&\small{Political Party} & \small{Wealth} & \small{Ethnicity} \\
\midrule
{\tt linkedin.com} ($postGrad$)     & {\tt linkedin.com} ($200KPlus$)     &\MakeUppercase{Maps} ($L$) &\MakeUppercase{YouTube} ($50KLess$) & \MakeUppercase{Yelp} ($A$) \\
\MakeUppercase{Play Store} ($someCollege$)     & \MakeUppercase{GameStop} ($20KLess$)         &{\tt reddit.com} ($D$) & \tt{buzzfeed.com} ($50KLess$)  & \MakeUppercase{Safeway}  ($A$)\\
\MakeUppercase{YouTube} ($someCollege$)         & \MakeUppercase{Snapchat} ($20KLess$)      &{\tt complex.com} ($D$) &\MakeUppercase{Instagram} ($50KLess$)  & \MakeUppercase{LINE}  ($A$) \\
\MakeUppercase{LinkedIn} ($postGrad$)     & \MakeUppercase{YouTube} ($20KLess$)         &facebook.com ($D$) &\MakeUppercase{Ameritrade} ($1000KPlus$) & \tt{slickdeals.net}  ($A$)\\
{\tt jw.org} ($high-school$)         & \MakeUppercase{Tumblr} ($20KLess$)        &\MakeUppercase{Flipboard} ($D$) &\MakeUppercase{Facebook} ($50KLess$)  & \MakeUppercase{YouTube} ($H$)\\
{\tt google.com} ($postGrad$)     & \MakeUppercase{Facebook} ($30Kto50K$)       &{\tt imgur.com} ($R$)  &\MakeUppercase{Pinterest} ($50KLess$)  & \MakeUppercase{Pinterest} ($W$)\\
\MakeUppercase{MI Mobile} ($high-school$)    & \MakeUppercase{Gmail} ($150Kto200K$)        &\MakeUppercase{Color Note} ($L$) &\MakeUppercase{Snapchat} ($50KLess$)  & \tt{wetpaint.com} ($AA$) \\
{\tt att.com} ($tradeSchool$)         & {\tt adadvisor.net} ($200KPlus$)     &{\tt iflscience.com} ($D$) &\MakeUppercase{Amazon} ($100Kto250K$) & \MakeUppercase{WeChat}  ($A$)\\
{\tt gamestop.com} ($high-school$)     & {\tt google.com} ($200KPlus$)     &{\tt diply.com} ($L$) &\tt{facebook.com} ($50KLess$)  & \MakeUppercase{WhatsApp} ($H$)\\
\MakeUppercase{Facebook} ($someCollege$)         & \MakeUppercase{Play Store} ($30Kto50K$)     &\MakeUppercase{NFL Fantasy Football} ($L$) &\MakeUppercase{NYTimes} ($1000KPlus$) & \MakeUppercase{Bible} ($AA$) \\
\\
\midrule
\rowcolor{light-gray} \small{Gender} & \small{Exercise} & \small{Age} & \small{Marital Status} & \small{Parent} \\
\midrule
\MakeUppercase{Pinterest} ($F$)                      & \MakeUppercase{Fitbit} ($Y$) & \MakeUppercase{Snapchat} ($18to24$) & \MakeUppercase{Gmail} ($S$)&\MakeUppercase{Snapchat} ($N$))\\
\MakeUppercase{Cartwheel by Target} ($F$)             & \MakeUppercase{Maps} ($Y$) &\MakeUppercase{Instagram} ($18to24$) & \MakeUppercase{NCPMobile}($W$)&\MakeUppercase{Gmail} ($N$)\\
\MakeUppercase{Facebook} ($F$)                       & \MakeUppercase{MapMyRun} ($Y$) &{\tt buzzfeed.com} ($25to34$) & \MakeUppercase{Flipboard}($W$)&\tt{elitedaily.com} ($N$)\\
{\tt playbuzz.com} ($F$)                  & {\tt active.com} ($Y$)&\MakeUppercase{YouTube} ($25to34$) &\MakeUppercase{NCP}($W$) &\MakeUppercase{Yelp} ($N$)\\
\MakeUppercase{Instagram} ($F$)                         & \MakeUppercase{Angry Birds} ($N$) &\MakeUppercase{Facebook} ($25to34$) &\MakeUppercase{Snapchat} ($S$) &\tt{buzzfeed.com} ($N$)\\
{\tt buzzfeed.com} ($F$)                    & \MakeUppercase{MyFitnessPal} ($Y$)&\MakeUppercase{Pinterest} ($25to34$) &\tt{google.com} ($S$)&\MakeUppercase{Tumblr} ($N$)\\
\MakeUppercase{Sports Center} ($M$)                    & {\tt yelp.com} ($Y$) &{\tt google.com} ($18to24$) &\tt{jetblue.com} ($M$)&\MakeUppercase{Instagram} ($N$)\\
\MakeUppercase{Play Store} ($M$)                    & \MakeUppercase{Yelp} ($Y$)&\MakeUppercase{Tumblr} ($18to24$) &\MakeUppercase{Chase} ($D$)&\MakeUppercase{Swarm} ($N$)\\
{\tt gap.com} ($F$)                   & {\tt yahoo.com} ($N$)&{\tt facebook.com} ($25to34$) &\MakeUppercase{YouTube} ($S$) &\MakeUppercase{Uber} ($N$) \\
{\tt target.com} ($F$)                     & \MakeUppercase{ESPN} ($Y$)&{\tt aarp.org} ($55to64$)   &\MakeUppercase{Facebook} ($LT$) &\MakeUppercase{Maps} ($N$)\\
\\
\midrule
\rowcolor{light-gray} \small{Weight} & \small{Smoker}&&&\\
\midrule
\MakeUppercase{Facebook} ($Y$) &\MakeUppercase{Snapchat} ($N$)\\
\tt{express.com} ($N$) &\tt{sprintpcs.com} ($Y$)\\
\MakeUppercase{Maps} ($N$) &\tt{mailchimp.com} ($N$)\\
\MakeUppercase{WordsWithFriends} ($Y$) &\tt{google-analytics.com} ($N$)\\
\MakeUppercase{Amazon Kindle} ($Y$) &\tt{psychcentral.com} ($Y$)\\
\MakeUppercase{Lose It!} ($Y$) &\MakeUppercase{Trip Advisor} ($N$)\\
\MakeUppercase{Yahoo! Mail} ($Y$) &\tt{about.com} ($N$)\\
\tt{bodybuilding.com} ($N$) &\tt{18andabused.com} ($Y$)\\
\MakeUppercase{Uber} ($N$) &\MakeUppercase{Sports Center} ($N$)\\
\MakeUppercase{Gmail} ($N$) &\tt{google.com} ($N$)\\

\bottomrule
\end{tabularx}
\caption{Top ten websites and applications emerging as the best predictors for each of the demographic attributes from mobile records (fusion of Web browsing and Application usage). Uppercase style indicates an \MakeUppercase{Application} while typewriter style indicates a {\tt{Web Domain}}.  The signs next to each behaviour refer to whether the specific behaviour was employed by the model to classify an individual to a specific demographic group.}
\label{tab:impo_DEMO_MAPP}
\end{sidewaystable*}

\section{Discussion}

The obtained results for moral foundations and human values suggest that online behaviours are potentially informative of the individuals' worldviews and ideals.
The poor to medium prediction scores are probably related to the complexity of the attributes that we are trying to infer; in fact, morals are often expressed in subtle ways in everyday life, and only occasionally in a more intense way under specific circumstances, making them more difficult to assess from digital behaviours with respect to the well-studied topic of personality traits recognition.
Taking as an example the Big-Five model (\citealp{Goldberg1990}) for personality assessment, 
\cite{Kosinski2013}, one of the most influential studies on the topic, obtained Pearson correlations in the range [0.29-0.43] on their regression models for the prediction of Big-Five traits.
A major issue for automatic personality recognition is insufficient benchmark data and a consensus on the reported metrics and labelling strategies, leading to the lack of a direct comparison of the obtained results across the state of the art studies.
Despite that, a common finding that emerge is that the best performing attribute in the state-of-the-art is the extroversion trait commonly associated with out-going people who engage more in social interactions.
These behaviours are in general quantifiable; the assessment of intense face to face interaction and communication patterns is straightforward by a variety of digital data~(\citealp{Vinciarelli2014,Finnerty2016}).
Conversely, fairness as defined in the Moral Foundation Theory, may have many facets and ways to be expressed depending on the event that triggered the behaviour, resulting in complex dependencies between the observed data and the respective attribute value.

Taking a deeper look into the most predictive features of the moral foundations, ranked according to their Gini importances (Table~\ref{tab:impo_MFT_MAPP}), interesting insights emerge.
{\em The Huffington Post} and {\em ThinkProgress} which are usually considered progressive sources of news and information, that perceive the notion of authoritarian displays of power and arbitrary order as negative (\citealp{Mitchel2014}), arise as one of the main indicators of subversion.
Visiting {\tt change.org} acts as an important indicator of concern for equity and others' suffering since it organises around topics meant to invoke social change, especially for those experiencing victim status.
{\tt TheAnimalRescue.com} is evocative of nurturing and caring sentiments, and emerges as one of the main indicators for care.
Individuals who consider themselves as pure can be identified by the frequent use of applications like the {\em Bible} or the {\em Gospels}, and they tend to follow {\em Fox News} rather than {\em The New York Times}.
For the individualist and binding foundations (Table~\ref{tab:impo_MFT_IB}), the websites and applications that emerge as differentiators are strongly reflecting the choice of information sources, for instance {\tt huffingtonpost.com} for the individualists and {\tt foxnews.com} for the binders.
These findings are in line with existing literature studies on news selectivity and mechanisms of political opinion formation (\citealp{Yeo2015}). Moreover, according to a year-long Pew Report regarding political polarisation and the sources people usually get informed from,  people ``with ideological views on the left and right have information streams that are distinct from those of individuals with more mixed political views and very distinct from each other'' (\citealp{Mitchel2014}).

Regarding the demographic attributes' prediction, the ``gender'' is predicted with the highest AUROC score 90\%. The study by \cite{Kosinski2013} reported an AUROC of 93\%, when employing the users' ``Likes'' on the Facebook platform as predictors and is the only one that outperforms our score.
The prediction score for gender is of great importance not because of the high accuracy achieved but because it is the most broadly studied attribute in the literature and one of the few that can be employed assess the validity and potentials of our study with respect to the state-of-the-art literature. 
Table~\ref{tab:demo_predict} reports the findings of the state-of-the-art studies against which we can directly compare our findings since they reported also the AUROC metric on gender, age, parenthood and marital status from various sources of digital data.
For the ``age'' attribute the prediction scores vary according to the number of classes considered; in our study we considered six classes and achieved 71\% which is a good performance. To compare our methodology against the study of~\cite{Malmi2016}, we re-trained the predictive model for the ``age'' attribute as single-class classification;  dividing our sample in approximately two balanced subsets, under and over 49 years old. The AUROC in this case was 86\% which is slightly higher than the performance of~\cite{Malmi2016}.
The ``smoker'' attribute was predicted with a higher score using Facebook likes 73\% in~\cite{Kosinski2013} against 64\% in our case.
It is not straightforward to perform a direct comparison of the remaining demographic attributes with studies close to ours since they often opted for binary classifiers for all labels, while in our scenario we employed multi-class classification.
In general,  multi-class classification is a much harder task; taking the political party prediction as an example, we remark our reported AUROC score of 60\% increases to 70\% if we focus only on the two most populated classes (``Democrats'' and ``Republicans''). The latter score is higher than the AUROC of 63\% reported in~\cite{Youyou2015} for the binary prediction of the political orientation of Facebook users, but lower than the 85\% value in \cite{Kosinski2013}.
Indicatively, considering a binarisation strategy for our multi-class demographic attributes we have the following scores: 
education 63\%, ethnicity 77\%, income 70\%, marital status 65\%, and wealth 71\% (see Supplementary Information 2,
Table 2
 for details regarding the binarisation strategy).

Shifting our attention to the most important indications of demographic attributes; among the patterns for education levels, {\tt Linkedin.com} emerges as the most significant indicator of higher education, and \MakeUppercase{Facebook} as an indicator of college education. Interestingly, the website of the Jehovah's Witnesses Church is an indicator of a lower educational level (high-school); these findings are consistent with studies on educational trends and social media usage, as reported in the Gallup Poll results in the US (\citealp{Newport2011}) and on educational trends and religion (\citealp{Stark1997}).
The top gender indicators {\em Pinterest} and {\em Cartwheel by Target} were also pointed out as representative of gender in another study (\citealp{Malmi2016}). As for the prediction of age, the social network preferences are the most predictive cues, for instance  \MakeUppercase{Snapchat} which is known to be very popular in the age group of $18to24$.
Indicatively, {\em LinkedIn} and {\em adadvisor.net} emerged as the most prevalent predictors of income for the higher income class, while the most predictive applications and websites for the ``exercise'' were related to fitness applications like \MakeUppercase{Fitbit} and  \MakeUppercase{Mapmyrun}. These indicators are cross-referenced by other studies contributing to the validity of our approach; however, their exhaustive interpretation is out of our scope.

Shedding light on the limitations of this study,  as a first point we must acknowledge the nature of our data.
Our web browsing data (of both desktop and mobile) consist of only the higher level domains which are for sure not as informative as the entire URL. For example, from our data, we know that two individuals visited {\em wikipedia.com} and spend the same amount of time reading an article, however, the topics of the articles are unknown to us.
From the quantitative results obtained, we note the performance of the moral foundations' and human values' prediction (AUROC scores between $7\%$ and $17\%$ above the AUROC of a random classifier) is much lower than the average performance for demographic attributes (AUROC scores between $9\%$ and $40\%$ above the AUROC of a random classifier). 
This observation might be due to the fact that moral and human values are often expressed in specific occasions, perhaps not evident in the everyday digital activity, which instead can provide straightforward information clearly depicting a specific demographic attribute such as ``gender''.
Moreover, these attributes are often expressed in verbal rather than non-verbal manner, for instance, direct answers to blog discussions or personal opinions on the Twitter platform (\citealp{Mooijman2018}), rather than the fact that the individual visited a certain blog.
All these elements, together with larger volumes of information in terms of a period of digital observation would benefit the performance of Random Forest classifiers.
This is evident also from the exploitative study presented in Section~\ref{sec:effects} where we showed that the AUROC score increases as we include more active users (users that visited plenty of domains).
Another point that must be acknowledged is that even if our cohort is closely representative to the US census, we could not account for potential and unavoidable self-selection bias that might have occurred during the recruitment phase. 
Moreover, even though the data collection period was quite long, we can not account for potential changes in the way our participants navigate the web or use the applications due to the fact that they knew their activity was scrutinised.

Overall, the proposed approach gives insights into the population's behaviour, culture and preferences, providing in-depth insights in the internal decision-making steps of the inference procedure embedded in the predictive models.
Exploiting the effect of the quality, nature and/or quantity of the initial data, we aimed at pointing out the importance of understanding how algorithmic and/or data collection choices may impact the findings; issues that are often overlooked.
Understandably, there are many possible sources of biases to account for, however, it is important to raise awareness and ensure the transparency of the processes, especially when the outcome of these models is to serve as evidence for decision-making in another field or influence the policymaking.

\begin{table}[!h]
\centering
\begin{tabular}{ll}
\toprule
\rowcolor{light-gray} \small{Related Study} &\small{Gender (AUC)}\\
\midrule
Facebook Likes \citealp{Kosinski2013} & 93\% \\
Search Queries \citealp{Bi2013}  & 80\% \\
Client web browsing history \citealp{Goel2012} & 85\% \\
Apps - Category and Content \citealp{Seneviratne2015} & 74\% \\  
Location check-ins \citealp{Zhong2015} & 85-86\%\\
User Applications \citealp{Malmi2016} & 90\%\\
Smartphone Call Logs \citealp{Ying2012} & 85\%\\
Smartphone Call Logs \citealp{Felbo2015} & 79.7\%\\
Social Networks \citealp{Dong2014} & 80\%\\
Web browsing \citealp{Hu2007} & 50\%\\
\midrule
\rowcolor{light-gray} \small{Related Study} &\small{Marital Status (AUC)}\\
User Applications \citealp{Malmi2016} & 79\%\\
Smartphone Call Logs \citealp{Ying2012} & 79\%\\
\midrule
\rowcolor{light-gray} \small{Related Study} &\small{Parenthood (AUC)}\\
User Applications \citealp{Malmi2016} & 68\%\\
\midrule
\rowcolor{light-gray} \small{Related Study} &\small{Age (AUC)}\\
Web browsing \citealp{Hu2007} & 50\% (5 classes)\\
Smartphone Call Logs \citealp{Ying2012} & 77\% (2 classes)\\
Social Networks \citealp{Dong2014} & 73\% (3 classes)\\
Smartphone Call Logs \citealp{Felbo2015} & 63\%(3 classes)\\
User Applications \citealp{Malmi2016} & 85\% (2 classes)\\
\bottomrule
\end{tabular}
\caption{Related studies on demographic attribute prediction which reported their results in terms of AUC metric which allows for a direct comparison with our results.}
\label{tab:demo_predict}
\end{table}

\section{Conclusions}

Exploiting digital trails of human behaviour has found numerous applications in the contexts of learning analytics~(\citealp{gray2014review, nistor2016newcomer}), measurement-based care~(\citealp{scott2015using}), the promotion of well-being~(\citealp{luhmann2017using}) and crime prevention~(\citealp{almagor2014people}).
We assessed the predictive power of low level digital behavioural sequences on complex psychological attributes like moral traits, human values and a series of advanced demographic attributes through a cross-validated machine learning classification framework.
Previous work showed the possibility of inferring demographic and psychometric attributes but most studies are based on platform-specific digital information and without a possibility to obtain ground-truth.
In the present study, the cohort engaged is a sample of the US population, closely representative to the US Census (see subsection~\ref{sec:Repr}) with respect to major demographic variables, and not affiliated to a specific web platform or application.
For this reason, our design avoids many cultural and demographic biases inherent to the users of specific platforms (\citealp{Golder2014}). 
Since our recruited sample is not tied to a specific social network, it is not directly subject to algorithmic manipulation and exposure to content (\citealp{Kramer2014}).
Moreover, its validation was based on self-reported information provided through a concrete survey designed for the scopes of this study.

Our findings suggest that digital traces can be informative of the demographics and can be used to sketch a portrait of emerging cyber-cultures.
Shifting to morals and human values the task becomes much more complex and the prediction accuracy drops significantly,
calling for further investigation. 
These findings are result of an exploitation of different digital data sources and modalities (web browsing,  mobile browsing, application usage) which
allowed us to perform a comparative study on the predictive power of each modality as well as their combination. 
Overall, deviations for performance between models inferencing from web browsing and/or smartphone usage were found to be minimal for all attributes.

Undoubtedly machine learning is an essential tool for understanding patterns in human culture and behaviour, exposing stereotypes inherent in our everyday lives, sometimes uncomfortable to acknowledge without such hard evidence.
This study highlights the possibility of learning complex demographic and psychological attributes by assessing web browsing data and/or mobile usage while evidencing the extent to which certain attributes can be predicted.
Providing a balanced perspective of the risks associated to readily available data like top level domains of the web browsing history, 
our quantitative results contribute to pointing out the ease in predicting demographic attributes including gender, and ethnicity from all available data sources. 
Further research is required in this direction for a deeper understanding of bias embedded in the predictive models and decision-making algorithms which may lead to involuntary discrimination (\citealp{Van2016,Zliobaite2015,Dwork2012,Edelman2017}).

In spite of this, the psychometric attribute prediction case is slightly different;  accurate prediction of complex psychometric attributes
was still far from fully disclosing psychological profiles calling for further investigation.
With prediction performances ranging from 60\% to 70\%, we highlight both the potentials and the limitations of such approaches, providing with a realistic dimension of the possibilities for personalised web services as well as the privacy and surveillance concerns that keep raising.
Considering that smartphones are among the most widely deployed technologies of human history, the capability of automatically predicting demographic attributes and human/moral values from multi-modal passively collected data is potentially a key enabler for delivering better targeted and more effective interventions at the population scale or nowcasting important social issues like poverty.






\section{Competing interests}

The authors declare that they have no competing interests.

%

\end{document}